\documentclass{aa}  

\usepackage{graphicx}
\usepackage{txfonts}
\usepackage{orcidlink}
\usepackage{booktabs}
\usepackage{upgreek}
\usepackage{threeparttable}
\usepackage[T1]{fontenc}
\usepackage{comment}
\DeclareRobustCommand{\VAN}[3]{#2}
\let\VANthebibliography\thebibliography
\def\thebibliography{\DeclareRobustCommand{\VAN}[3]{##3}\VANthebibliography}
\usepackage{graphicx}	
\usepackage{amsmath}	
\usepackage{tablefootnote}
\usepackage{url}
\bibpunct{(}{)}{;}{a}{}{,} 
\usepackage{txfonts}
\usepackage{lscape}
\usepackage{rotating}
\usepackage{hyperref}
\hypersetup{colorlinks=true,linkcolor=blue,citecolor=blue,filecolor=blue}

\begin{document}

\title{CHILLING: Continuum Halos in LVHIS Local Irregular Nearby Galaxies}
\subtitle{Radio continuum spectral behavior of dwarf galaxies}
\titlerunning{CHILLING - RC spectral behavior of dwarf galaxies}

   \author{Sam Taziaux \inst{\ref{rub},\ref{rapp}, \ref{csiro}}\orcidlink{0009-0001-6908-2433}
          \and
          Megan C. Johnson \inst{\ref{NSF}}\orcidlink{0000-0003-2030-3394}
          \and
           Onic I. Shuvo \inst{\ref{baltimore}}\orcidlink{0000-0003-4727-2209}
          \and 
         Dominik J. Bomans \inst{\ref{rub}, \ref{rapp}}\orcidlink{0000-0001-5126-5365}
         \and
          Christopher J. Riseley \inst{\ref{rub}, \ref{rapp}}\orcidlink{0000-0002-3369-1085}
          \and
         Timothy J. Galvin \inst{\ref{csiro}}\orcidlink{0000-0002-2801-766X}
         \and 
           Alec J. M. Thomson \inst{ \ref{ska}}\orcidlink{0000-0001-9472-041X}
          \and
          Peter Kamphuis \inst{\ref{rub}}\orcidlink{0000-0002-5425-6074}
          \and
          Amy Kimball \inst{\ref{NRAO}}\orcidlink{}
          \and
          Amanda Kepley \inst{\ref{charlotteville}}\orcidlink{0000-0002-3227-4917}
          \and 
          Michael Stein \inst{\ref{rub}}\orcidlink{0000-0001-8428-7085}
          \and
           George H. Heald \inst{\ref{ska}}\orcidlink{0000-0002-2155-6054}
          \and
          Nicholas Seymour \inst{\ref{curtin}}\orcidlink{}
          \and 
         Joe A. Grundy \inst{\ref{csiro}, \ref{curtin}}\orcidlink{}
          \and 
          Björn Adebahr \inst{\ref{rub}}\orcidlink{0000-0002-5447-6878}
            \and
          Ralf-Jürgen Dettmar \inst{\ref{rub}, \ref{rapp}}\orcidlink{0000-0001-8206-5956}
          }

   \institute{Ruhr University Bochum, Faculty of Physics and Astronomy, Astronomical Institute (AIRUB), Universitätsstraße 150, 44801 Bochum, Germany
              \email{sam.taziaux@rub.de} \label{rub}
        \and 
             Ruhr Astroparticle and Plasma Physics Center (RAPP Center) \label{rapp}
        \and 
            CSIRO Space and Astronomy, PO Box 1130, Bentley WA 6102, Australia \label{csiro}
        \and 
        National Science Foundation, 2415 Eisenhower Avenue, Alexandria, VA 22314, USA \label{NSF}
        \and 
            Department of Physics, University of Maryland Baltimore County, 1000 Hilltop Circle, Baltimore, MD 21250, USA \label{baltimore}
        \and
            SKA Observatory, SKA-Low Science Operations Centre, 26 Dick Perry Avenue, Kensington WA 6151, Australia \label{ska}
        \and
            National Radio Astronomy Observatory, 1011 Lopezville Road, Socorro, NM 87801, USA \label{NRAO}
        \and
            National Radio Astronomy Observatory, 520   Edgemont Road, Charlottesville, VA 22903, USA \label{charlotteville}
        \and 
        International Centre for Radio Astronomy Research, Curtin University, Bentley, WA, Australia \label{curtin}
            }

\date{Accepted XXX. Received YYY; in original form ZZZ}



\abstract
{Dwarf galaxies, due to their shallow gravitational potentials, provide critical environments for studying feedback mechanisms from star formation and its impacts on dwarf galaxy evolution. In particular, radio continuum (RC) observations offer valuable insights into cosmic ray dynamics, which play a significant role in shaping these processes. }
{This study investigates the detectability and spectral characteristics of RC emission in a sample of 15 dwarf galaxies (11 gas-rich, star forming dwarfs and 4 blue compact dwarfs) spanning a broad range of stellar masses and star formation histories.}
{Using multi-band RC data (L/S-, C-, and X-band) from the Australia Telescope Compact Array, we analyse the physical conditions responsible for RC emission and explore the dominant emission mechanisms within these systems. }
{RC emission is detected in 11 out of the 15 galaxies. Our results indicate that RC emission correlates strongly with star formation rate, far-infrared, and stellar mass, while dynamic parameters such as H{\sc i} and rotational velocity exhibit no significant correlation with RC detectability.
Spectral analysis reveals that the RC spectral energy distribution in these galaxies frequently deviate from a simple power-law behavior, instead displaying curvature that suggests more complex underlying physical processes. Statistical model comparison confirms that a single power-law model is inadequate to capture the observed spectral shapes, emphasising the necessity of more sophisticated approaches.
Additionally, the observed radio–far-infrared correlation indicates that cosmic ray electrons in lower-mass dwarf galaxies cool more rapidly than they can escape (e.g. via galactic winds), resulting in a measurable RC deficit.}
{}

   \maketitle
%




\section{Introduction}
Galactic winds and outflows\footnote{In this article we adhere to the definition of \citet[][]{Veilleux_2005}, who defines an "outflow" as  any material moving away from a central object, whereas "wind" specifically describes gas expelled from a galaxy's potential, typically driven by mechanisms
such as radiation pressure and incorporating both diffuse flows and jets.} are crucial in shaping the evolutionary histories of galaxies. These processes, driven by the energy and momentum from star formation, influence the interstellar medium (ISM) by enhancing turbulence and providing additional pressure support. This additional pressure can counteract gravitational collapse, thereby regulating further star formation. Moreover, outflows can expel gas and metals from the ISM, potentially enriching the circumgalactic medium or even escaping into the intergalactic medium if the outflow velocity exceeds the escape velocity of the host galaxy \citep[][]{Veilleux_2005}. 

Dwarf galaxies serve as essential laboratories for investigating feedback mechanisms from star formation, particularly the role of star formation in driving winds and outflows. Due to their shallow gravitational potentials, even moderate starburst events can generate significant winds \citep[e.g.][]{chyzy_magnetized_2016} or outflows \citep[e.g.][]{Devine_1999,Strickland_2009, Adebahr_2013, Dirks_2023}. Understanding these feedback mechanisms is vital to comprehending the interplay between star formation and dwarf galaxy evolution \citep[e.g.][]{Dekel_1986,Stevens_2002,chyzy_magnetized_2016}.
Most studies on dwarf galaxies in the literature have focused on their integrated radio continuum (RC) properties \citep[e.g.][]{klein_2018}. However, resolved RC observations of these galaxies enable a better understanding of the evolution and outflow characteristics of these low-mass galaxies. Only a few studies have utilised resolved RC and spectral index observations to analyse dwarf galaxies in detail \citep[e.g.][]{chyzy_regular_2000, kepley_role_2010, Basu_2017, Kepley_2011, hindson_radio_2018, taziaux_2025}. These RC observations have shown a mean non-thermal spectral index ($S\propto \nu^{\alpha_{\rm nth}}$) of $-0.6$ that steepens at the galaxy outskirts to $-1.1$ \citep[e.g.][]{chyzy_regular_2000, kepley_role_2010, Kepley_2011, westcott_spatially_2018, taziaux_2025}. In the resolved starforming knots, the spectral index can reach a flatter spectrum of $-0.3$ or even be inverted \citep[e.g.][]{Basu_2017, taziaux_2025}, depending on the optical depth of the galaxy region.

Radio observations also provide a crucial tool for studying cosmic ray (CR) transport and the dominant energy loss processes (synchrotron, inverse-Compton or bremsstrahlung), with relativistic electrons serving as tracers \citep[e.g.][]{Lacki_2010, Werhahn_2021, Pfrommer_2022, heesen_nearby_2022}. CRs propagate from supernova remnants into the galactic halo through advection, diffusion, and streaming \citep[e.g.][]{Strong_2007, Stein_2019, Thomas_2020}. In this process, CRs excite Alfvén and whistler waves via resonant plasma instabilities, which scatter CRs and regulate their drift speed along the magnetic field \citep{Kulsrud_1969, Shalaby_2021, Shalaby_2023, Lemmerz_2024}. Through resonant scattering, CRs transfer momentum to the thermal gas, exerting pressure on the ambient plasma and driving galactic winds, a key aspect of CR hydrodynamic models \citep{Zweibel_2013, Pfrommer_2017, Thomas_2019}.

Hydrodynamic simulations indicate that CR-driven winds significantly influence mass-loss rates, gas distribution, and wind formation \citep[e.g.][]{Breitschwerdt_1991, Uhlig_2012, Pakmor_2016, Girichidis_2016, Recchia_2017, Dashyan_2020, Thomas_2023}. \citet{Thomas_2024} demonstrated that CRs, due to their long cooling times and strong plasma coupling, drive denser winds with higher mass-loading factors, efficiently redistributing gas into the halo and regulating star formation.

This study further investigates the properties of these winds and outflows, particularly their velocity into the galactic halo, as driven by stellar feedback in low metallicity regimes. Because CR protons (CRPs) primarily lose energy through hadronic interactions with dense gas, they predominantly illuminate high-density regions \citep[][]{Pfrommer_2004, Pfrommer_gamma_2017, Werhahn_2023}. In contrast, lower-density outflows are best traced via synchrotron emission from CR electrons (CREs). Modelling CREs alongside CRPs is essential, as their different loss timescales cause deviations in their energy spectra \citep{Ruszkowski_2023}, necessitating a separate spectral treatment for accurate predictions of galactic radio emission \citep{Chiu_2024}.

Beyond their role in regulating outflows, CRs also contribute to the evolution of large-scale ordered magnetic fields in galaxies. Observations and simulations of superbubbles in nearby spiral galaxies suggest that starburst-driven outflows can advect ordered magnetic field lines from the inner disk regions to the outer halo \citep[e.g.][]{kepley_role_2010, Chyzy_2011, Kepley_2011}. This interplay between galactic winds, magnetic field evolution, and cosmic ray transport is particularly relevant for dwarf galaxies, where such effects can be more pronounced due to lower gravitational binding energy.

Despite theoretical and numerical advancements, direct observational evidence of CR-driven winds remains challenging due to the (i) sensitivity limits of current radio facilities, which hinder the detection of faint halo emissions, and (ii) lack of robust modelling frameworks for various CR propagation mechanisms. RC surveys targeting dwarf galaxies are crucial for addressing these challenges and testing theoretical predictions of cosmic ray transport and their role in galactic feedback. 

In this paper, we further explore the detectability of radio emission in dwarf galaxies, the dependence of energy loss mechanisms, and the modelling of cosmic ray transport across a larger sample of dwarf galaxies spanning a wide range of masses and star formation histories.
We describe our sample and the data reduction in Sect.~\ref{dataset}. In Sect.~\ref{detect}, we present detectability of radio emission across various parameter and the basic properties of RC will be shown in Sect.~\ref{radio}. Sect.~\ref{rcfir} will focus on the RC--FIR correlation. We discuss our observations in the context of what is known so far in Sect~\ref{discussion} with a summary and conclusion presented in Sect.~\ref{concl}.

\begin{figure*}
    \centering
    \begin{minipage}[b]{0.24\linewidth}
        \includegraphics[width=\linewidth]{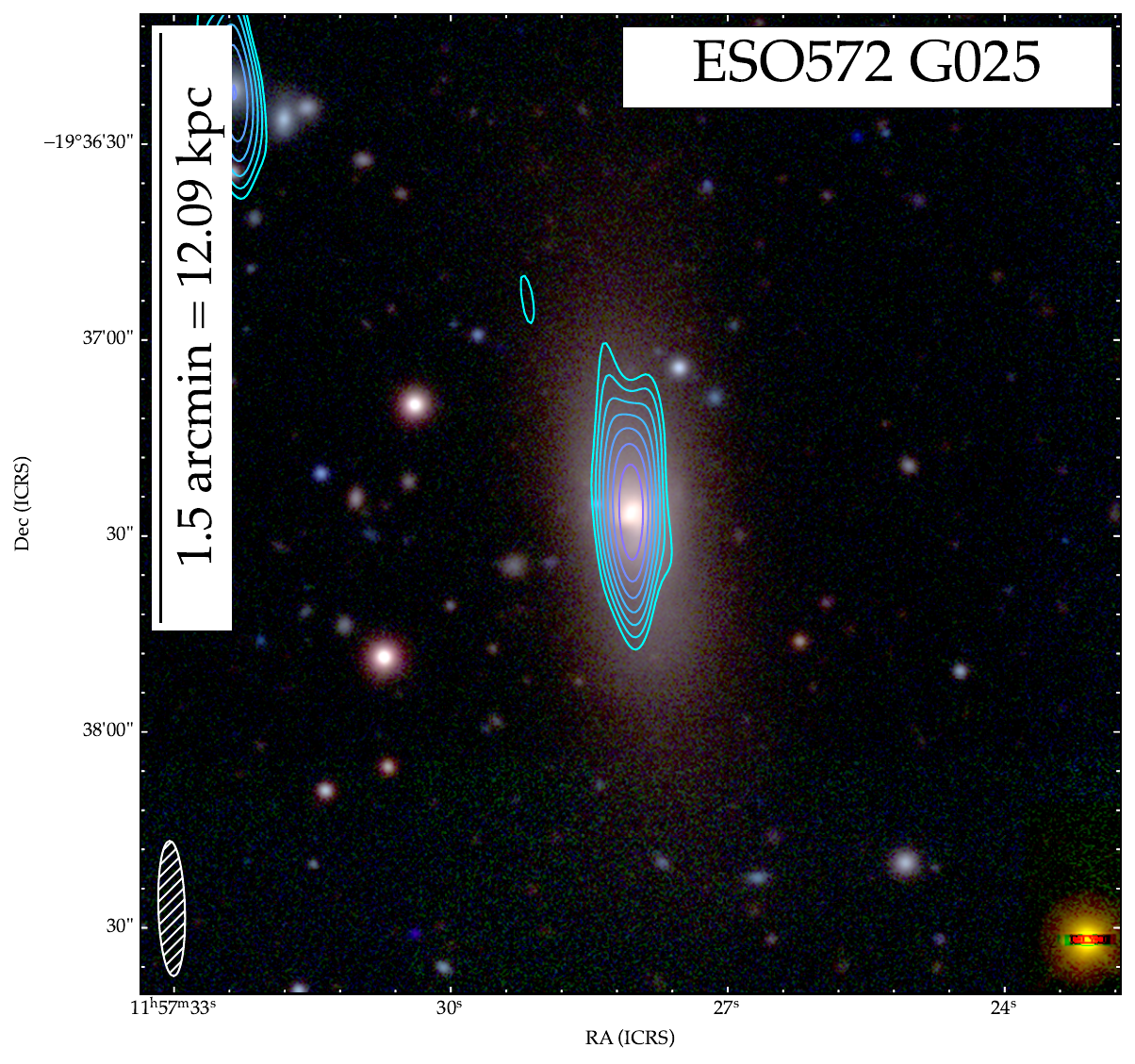}
    \end{minipage}
    \begin{minipage}[b]{0.24\linewidth}
        \includegraphics[width=\linewidth]{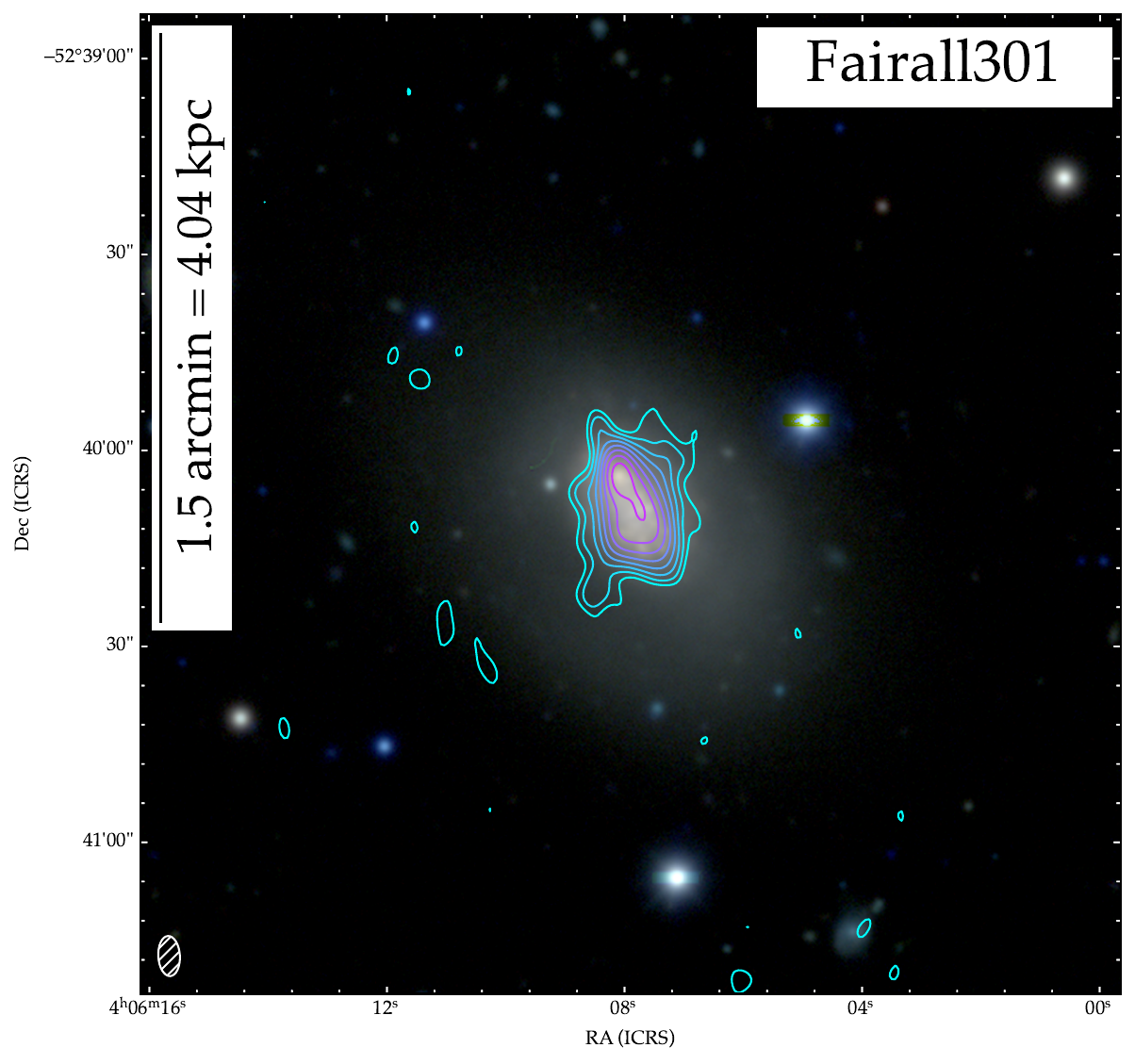}
    \end{minipage}
    \begin{minipage}[b]{0.24\linewidth}
        \includegraphics[width=\linewidth]{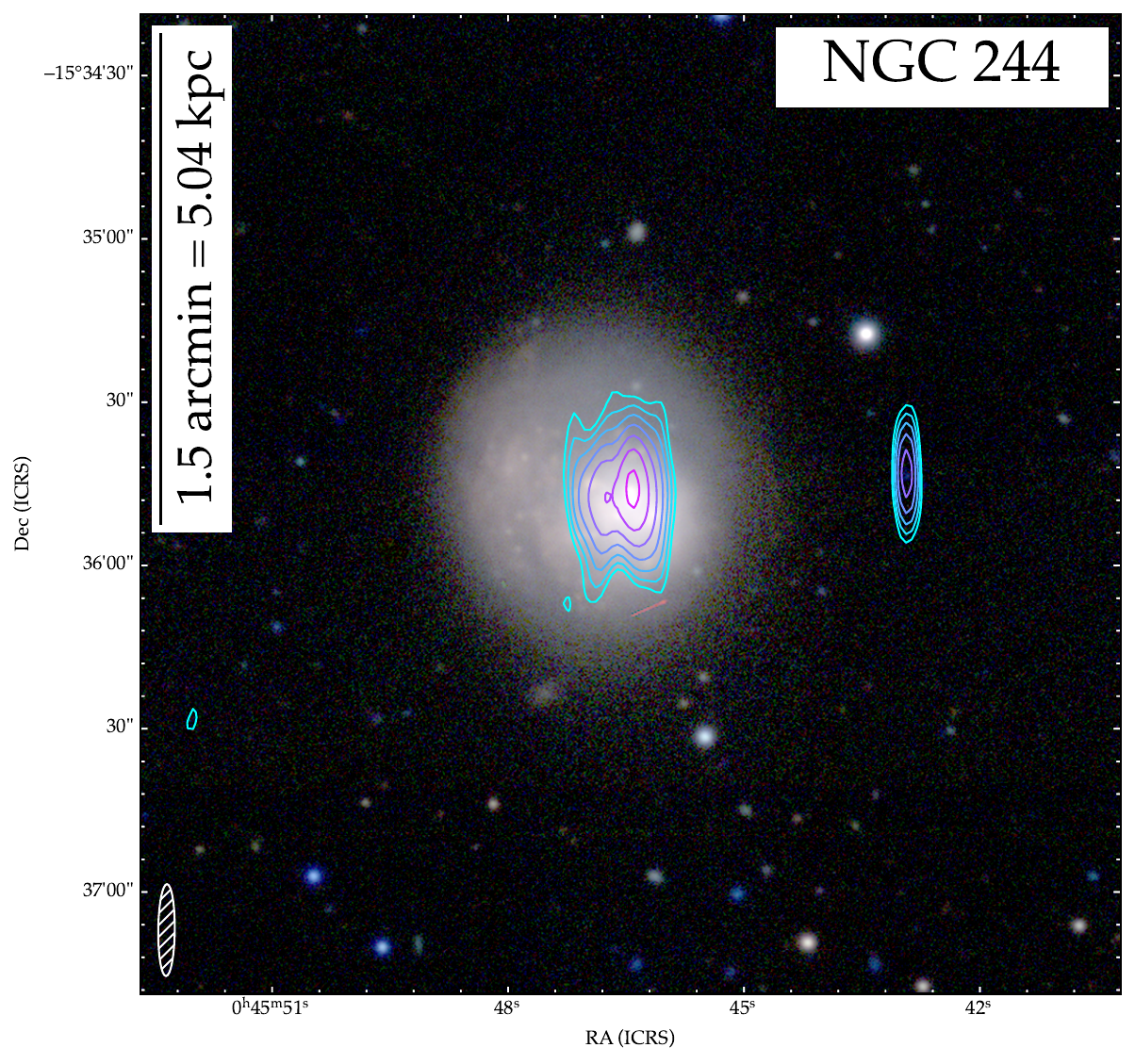}
    \end{minipage}
    \begin{minipage}[b]{0.24\linewidth}
        \includegraphics[width=\linewidth]{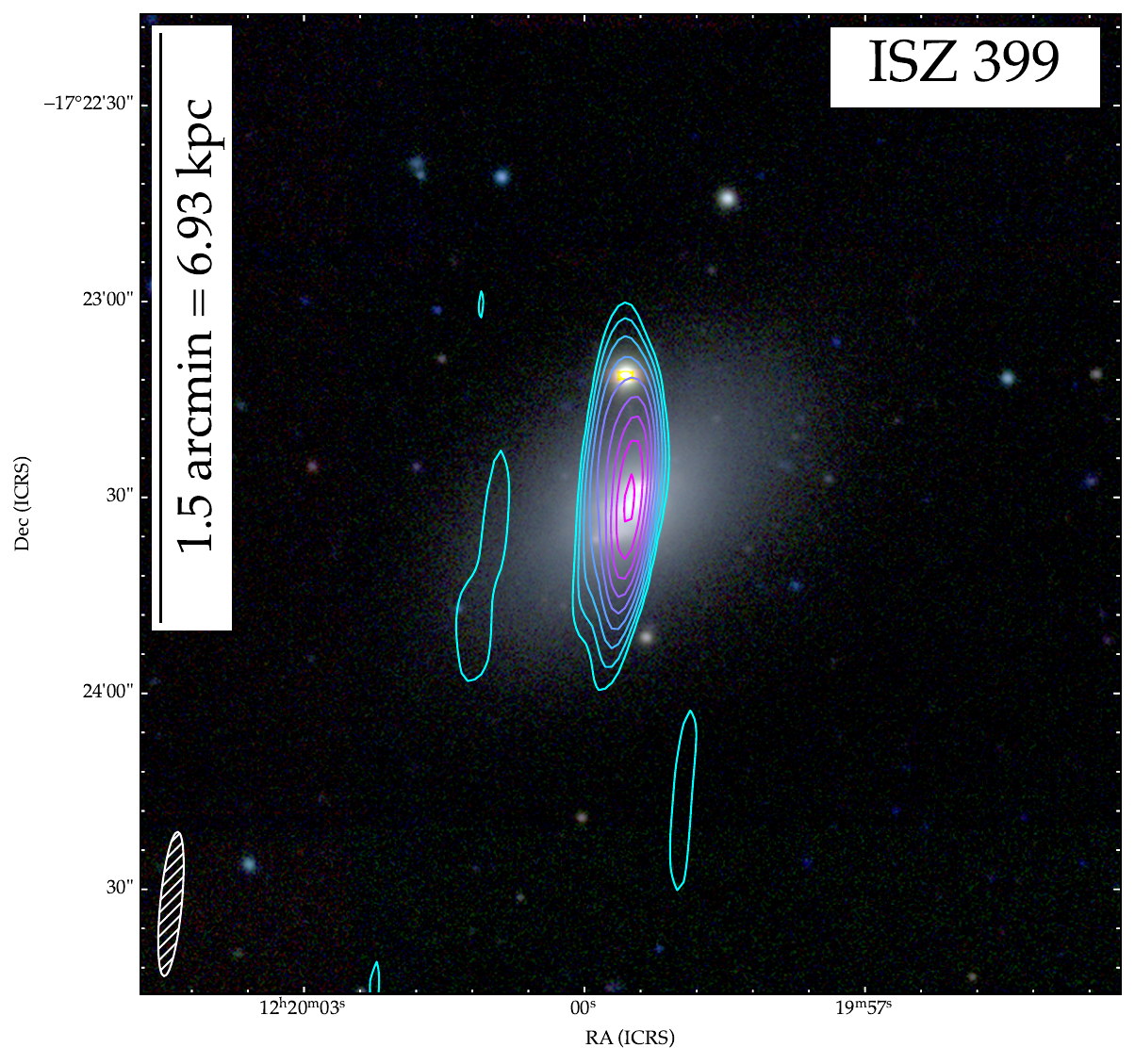}
    \end{minipage}
    \begin{minipage}[b]{0.24\linewidth}
        \includegraphics[width=\linewidth]{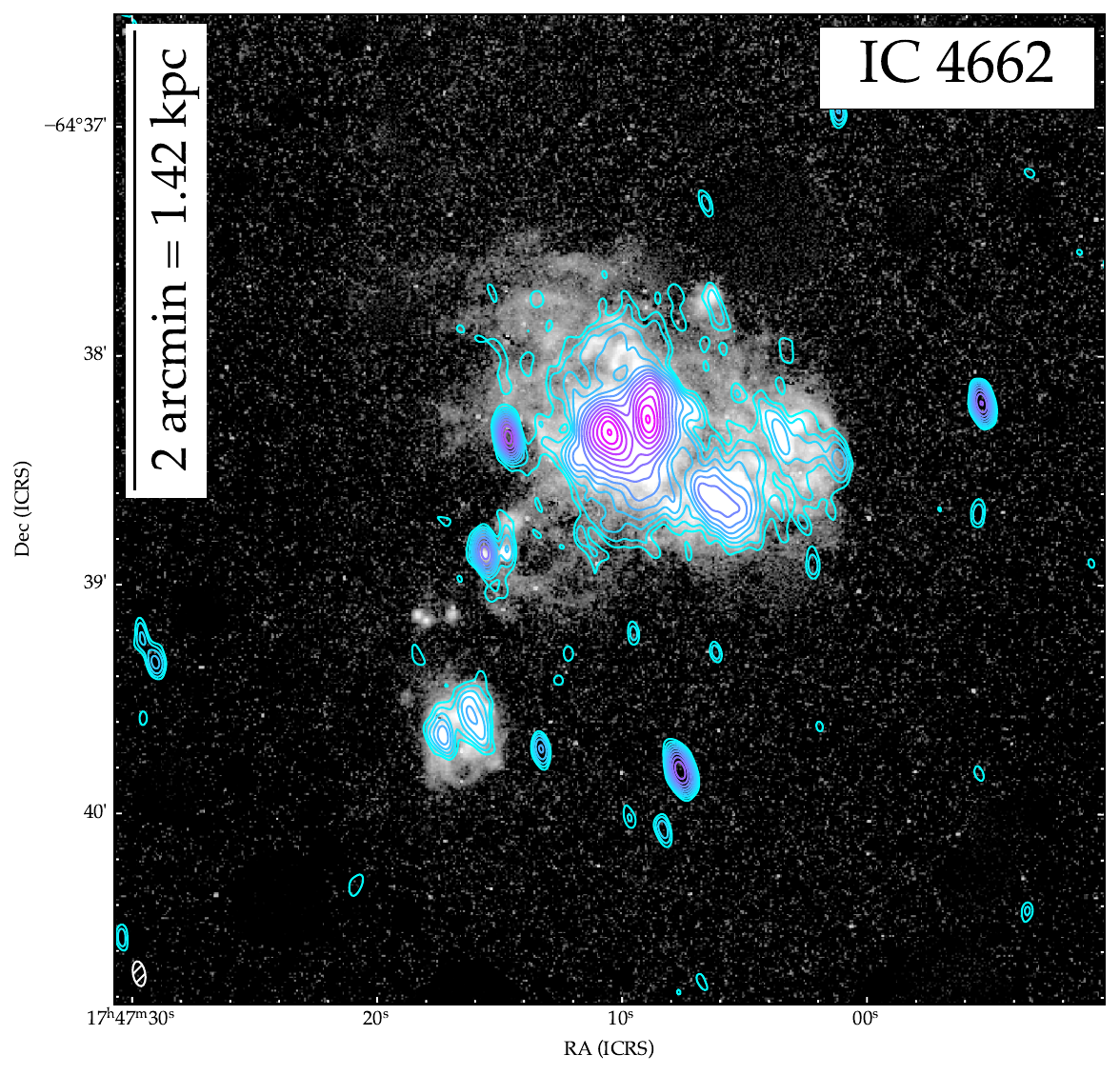}
    \end{minipage}
    \begin{minipage}[b]{0.24\linewidth}
        \includegraphics[width=\linewidth]{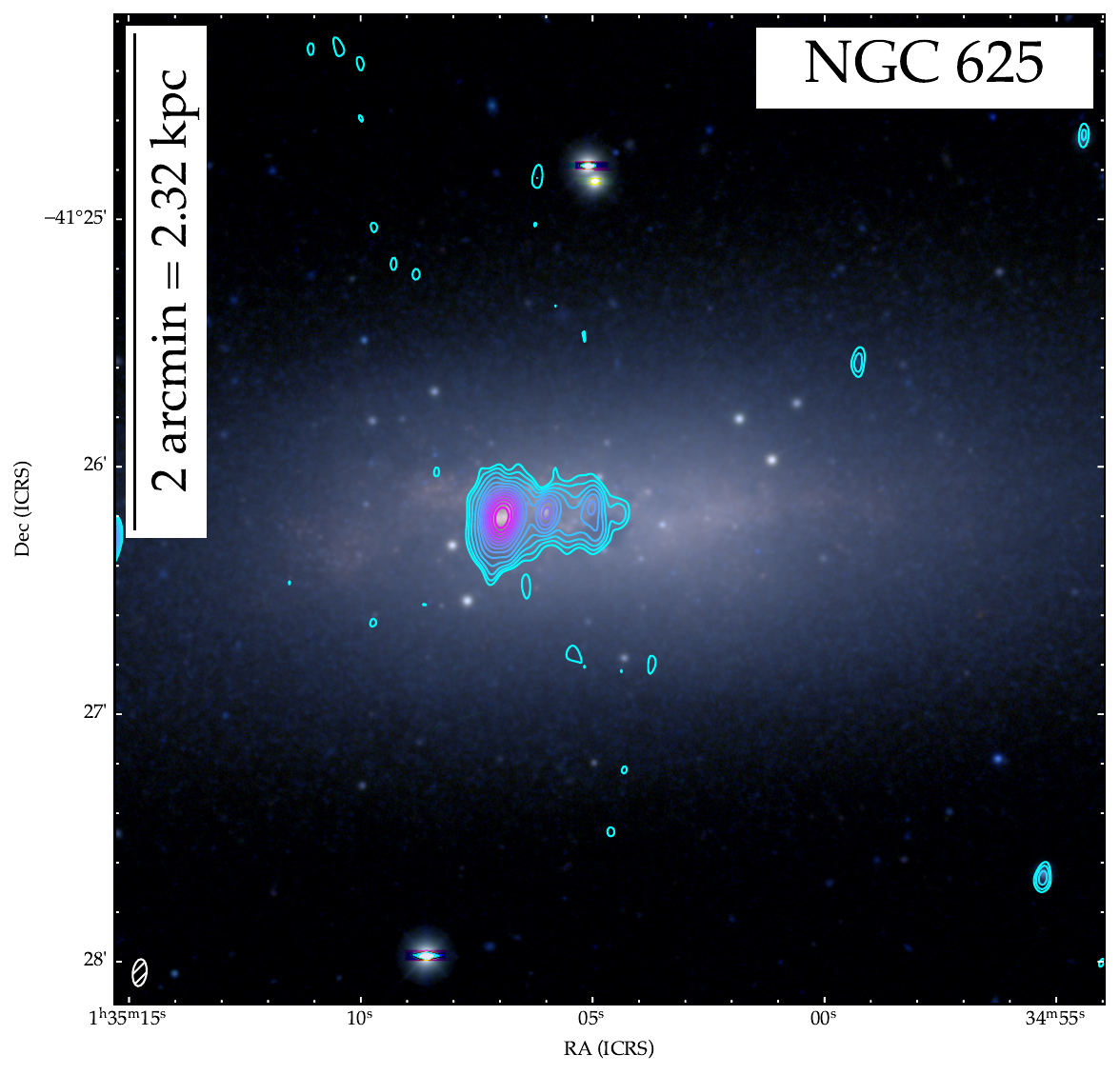}
    \end{minipage}
    \begin{minipage}[b]{0.24\linewidth}
        \includegraphics[width=\linewidth]{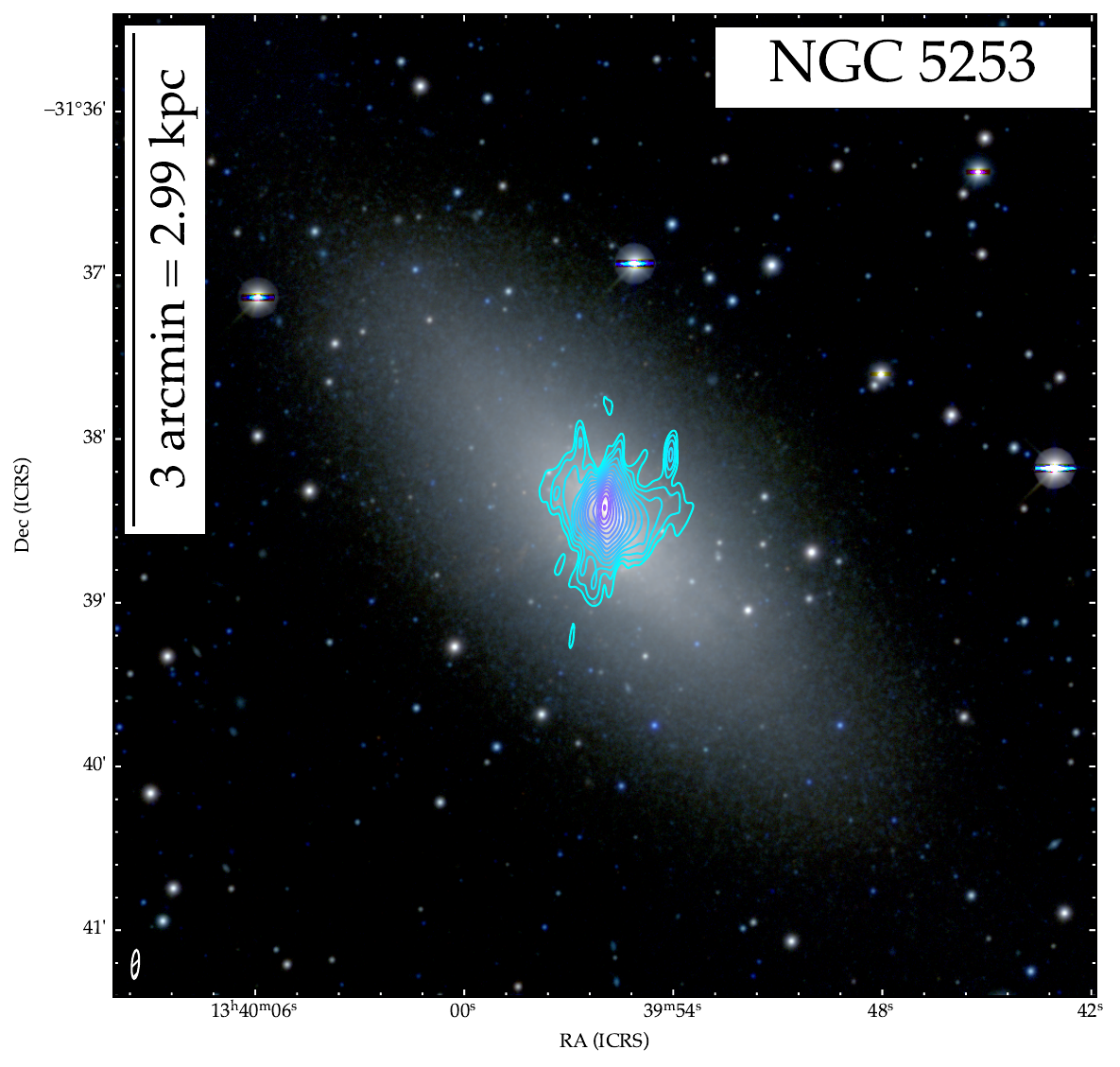}
    \end{minipage}
    \caption{Colour-composite images from the DESI Legacy Imaging Surveys of the RC detected CHILLING sample, showing more diffuse emission, with overlaid ATCA radio emission contours starting at $3\,\sigma$ and increasing by a factor of $\sqrt{2}$ at a central frequency of 2.1\,GHz. 
    For IC\,4662, we use the H$\alpha$ map from \citet[][]{Hunter_2004} as the declination is too low for the DESI Legacy Imaging Surveys.
    The scale in the top left corner is calculated using the distance to the source. The beam is shown in the bottom left corner. The noise level $\sigma_{\rm 2.1\,GHz}$ can be taken from Table~\ref{obs_params}.}
    \label{legacy_diffuse}
\end{figure*}

\begin{figure*}
    \centering
    \begin{minipage}[b]{0.24\linewidth}
        \includegraphics[width=\linewidth]{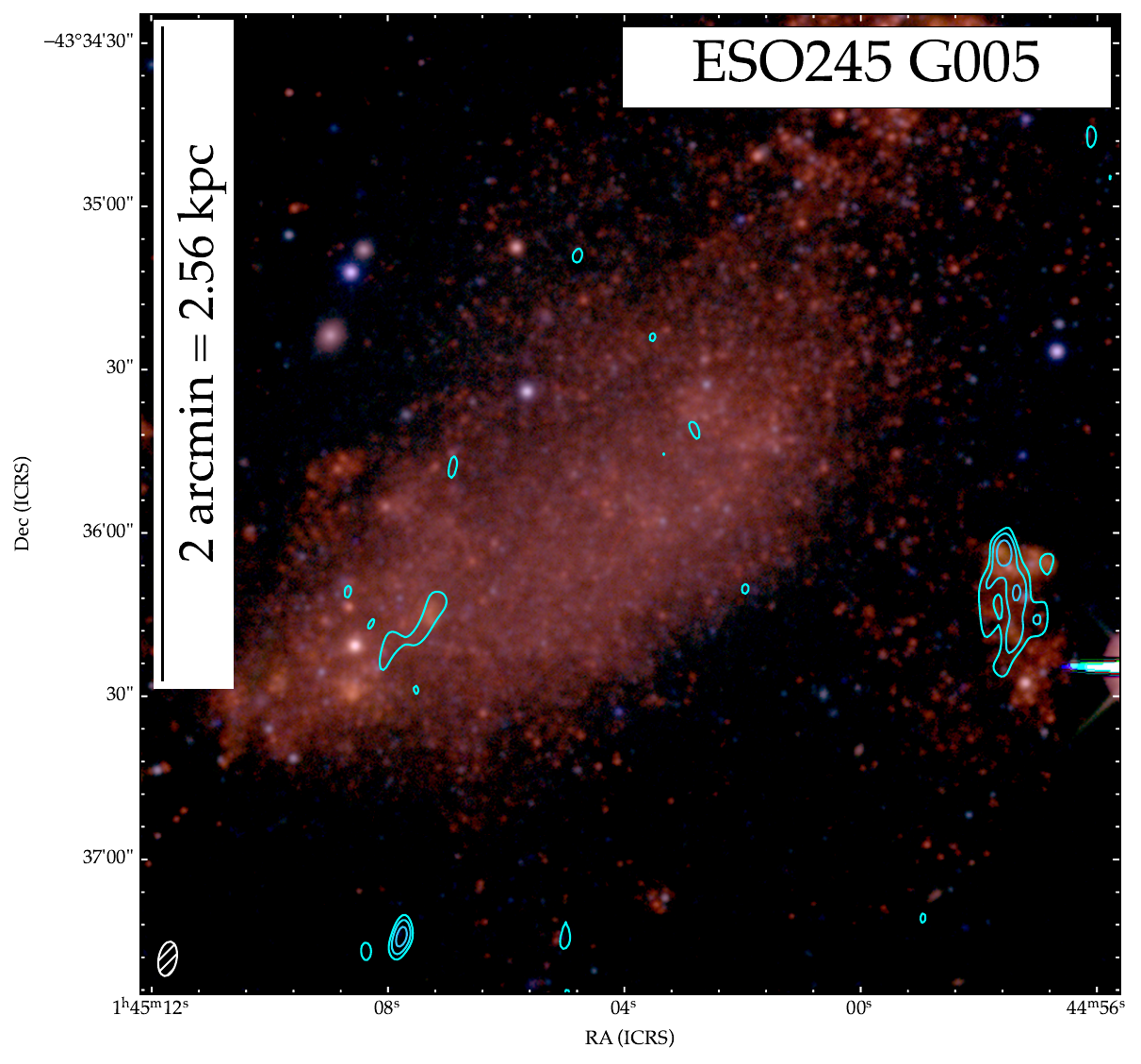}
    \end{minipage}
    \begin{minipage}[b]{0.24\linewidth}
        \includegraphics[width=\linewidth]{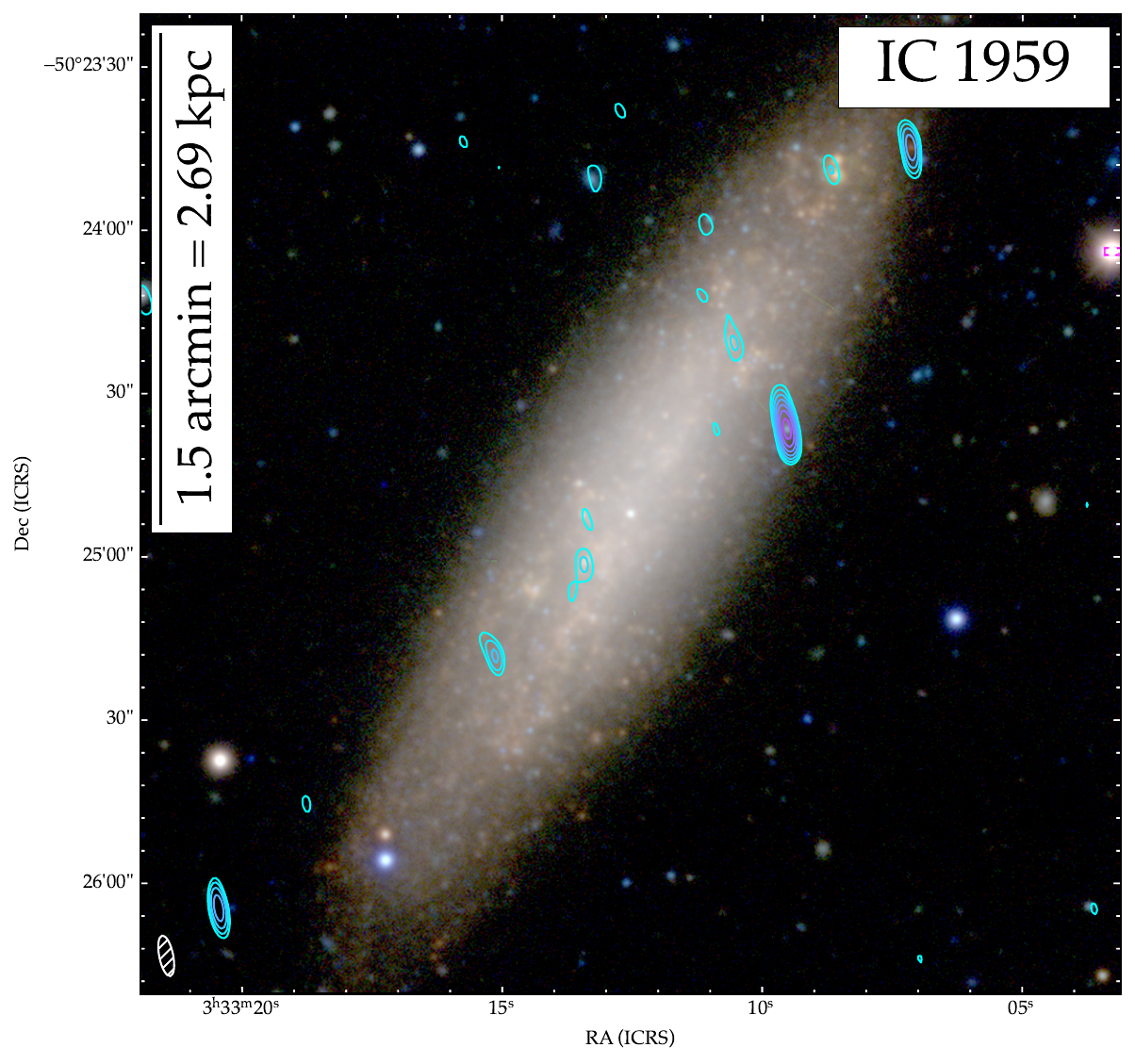}
    \end{minipage}
    \begin{minipage}[b]{0.24\linewidth}
        \includegraphics[width=\linewidth]{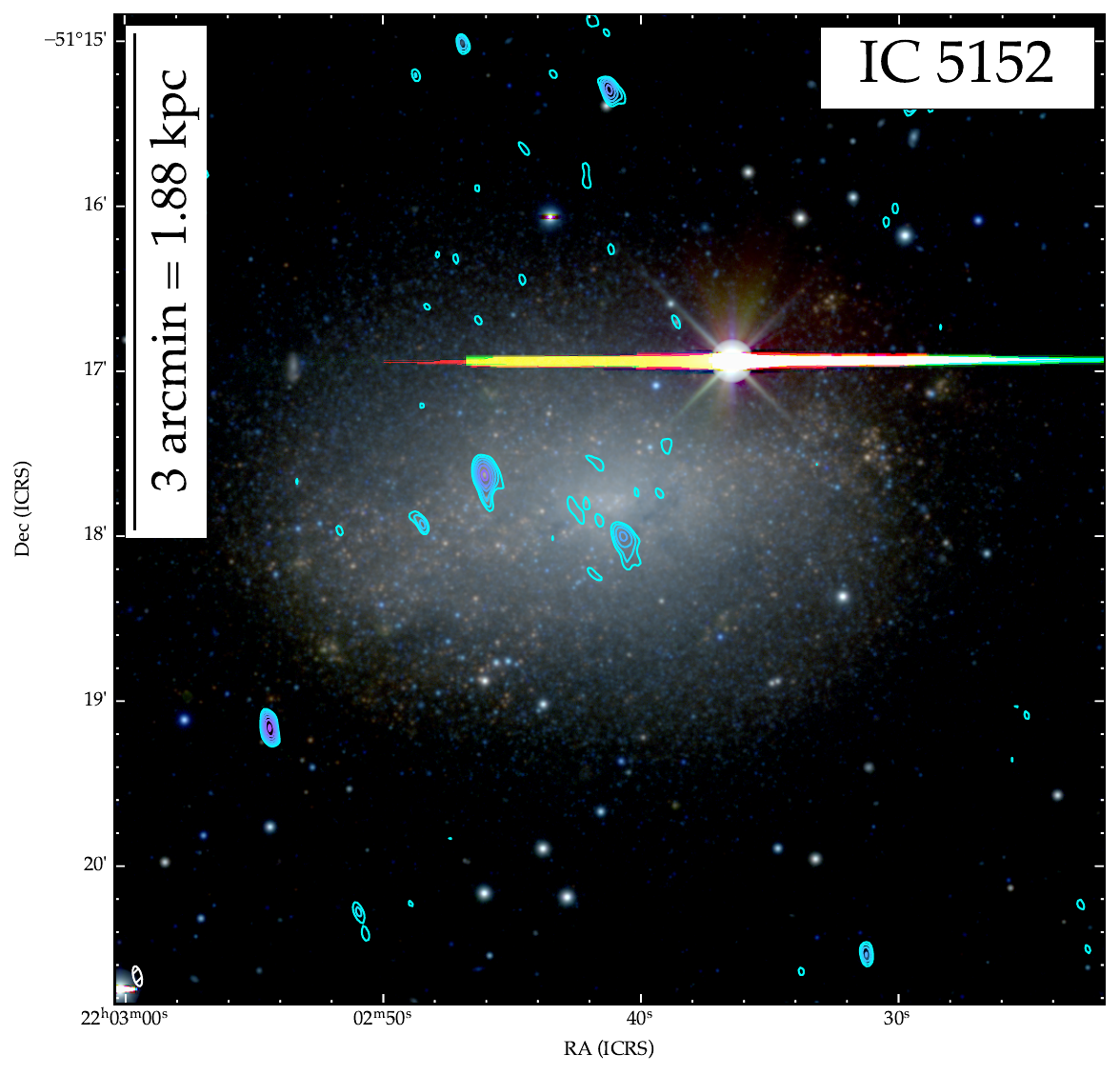}
    \end{minipage}
    \begin{minipage}[b]{0.24\linewidth}
        \includegraphics[width=\linewidth]{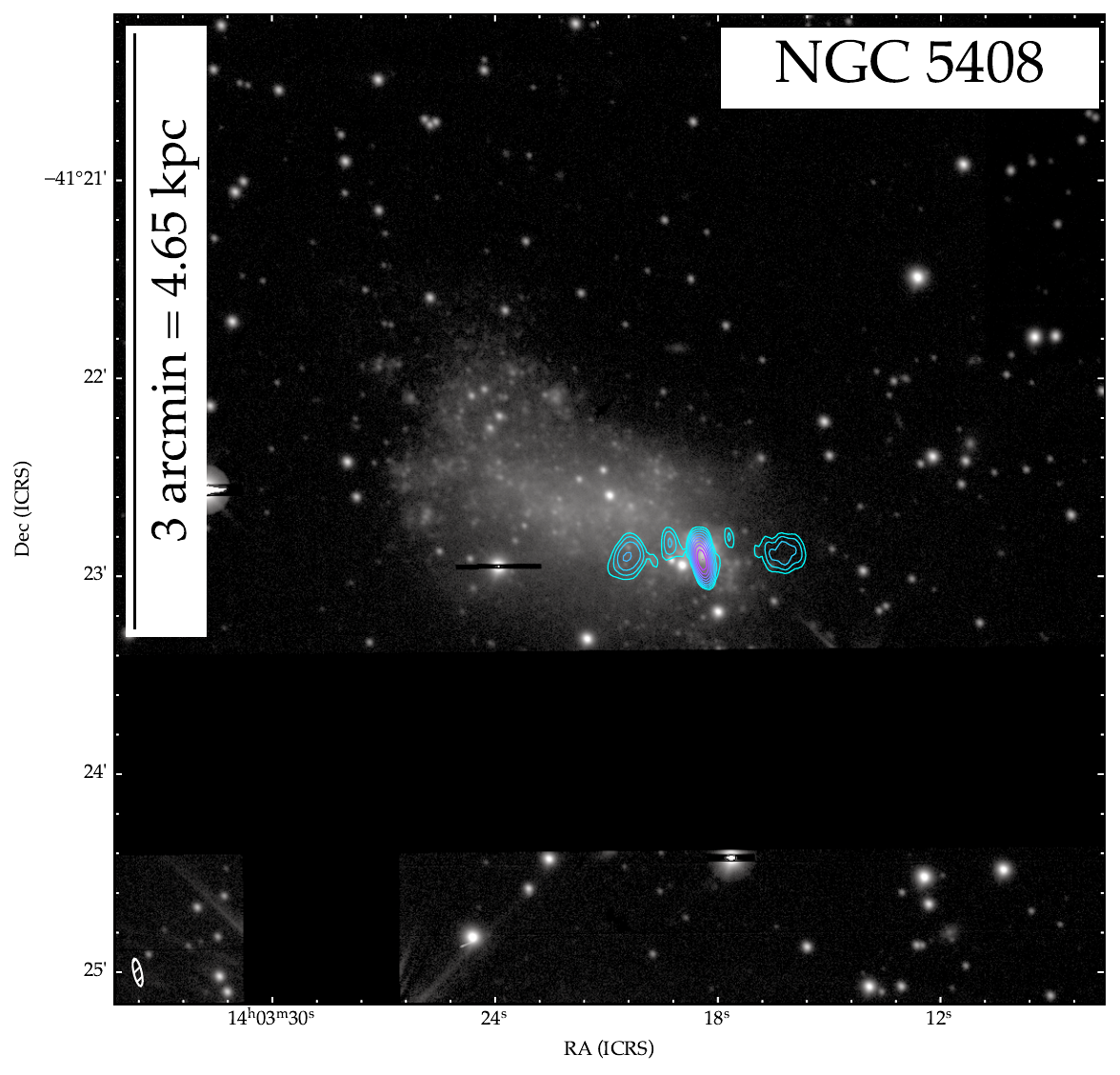}
    \end{minipage}
    \caption{Colour-composite images from the DESI Legacy Imaging Surveys of the RC non-detected CHILLING sample, showing more compact emission, with overlaid 2.1\,GHz ATCA radio emission contours starting at $3\,\sigma$ and increasing by a factor of $\sqrt{2}$. 
    The scale in the top left corner is calculated using the distance to the source. The beam is shown in the bottom left corner. The noise level $\sigma_{\rm 2.1\,GHz}$ can be taken from Table~\ref{obs_params}.}
    \label{legacy_compact}
\end{figure*}

\begin{figure*}
    \centering
    \begin{minipage}[b]{0.24\linewidth}
        \includegraphics[width=\linewidth]{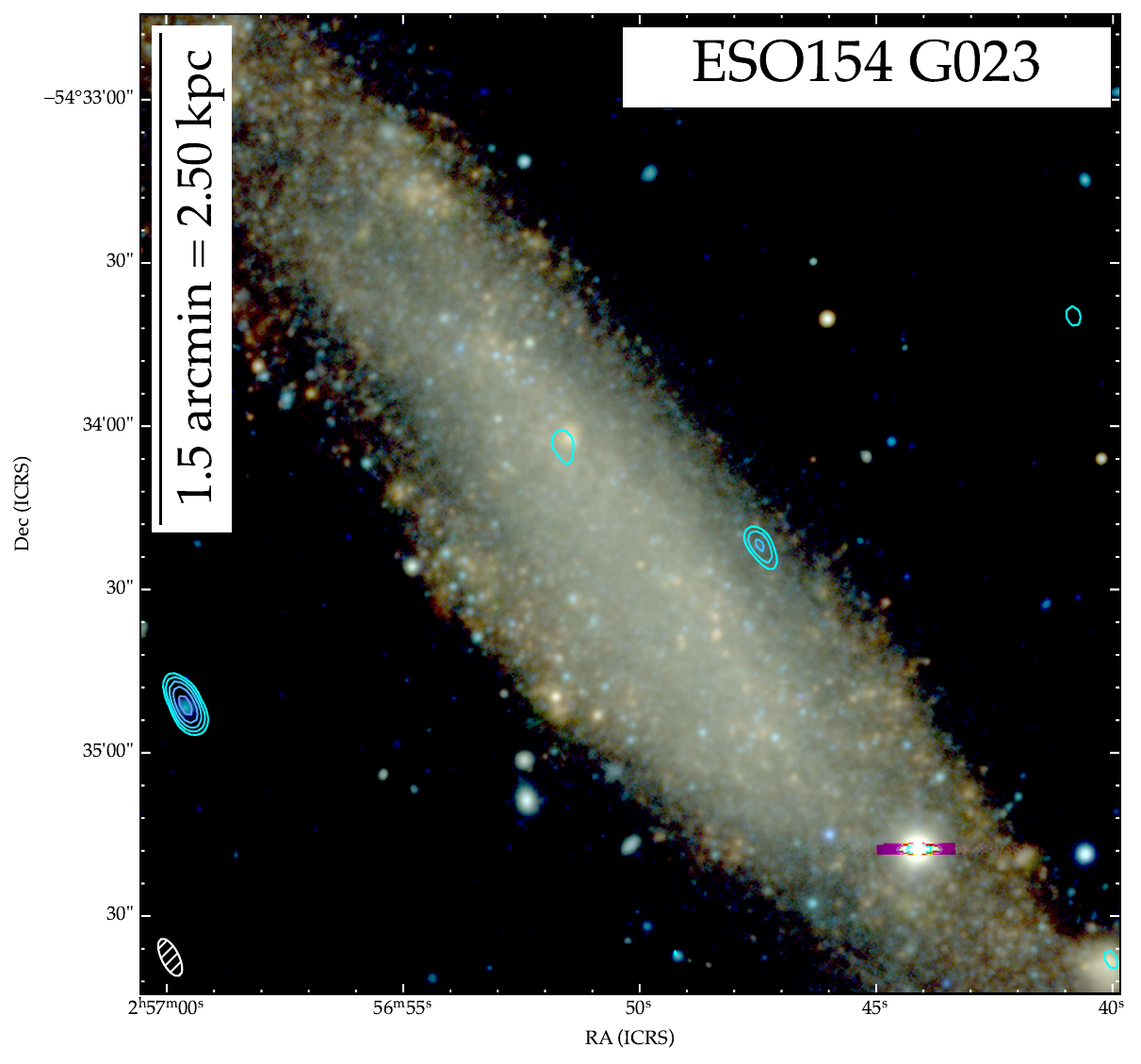}
    \end{minipage}
    \begin{minipage}[b]{0.24\linewidth}
        \includegraphics[width=\linewidth]{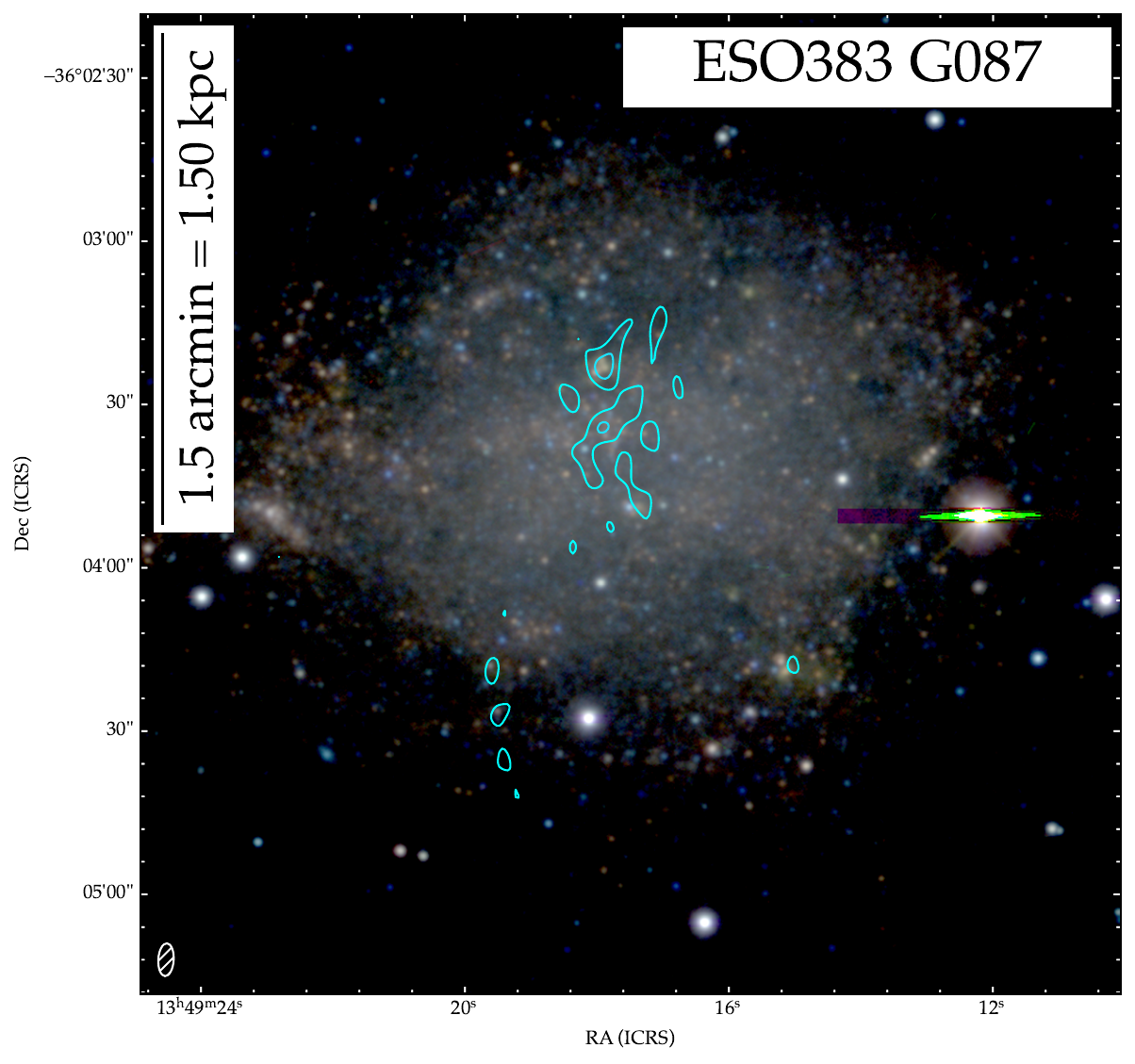}
    \end{minipage}
    \begin{minipage}[b]{0.24\linewidth}
        \includegraphics[width=\linewidth]{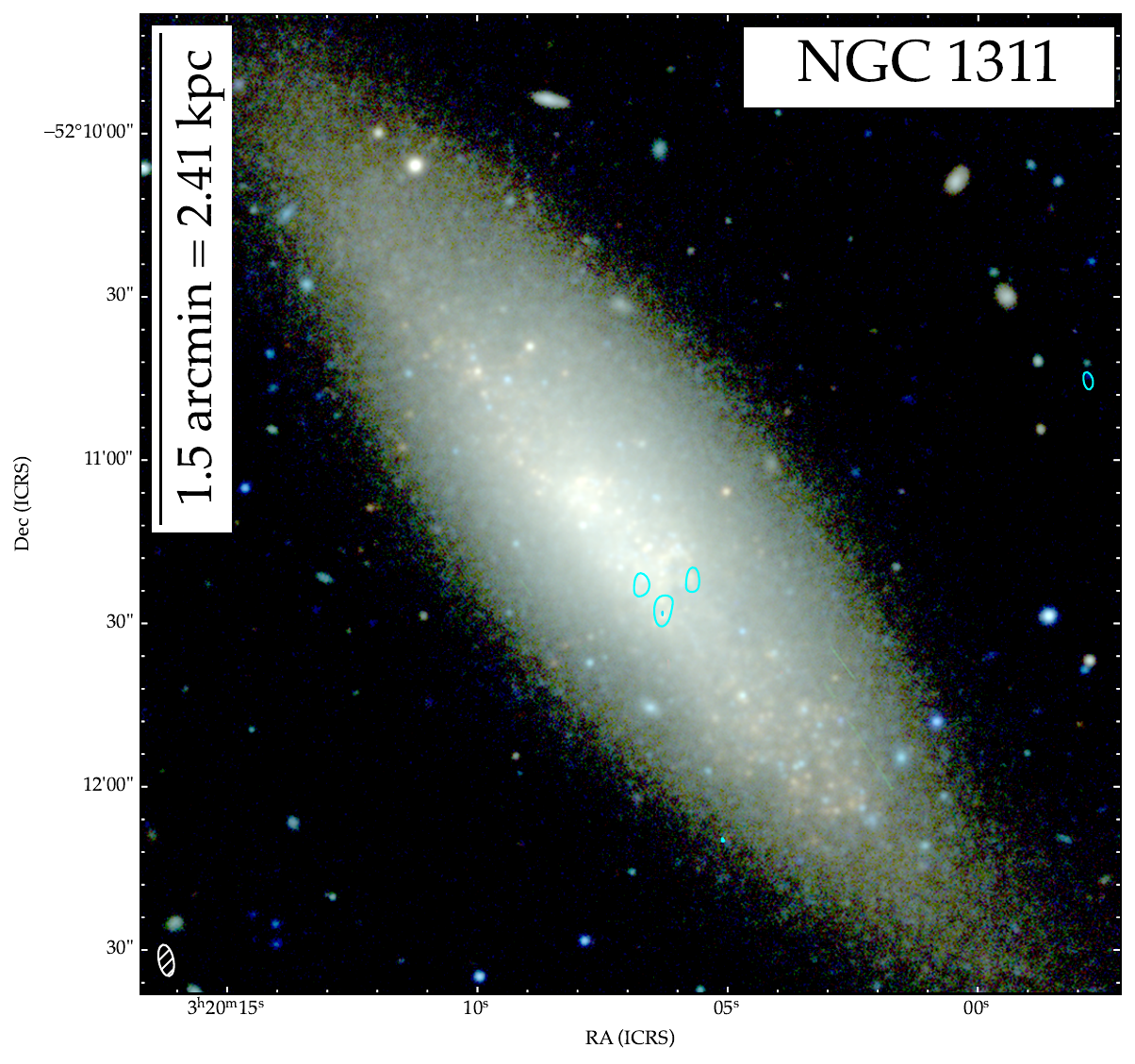}
    \end{minipage}
    \begin{minipage}[b]{0.24\linewidth}
        \includegraphics[width=\linewidth]{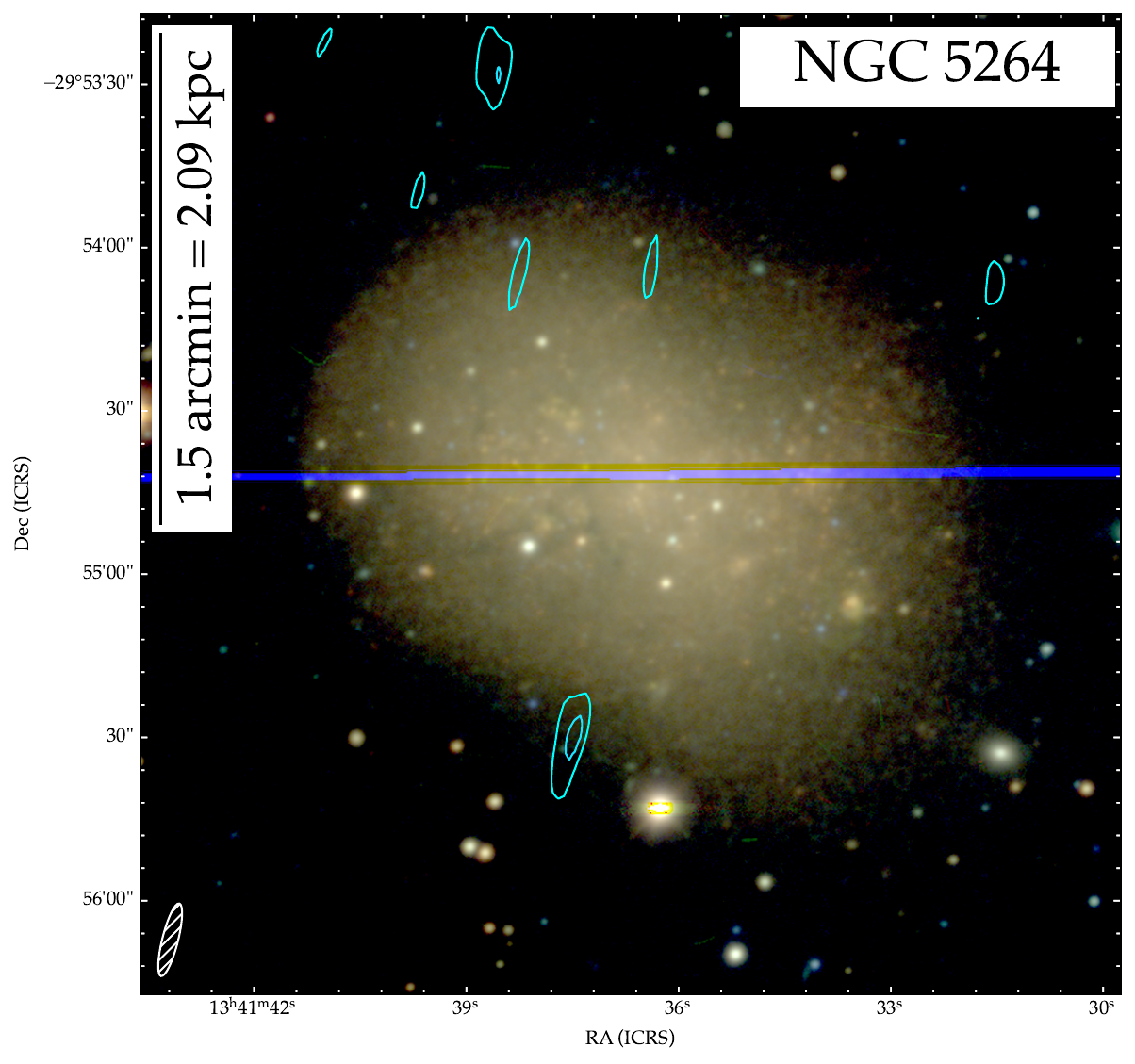}
    \end{minipage}
    \caption{Colour–composite images from the DESI Legacy Imaging Surveys of the CHILLING sample with non-detected or marginally detected RC emission, overlaid 2.1\,GHz ATCA radio emission contours starting at $3\,\sigma$ and increasing by a factor of $\sqrt{2}$. The scale in the top left corner is calculated using the distance to the source. The beam is shown in the bottom left corner. The noise level $\sigma_{\rm 2.1\,GHz}$ can be taken from Table~\ref{obs_params}.
}
    \label{legacy}
\end{figure*}

\begin{table*}[h!]
\centering
\caption{CHILLING sample properties including the coordinates, distance, absolute magnitude in B-band, star formation rate, H{\sc i} mass, rotational velocity, and the near and far-infrared luminosities of each dwarf galaxy}
\label{properties}
\begin{threeparttable}
\begin{tabular}{lccccccccc}
\toprule
Name & RA & Dec & $D$ & $M_{\mathrm{B}}$ & $\log\left[\frac{\mathrm{SFR_{H\alpha}}}{\mathrm{M_{\odot} \, \mathrm{yr}^{-1}}}\right]$ & $\log\left[\frac{M_{\mathrm{H\textsc{i}}}}{M_{\odot}}\right]$ & $\varv_{\mathrm{rot}}$ & $L_{3.4\upmu\mathrm{m}}$ & $L_{60\upmu\mathrm{m}}$ \\
 &  &  & (Mpc) & (mag) &  &  & (${\rm km\,s^{-1}}$) & (mag) & (Jy)  \\
\midrule
ESO154-G023 & 02$^{\rm h}$56$^{\rm m}$50.38$^{\rm s}$ & $-$54$^{\rm d}$34$^{\rm m}$17.10$^{\rm s}$ & 5.76 & $-16.45$ & $-1.32$ & 9.14 & 53.0 & 15.21 & 0.32 \\
ESO245-G005 & 01$^{\rm h}$45$^{\rm m}$03.73$^{\rm s}$ & $-$43$^{\rm d}$35$^{\rm m}$52.93$^{\rm s}$ & 4.43 & $-15.68$ & $-1.59$ & 8.6 & 51.0 & 15.99 & 0.22 \\
ESO383-G087 & 13$^{\rm h}$49$^{\rm m}$17.50$^{\rm s}$ & $-$36$^{\rm d}$03$^{\rm m}$48.40$^{\rm s}$ & 3.45 & $-16.83$ & $-1.49$ & 7.82 & 18.0 & 14.05 & 1.21 \\
ESO572-G025$^\dagger$ & 11$^{\rm h}$57$^{\rm m}$28.03$^{\rm s}$ & $-$19$^{\rm d}$37$^{\rm m}$26.60$^{\rm s}$ & 27.88 & $-17.50$ & $-0.27$ & 8.36 & -- & 12.88 & 0.51 \\
Fairall301$^\dagger$  & 04$^{\rm h}$06$^{\rm m}$07.92$^{\rm s}$ & $-$52$^{\rm d}$40$^{\rm m}$06.30$^{\rm s}$ & 9.30 & $-15.92$ & $-1.11$ & -- & 52.3 & 12.38 & 0.90 \\
IC1959      & 03$^{\rm h}$33$^{\rm m}$12.59$^{\rm s}$ & $-$50$^{\rm d}$24$^{\rm m}$51.30$^{\rm s}$ & 6.19 & $-16.07$ & $-1.38$ & 8.38 & 57.0 & 13.90 & 0.51 \\
IC4662      & 17$^{\rm h}$47$^{\rm m}$08.86$^{\rm s}$ & $-$64$^{\rm d}$38$^{\rm m}$30.33$^{\rm s}$ & 2.55 & $-15.61$ & $-1.12$ & 8.24 & 41.0 & 11.99 & 8.82 \\
IC5152      & 22$^{\rm h}$02$^{\rm m}$41.51$^{\rm s}$ & $-$51$^{\rm d}$17$^{\rm m}$47.20$^{\rm s}$ & 1.96 & $-15.56$ & $-2.23$ & 8.01 & 43.0 & 13.77 & 2.46 \\
ISZ399$^\dagger$      & 12$^{\rm h}$19$^{\rm m}$59.51$^{\rm s}$ & $-$17$^{\rm d}$23$^{\rm m}$31.00$^{\rm s}$ & 15.94 & $-17.15$ & $-0.81$ & -- & -- & 10.71 & 2.86 \\
NGC244$^\dagger$      & 00$^{\rm h}$45$^{\rm m}$46.43$^{\rm s}$ & $-$15$^{\rm d}$35$^{\rm m}$48.80$^{\rm s}$ & 11.60 & $-16.25$ & $-1.11$ & 8.73 & -- & 12.05 & 0.53 \\
NGC625      & 01$^{\rm h}$35$^{\rm m}$04.63$^{\rm s}$ & $-$41$^{\rm d}$26$^{\rm m}$10.30$^{\rm s}$ & 4.02 & $-16.50$ & $-1.20$ & 8.00 & 34.0 & 12.12 & 5.73 \\
NGC1311     & 03$^{\rm h}$20$^{\rm m}$06.96$^{\rm s}$ & $-$52$^{\rm d}$11$^{\rm m}$07.90$^{\rm s}$ & 5.55 & $-15.41$ & $-1.55$ & 8.04 & 36.0 & 13.46 & 0.39 \\
NGC5253     & 13$^{\rm h}$39$^{\rm m}$55.96$^{\rm s}$ & $-$31$^{\rm d}$38$^{\rm m}$24.38$^{\rm s}$ & 3.44 & $-17.05$ & $-0.64$ & 7.91 & 31.0 & 9.25 & 30.51 \\
NGC5264     & 13$^{\rm h}$41$^{\rm m}$36.68$^{\rm s}$ & $-$29$^{\rm d}$54$^{\rm m}$47.10$^{\rm s}$ & 4.79 & $-16.02$ & $-1.92$ & 7.70 & 15.0 & 13.72 & 0.30 \\
NGC5408     & 14$^{\rm h}$03$^{\rm m}$20.91$^{\rm s}$ & $-$41$^{\rm d}$22$^{\rm m}$39.70$^{\rm s}$ & 5.32 & $-16.73$ & $-0.83$ & 8.48 & 30.0 & 13.06 & 2.82 \\
\bottomrule
\end{tabular}
\tablefoot{$^\dagger$ Blue compact dwarf galaxies (BCDs); the rest are irregular dwarfs from the LVHIS sample.}
\end{threeparttable}
\end{table*}

\section{Observation and data reduction}
\label{dataset}
\subsection{CHILLING sample}

\begin{table*}
\caption{ATCA RC observations of CHILLING galaxies, with its configuration and the total on-source integration time}
\centering
\begin{threeparttable}
\begin{tabular}{lcccccccc}
\toprule
Name &  &  & \begin{tabular}[c]{@{}c@{}}ATCA \\ configurations\end{tabular} &  &  & &  Total on-source time / hrs & \\
 &  &  &  &  &  & 2100\,MHz & 5500\,MHz & 9000\,MHz \\
\midrule
ESO154-G023 & 1.5A & 750B & H168 &  & & 7.17 & 6.85 & 6.85 \\
ESO245-G005 &  1.5B &  750C  &  & H214 &  &  10.53 & 7.59 & 7.59 \\
ESO383-G087 & 1.5B & 750C &  & H214 &  & 8.56 & 8.37 &  8.37 \\
ESO\,572-G025 & 1.5B &  &  & H214 &  & 4.92 & 4.60 & 4.60 \\
Fairall\,301 & 1.5C & 750C & H168 &  &  & 6.19 &  5.26 & 5.26 \\
IC1959 & 1.5C & 750B &  & H214 & & 5.25 & 5.27 & 5.27 \\
IC\,4662$^\dagger$ & 1.5C & 750C &  & H214 &  & 37.66 & 11.93 & 11.93\\
IC\,5152 & 1.5A & 750B &  & H214 &  & 5.18 & 4.60 & 4.60 \\
ISZ\,399 & 1.5B & 750C &  &  & EW352 & 4.42 & 4.27 &  4.27 \\
NGC\,244 & 1.5A & 750C &  & H214 &  & 7.18 & 6.57  & 6.57\\
NGC\,625 & 1.5A & 750C &  & H214 &  & 9.20  &  6.21 &  6.21 \\
NGC\,1311 & 1.5A & 750C &  & H214 &  & 5.59 & 3.39 & 3.39\\
NGC\,5253 & 1.5B & 750B &  & H214 & & 7.23 & 6.90 & 6.90 \\
NGC5264 & 1.5B & 750B &  & H214 &  & 4.22 & 5.36 & 5.36\\
NGC\,5408 & 1.5C & 750B & H168 & H214 & & 4.88 & 10.19 &  10.19 \\
\bottomrule
\end{tabular}
\label{config}

\tablefoot{$^\dagger$ Observation of IC\,4662 is a combination of the CHILLING observations and the observations running under project ID: C3531 with detailed explanation in the Sect.~\ref{observations}.}
\end{threeparttable}
\end{table*}

The Continuum Halos In LVHIS Local Irregular Nearby Galaxies (CHILLING) sample covers a range of galaxy properties, such as star formation rate, luminosities, and flux densities. The selected galaxies have an absolute magnitude of $M_\text{B} > -17$, IRAS 60\,$\upmu$m detections, and are located at a declination below about $-20^\circ$.
The CHILLING sample includes 11 dwarf galaxies from the Local Volume H{\sc i} Survey (LVHIS) project \citep[][]{Koribalski_2018}, as well as 4 blue compact dwarf galaxies (BCDs) that are further away and therefore not part of the main LVHIS sample. Table~\ref{properties} gives an overview of the properties of the  CHILLING sample.

\subsection{Observations}
\label{observations}
The CHILLING dwarf galaxies were observed with the ATCA (project ID: C3041; PI: M. Johnson). The observations were performed at 1.1--3.1\,GHz (L/S-band), 3.9--7.1\,GHz (C-band) and 8--11\,GHz (X-band) between 18-May-2015 and 16-June-2016. The configurations used for each target are listed in Table~\ref{config}. Observations of the target galaxies and a phase calibrator were alternated through the observing run, giving a total on-source observing time of about 5 to 11\,hrs for each galaxy (see Table~\ref{config}). Along with the target and phase calibrator observations, the standard ATCA primary flux calibrator 1934$-$638 and the secondary flux calibrator 0823$-$500 have been used. For the galaxy IC\,4662, we use additional observational data of on-source integration time of 29.41\,hrs at central frequencies 2.1\,GHz, 4.71\,hrs at 5.5 and 9\,GHz, which have been observed between 4-May-2023 and 26-August-2024 (project ID: C3531; PI: S. Taziaux).

\begin{table*}
    \centering
    \caption{Overview of the observable parameters of the CHILLING galaxies, showing the name, the flux and phase calibrator, the Briggs \texttt{robust} weighting, the noise level of the RC image, and the category of detection.}
    \begin{threeparttable}
    \begin{tabular}{lccccc}
    \toprule
    name & flux calibrator & phase calibrator  & \texttt{robust} &  $\sigma_{\rm 2.1/5.5/9\,GHz}$  & Detection Category \\
    &  & &  &   ($\upmu$Jy/beam)   &  \\
    \midrule
    ESO\,154-G0154 & 1934-638 & 0252-549 & 0.0 & 25/10/10   & No/Marginal \\ 
    ESO\,245-G005 & 1934-638 & 0153-410 & 0.3 & 15/6/6   &  Compact \\ 
    ESO\,383-G087 & 1934-638 & 1424-418 &  0.0 & 39/15/20  &  No/Marginal \\ 
    ESO\,572-G025 & 1934-638/0823-500$^\dagger$ & 1127-145 &  0.0 & 30/20/30  & Diffuse \\ 
    Fairall\,301  & 1934-638 & 0302-623 &  0.3 & 14/12/25   & Diffuse \\
    IC\,1959 & 1934-638/0823-500$^\dagger$ & 0252-549 &  0.0 & 20/10/10  & Compact \\
    IC\,4662 & 1934-638 & 1718-649 &  0.3 & 9/8/12  &  Diffuse\\
    IC\,5152 & 1934-638 & 2326-477 &  0.0 & 15/8/8  & Compact \\
    ISZ\,399 & 0823-500  & 1213-172 &  0.3 & 73/20/25  &  Diffuse \\
    NGC\,244 & 1934-638 & 0023-263 &  0.3 &  10/8/8  & Diffuse \\
    NGC\,625 & 1934-638 & 0201-440 &  0.3 & 15/8/8    & Diffuse \\
    NGC\,1311 & 1934-638 & 0302-623 &  0.0 & 27/15/10  &  No/Marginal \\
    NGC\,5253 & 1934-638  & 1421-490 &  0.0 & 48/35/50   & Diffuse \\
    NGC\,5264 & 1934-638 & 1255-316 &  0.0 & 80/20/13    & No/Marginal \\
    NGC\,5408 & 1934-638/0823-500$^\dagger$ & 1421-490 &  0.0 & 85/17/18  &  Compact  \\
    \bottomrule
    \end{tabular}
    \label{obs_params}
\tablefoot{$^\dagger$ These datasets make use of the secondary flux calibrator at 5.5 and 9\,GHz.}
\end{threeparttable}
\end{table*}

\subsection{Data reduction}
\label{datareduction}
We processed the observations using the Multichannel Image Reconstruction, Imaging Analysis, and Display package \citep[\texttt{miriad;}][]{miriad} following standard data calibration procedures using \texttt{mfcal} for bandpass calibration, \texttt{gpcal} to derive the gains and instrument leakage. To minimise the effects of radio frequency interference (RFI) during flux and phase calibration, the edges of the dataset have been flagged using \texttt{uvflag}, automated flagging was performed using \texttt{pgflag} to suppress interference in the target sources. Any remaining corrupted data identified throughout the calibration process was manually flagged, using the interactive flagging tool \texttt{blflag} for manual inspection and removal of contaminated data, to ensure data integrity.
Initially, after cross-calibration and excluding channels with H{\sc i} emission for L/S-band dataset, we employed an iterative imaging using \textsc{WS-clean} \citep{wsclean} and self-calibration (phase-only, frequency-independent, with a solution interval of 5\,min down to 60\,s) using \texttt{CASA} \citep[version 6.4.4.31;][]{casa} until image quality reached convergence with a Briggs weighting of $\texttt{robust} = -1, -0.5, -0.3, 0, 0.3$ to slowly reconstruct the diffuse emission. Imaging and self-calibration was performed independently for L/S-, C-, and X-bands and by splitting into 32 \texttt{-channels-out} to avoid bandwidth smearing larger than the synthesized beam, during imaging of the field of view.
Multifrequency and multiscale CLEANing \citep{hoegbom_1974} utilising interactive masks using \texttt{flint\_masking} (Galvin et al. in prep.)\footnote{\url{https://github.com/flint-crew/flint}} around visible sources to minimise artefacts and flux scattering and applying detection thresholds was used to retain genuine emission only. During the imaging step, we use \texttt{-join-channels}  to leverage the ATCA's full bandwidth. 
The continuum images for each galaxy were generated using different Briggs \texttt{robust} weighting to increase the quality, resulting in resolutions and noise levels shown in Table~\ref{obs_params}. In the last round of imaging, we employed joint imaging\footnote{Joint imaging means gridding and deconvolving the L/S-, C-, and X-bands jointly, resulting in a highly sensitive image with central frequency of 5.55 GHz, with a total bandwidth of 8.95 GHz} with \texttt{-channels-out} 32 images. As these images are only used in science for the SED fitting in Sect.~\ref{radio}, we do \texttt{-no-mf-weighting} in the last round of imaging to ensure not to include artificial spectral substructures. We use the integrated task \texttt{linmos} in \texttt{miriad} to correct for the primary beam to each 32 images of the dataset.

\section{Detectability dependency}
\label{detect}
\subsection{Definition of RC detectability}
Since the definition of what qualifies as an RC detection can vary, we divided our sample into three categories. 
In the first category, a galaxy is classified as a `diffuse' detection if spatially extended RC emission is present across the majority of the galaxy. Specifically, we require that emission is detected at $>3\sigma$ significance over at least $\sim70-80\,\%$ of the galaxy’s optical extent (as defined by the $R_{25}$ radius) and that the emission is more extended than the synthesized beam. These detections leave little doubt that the galaxy as a whole emits RC emission (see Fig.\ref{legacy_diffuse}).
A galaxy is classified as a `compact' detection if it shows at least one RC-bright H{\sc ii} region ($>3\sigma$ peak emission) but lacks diffuse emission across most of the galaxy. In this case, RC emission is localized to one or a few star-forming regions and covers $<50\,\%$ of the galaxy’s optical extent (see Fig.~\ref{legacy_compact}).
In the third category, a galaxy is classified as `no/marginal' detection if it shows no significant RC emission above $3\sigma$ within its optical extent, or only weak, patchy signals covering $<20\,\%$ of the optical radius that cannot be robustly distinguished from noise or imaging artefacts (see Fig.~\ref{legacy}).

\subsection{RC properties}
Only 11 out of these 15 dwarf galaxies show RC emission, with 7 showing diffuse emission, 4 showing compact emission coming from a starforming region and 4 are showing no or very weak emission. The detection thresholds are set to $3\sigma$. The complete L/S-, C-, and X-band RC maps for all 15 dwarf galaxies are provided in Appendix~\ref{RCmaps}, with galaxies showing diffuse emission in Fig.\ref{detections_diffuse}, compact emission in Fig.\ref{detections_compact}, and no or very weak emission in Fig.~\ref{non-detections}. We notice that all 4 of the BCDs are detected with diffuse emission while only 7 out of the 11 LVHIS dwarfs were detected. Only three of them, IC\,4662, NGC\,5253 and NGC\,625, show substructures as opposed to merely an unresolved structure. NGC\,5408 only shows RC emission on the south-western side, aligning with the observed high H{\sc i} velocity dispersion as described in \citet[][]{vanEymeren_2010}, possibly due to the gas outflows. 
Similar to NGC\,5408, ESO245-G005 only shows RC emission near star-forming regions and a depression at the center, a feature that is also seen in H{\sc i} \citep[][]{cote_2000}. 

All 4 BCDs are detected in our sample, compared to only 7 irregular dwarf galaxies. This likely reflects the intrinsically higher star formation rates of BCDs relative to typical star-forming dwarfs. However, an observational bias may also contribute as the selected BCDs are more distant, their star-forming regions appear more spatially concentrated on the sky, enhancing the detectability of their compact emission compared to the more nearby, diffuse dwarf galaxies.
If the radio continuum emission is associated with star formation, which we explore in this study, this suggests that the measured signal is dominated by emission from concentrated star-forming regions as opposed to more extended ones

\subsection{Classification of RC dependency}
\label{classification}
\begin{figure}
    \centering
    \includegraphics[width=\linewidth]{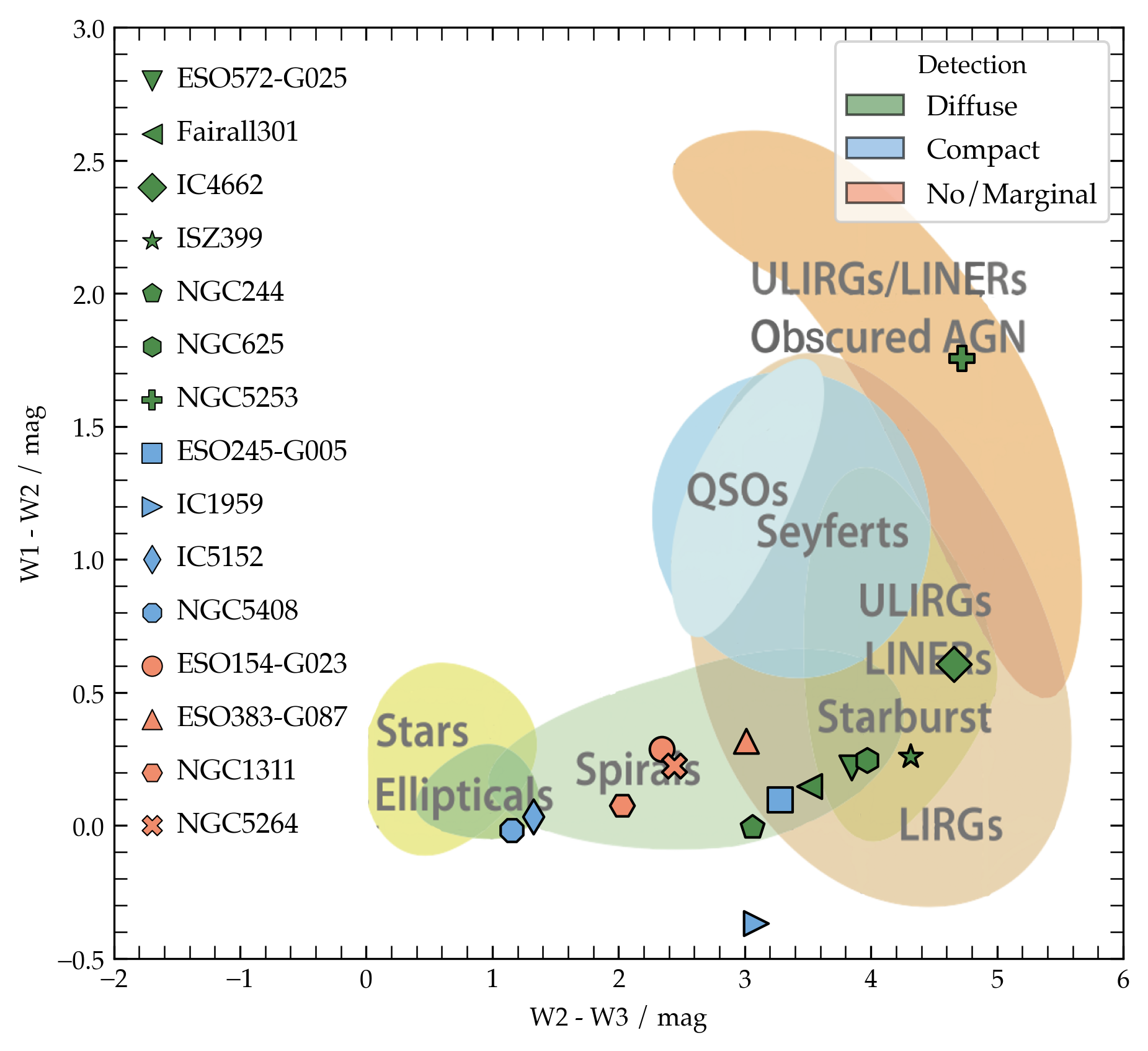}
    \caption{WISE color–color diagram showing in the background the locations of interesting classes of objects \citep[][]{Wright_2010}, while showing the CHILLING galaxies in dependency of their detectability category of RC emission. Dwarf galaxies with diffuse RC emission are shown in green, those with compact RC emission are shown in blue, and non-detections or weak detections are shown in orange. The WISE W1, W2 and W3 bands are $3.4\,\upmu$m, $4.6\,\upmu$m and $12\,\upmu$m, respectively.}
    \label{WISE-color}
\end{figure}

\begin{figure*}
    \centering
     \includegraphics[width=0.94\linewidth]{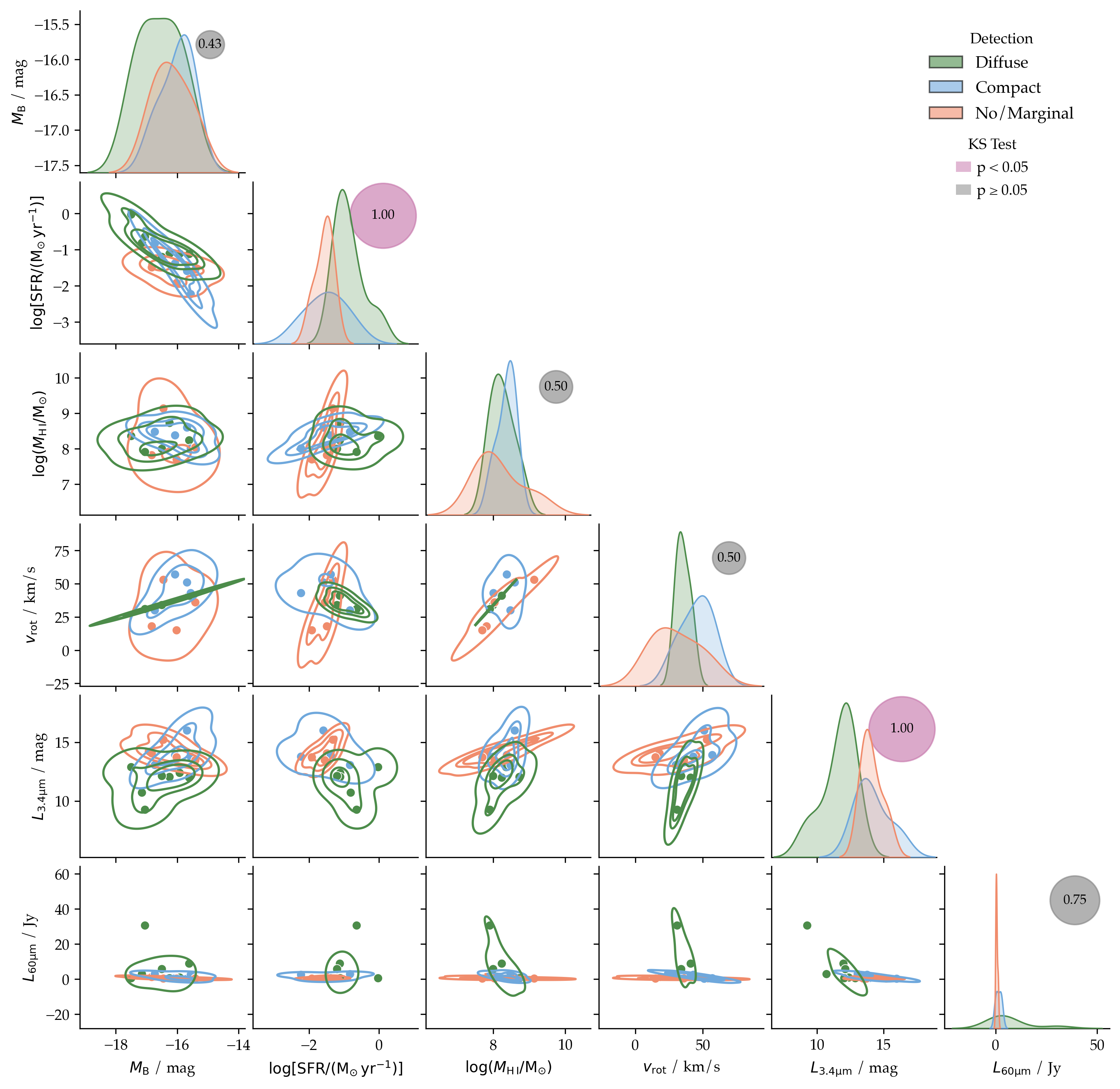}
    \caption{Correlation analysis of several galaxy parameters with the dependence of RC emission detection. It represents different physical properties, including absolute magnitude in the B-band $M_{\rm B}$, star formation rate (SFR), H{\sc i} mass M$_{\rm HI}$, maximum rotation velocity of the gas corrected for inclination ($\varv_{\rm rot}$),  mid-infrared magnitude at $3.4\,\upmu$m ($L_{\rm 3.4\upmu m}$) from WISE and far-infrared (FIR) luminosity at $60\,\upmu$m ($L_{\rm 60\upmu m}$) from IRAS. Galaxies with diffuse RC emission are shown in green, those with compact emission in blue, and galaxies with no or very weak detection in orange. The Kolmogorov–Smirnov (KS) test results are visualised as circles on the diagonal of the plot, with sizes ranging from 0 to 1; larger circles indicate greater differences between the distributions. The exact value from the KS-test are shown inside the circles. The color represents statistical significance based on the $p$-value: pink circles indicate significant differences ($p < 0.05$), while gray circles indicate non-significant results.}
    \label{detection}
\end{figure*}

To explore the factors influencing the detectability of RC emission in these 15 dwarf galaxies, we construct a WISE color-color diagram \citep[][]{Wright_2010} to examine the relationship between radio emission and the classification of these objects. 
As shown in Fig.~\ref{WISE-color}, 6 out of 7 dwarf galaxies that display diffuse RC emission are classified as starbursts, primarily located in the right section of the plot, while the dwarf galaxies, showing a more compact RC emission are either close to the border to be classified as a starburst or they are very left in the WISE color-color diagram.
Starburst galaxies, such as NGC\,5253, IC\,4662, and NGC\,625, exhibit extended radio emission that extends into the outskirts of the galaxies, while other galaxies show RC detections appearing  to align with the brightest star forming regions with the exception of a few tenuous structures at lower frequencies. Notably, both IC\,4662 and NGC\,5253 have higher W1-W2 values. NGC\,5253 falls within the region typically associated with obscured AGN, although it is currently known to host a super star cluster at its core \citep[e.g.][]{Smith_2020}, rather than a traditional AGN. This may suggest a trend, saying galaxies located further to the right in the diagram tend to exhibit stronger RC emission. These results suggest that the intensity of star formation, as quantified by the SFR and SFR surface density, are key factors governing the detection of RC emission in dwarf galaxies. This implies that not just the presence of star formation, but its localised strength, plays a critical role in producing detectable levels of RC emission.

\subsection{Parameter behaviour RC dependency}
\label{params_det}
To explore which factors, in addition to a galaxy’s classification as a starburst, may influence the presence of RC emission, we examine a set of physical parameters, which can be taken from Table~\ref{properties}.
These include the absolute B-band magnitude ($M_{\rm B}$), star formation rate (SFR), H{\sc i} mass (M$_{\rm HI}$), maximum rotation velocity of the gas corrected for inclination ($\varv_{\rm rot}$), and mid-infrared luminosity at $3.4\,\upmu$m ($L_{\rm 3.4\upmu m}$), which serves as a reliable proxy for the stellar mass \citep[e.g.][]{Jarrett_2023}. We also consider the far-infrared (FIR) luminosity at $60\,\upmu$m ($L_{\rm 60\upmu m}$) from IRAS.

From the diagonal panels in Fig.~\ref{detection}, it becomes clear that the most prominent difference between galaxies with and without RC detection lies in their SFRs, which is also linked to absolute magnitude. Regarding mass-related parameters, galaxies with detected RC tend to be brighter at $3.4\,\upmu$m magnitude, suggesting a higher stellar mass, whereas the H{\sc i} masses do not exhibit a significant difference between the two groups.
The FIR luminosity at $60\,\upmu$m displays a narrow range for non-detected galaxies, with most values clustering close to zero, while the detected galaxies show notably higher FIR fluxes. When examining the gas rotation velocity, which is physically linked to the total mass, we find that galaxies with no or weak RC detections tend to have lower rotation velocities, while those with compact RC detections show higher velocities. Dwarf galaxies with diffuse RC detections lie in between these two groups, although the differences are not statistically significant.

To further quantify the correlation between RC detection and the various physical parameters, we compute Kolmogorov-Smirnov \citep[KS;][]{KStest} test with its p-value for statistical significance. This method compares two distributions to determine whether they differ significantly. The p-value indicates the probability that the observed difference happened by chance, while a small p-value (typically below $0.05$) suggests a statistically significant difference. Looking at Fig.~\ref{detection}, significant differences were found for SFR and the $3.4\,\upmu$m magnitude, used as proxy for stellar mass (both with KS$=1.0$ and $p=0.0006$) indicating that the distributions of these parameters differ meaningfully between the three populations. For $60\,\upmu$m FIR luminosity (KS$=0.75$, $p=0.06$), the difference is slightly not statistical significant to see difference between these populations. These results suggest that both the SFR and stellar mass are linked to the detection of RC emission. This provides quantitative support for the idea that the intensity and spatial concentration of star formation are critical factors in generating detectable RC in dwarf galaxies, likely due to their role in driving synchrotron-emitting processes.
In contrast, properties such as absolute magnitude in B-band $M_{\rm B}$, H{\sc i} mass M$_{\rm HI}$ and rotation velocity $\varv_{\rm rot}$ showed no statistically significant difference ($p>0.05$), suggesting similar distributions between these populations. 

It is important to note that all of these correlations should be interpreted with caution. The analysis is based on a small sample of 15 dwarf galaxies, comprising 11 with RC detections and 4 without. This sample size is insufficient to draw statistically robust conclusions on its own about the entire dwarf galaxy population, although it is in agreement with previous studies \citep[e.g.][Taziaux in prep.]{Roychowdhury_2012, Schleicher_2016,hindson_radio_2018}

\section{Spectral model fitting}
\label{radio}
\subsection{Overview of the RC data}
For the SEDs, we include additional data from multiple surveys in addition to our ATCA observations. For NGC\,625 and Fairall\,301, we incorporate GaLactic and Extragalactic All-sky MWA survey eXtended (GLEAM-X) data \citep[][]{gleamx}, which cover frequencies from 87\,MHz to 221\,MHz. For all sources, we also include the Australian SKA Pathfinder (ASKAP) survey, called Rapid ASKAP Continuum Survey (RACS-Low) data \citep[][]{racs} at 885\,MHz.
For the ATCA dataset, we applied a $3\,\sigma$ clipping threshold and computed the integrated flux density, summing all emission over its resolved size with an optical aperture using \texttt{RadioFluxTools}\footnote{\url{https://gitlab.com/Sunmish/radiofluxtools}}, across 32 individual frequency slices. To ensure the reliability of the measurements, only slices with a noise level below $90\,\upmu$Jy/beam were considered. As the ATCA observations were conducted in snapshot mode, we adopt a standard calibration uncertainty of 10\,\%. The total uncertainty is then estimated as the combination of the statistical uncertainty, determined from the background noise and the specified beam area, and the assumed calibration uncertainty.
For the SED analysis, we only use the 7 dwarf galaxies which show extended and diffuse emission.

\begin{figure*}
    \centering
    \begin{minipage}[b]{0.32\linewidth}
        \includegraphics[width=\linewidth]{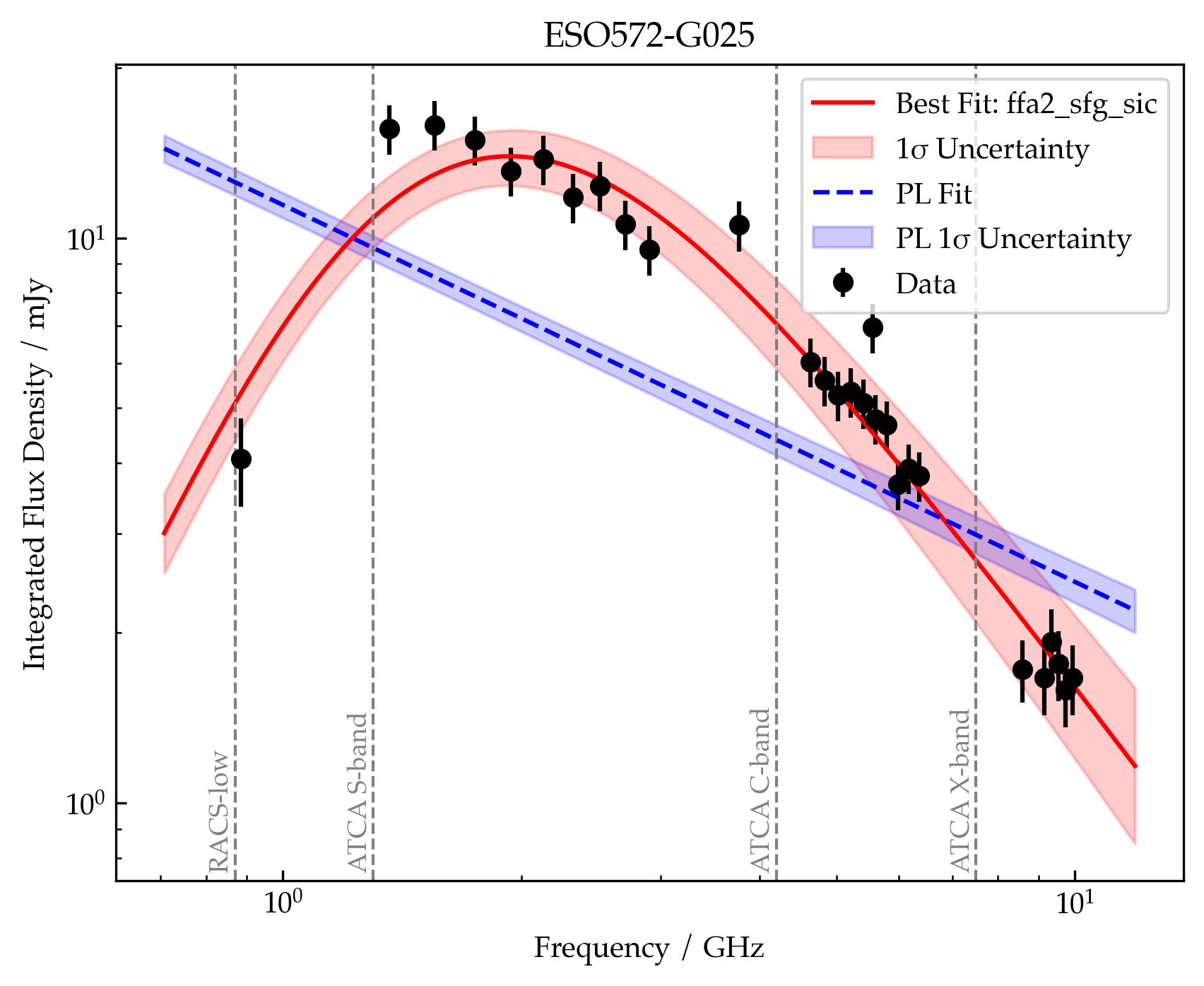}
    \end{minipage}
    \begin{minipage}[b]{0.32\linewidth}
        \includegraphics[width=\linewidth]{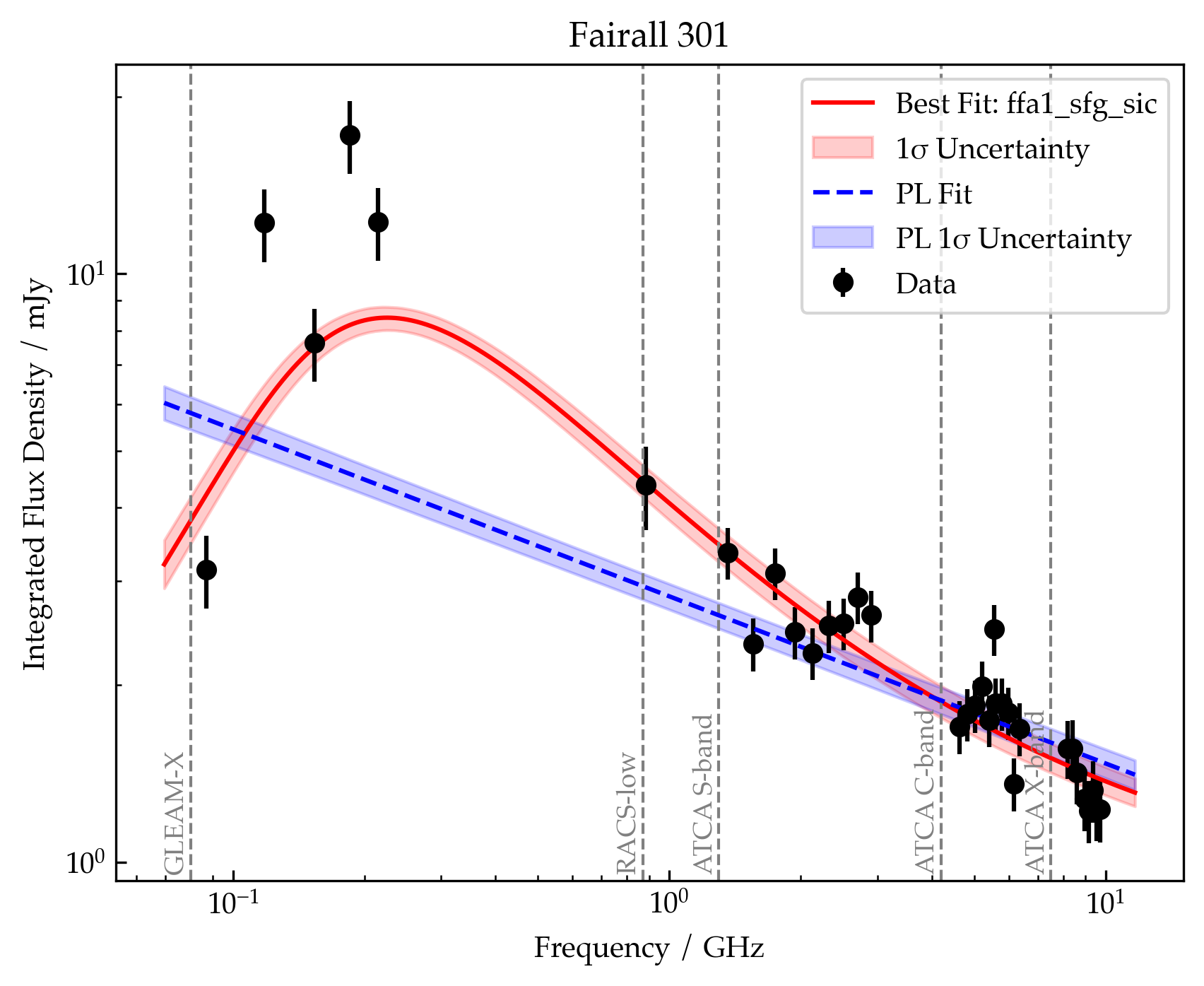}
    \end{minipage}
    \begin{minipage}[b]{0.33\linewidth}
        \includegraphics[width=\linewidth]{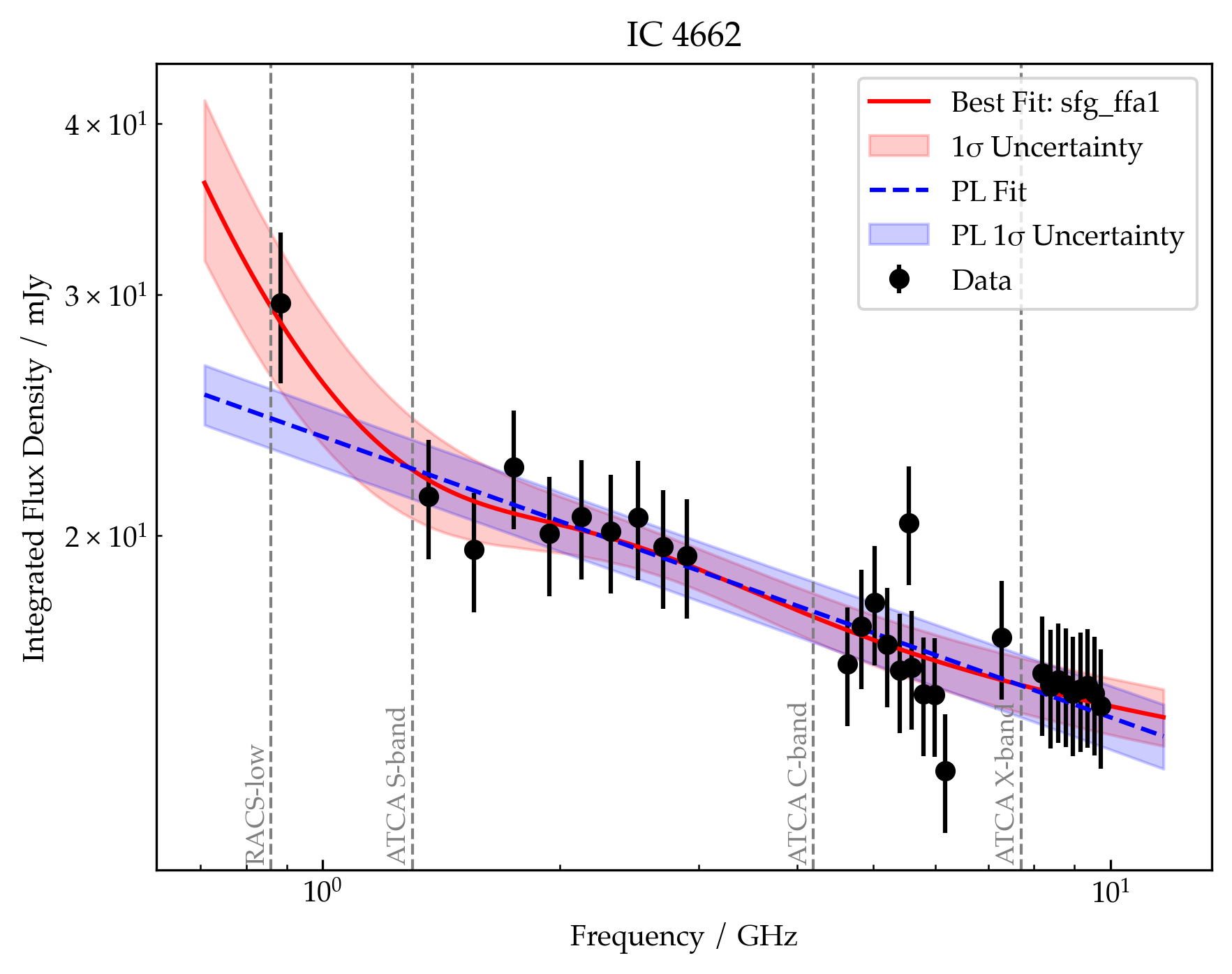}
    \end{minipage}
    \begin{minipage}[b]{0.32\linewidth}
        \includegraphics[width=\linewidth]{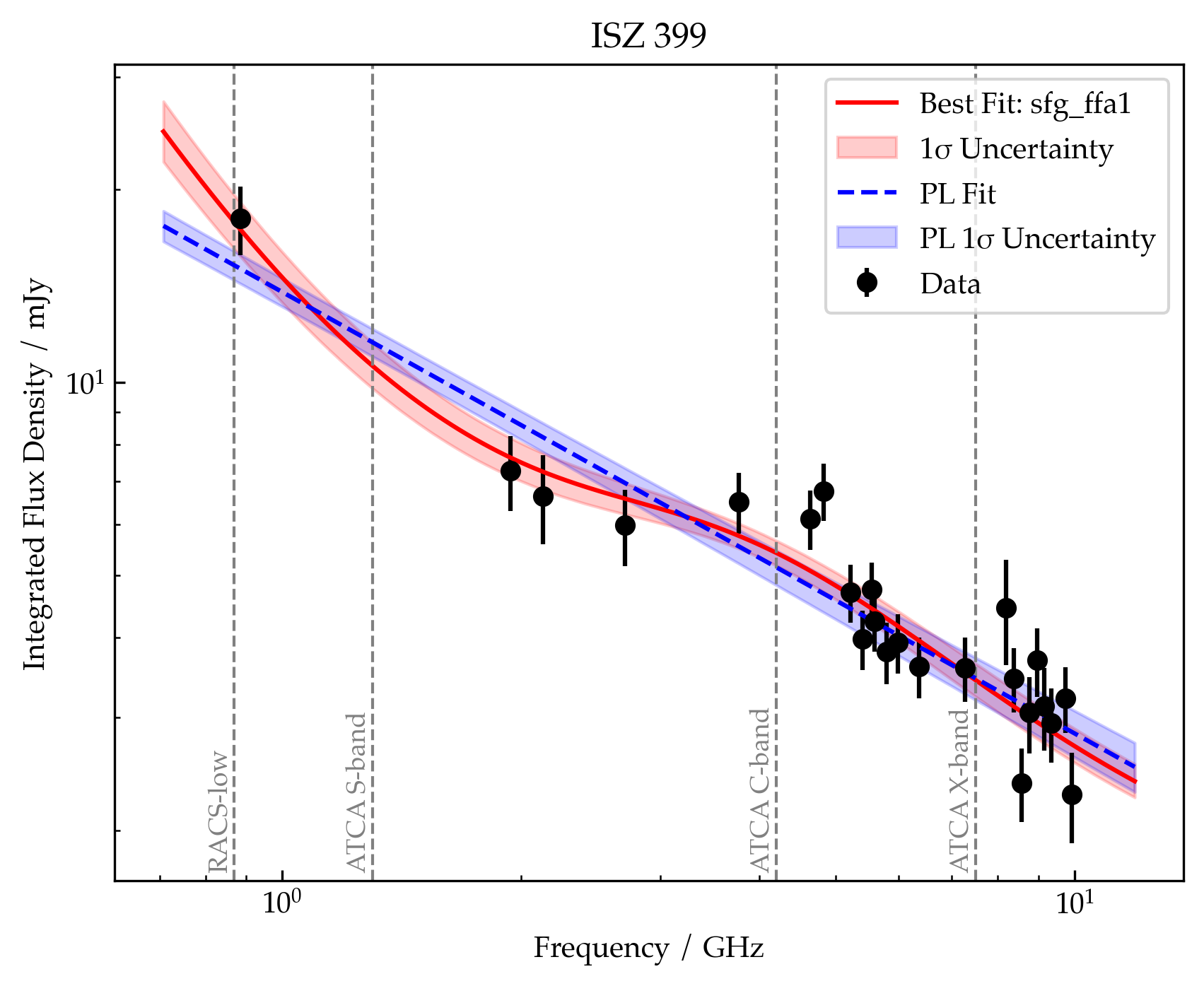}
    \end{minipage}
    \begin{minipage}[b]{0.33\linewidth}
        \includegraphics[width=\linewidth]{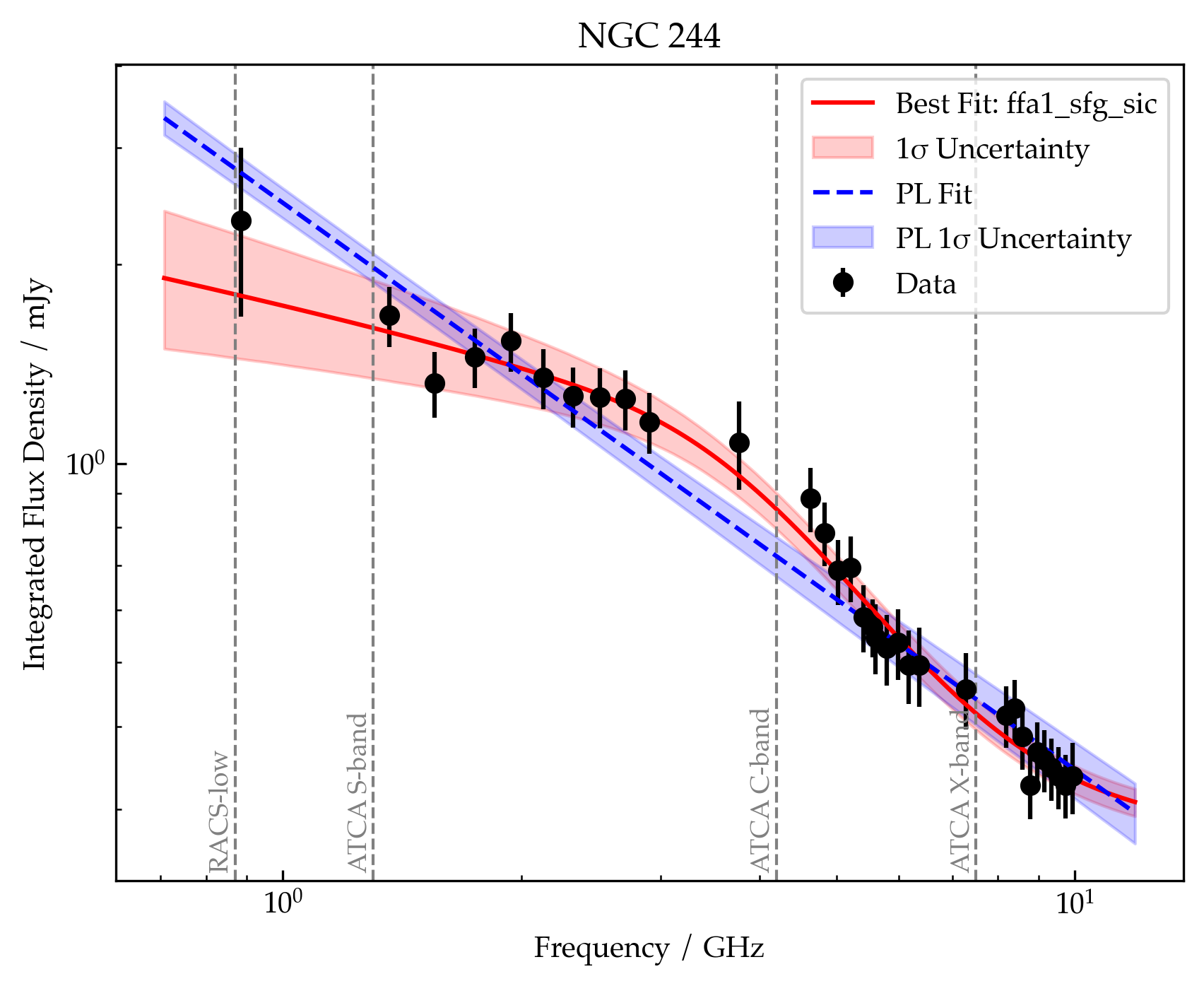}
    \end{minipage}
    \begin{minipage}[b]{0.33\linewidth}
        \includegraphics[width=\linewidth]{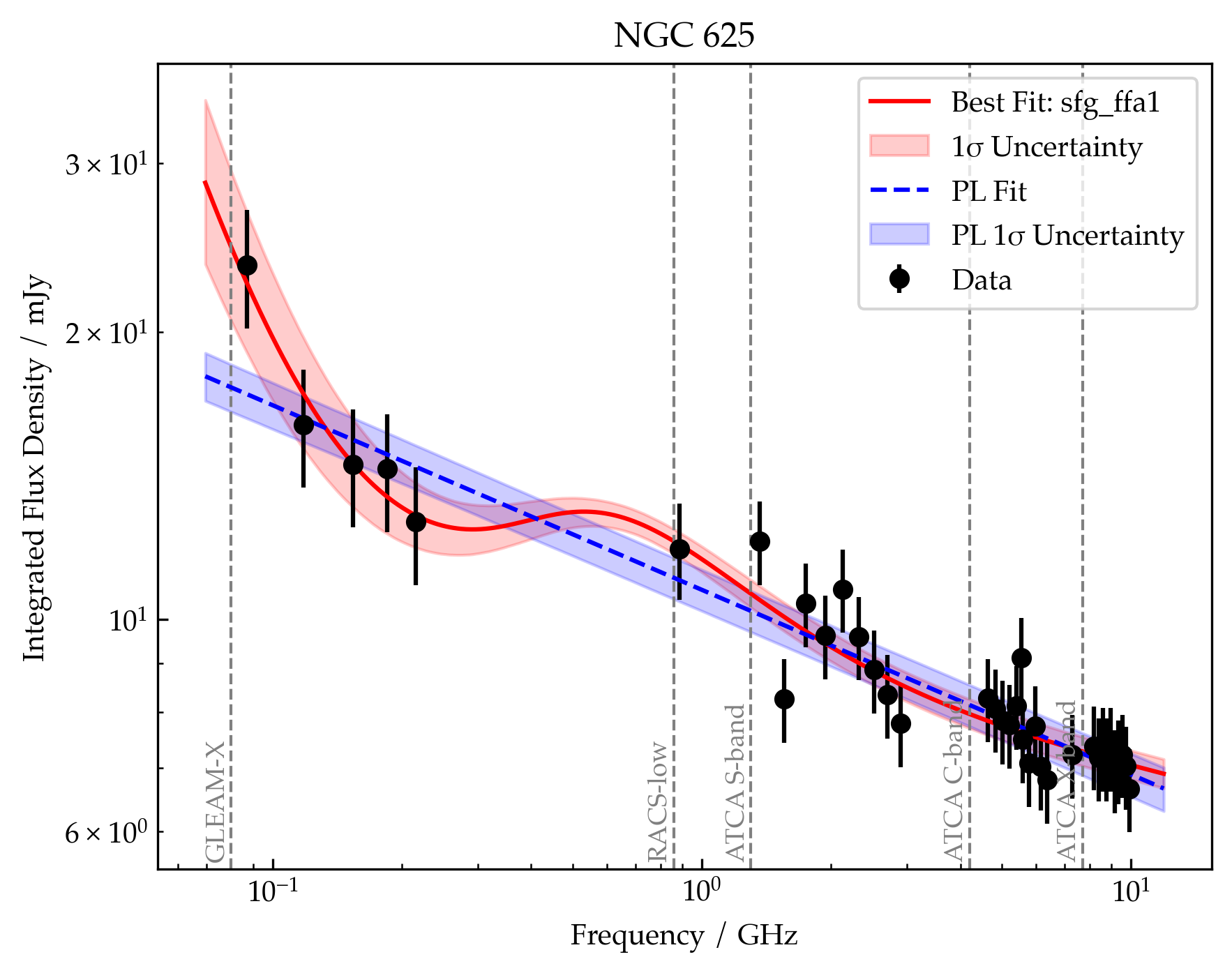}
    \end{minipage}
    \begin{minipage}[b]{0.33\linewidth}
        \includegraphics[width=\linewidth]{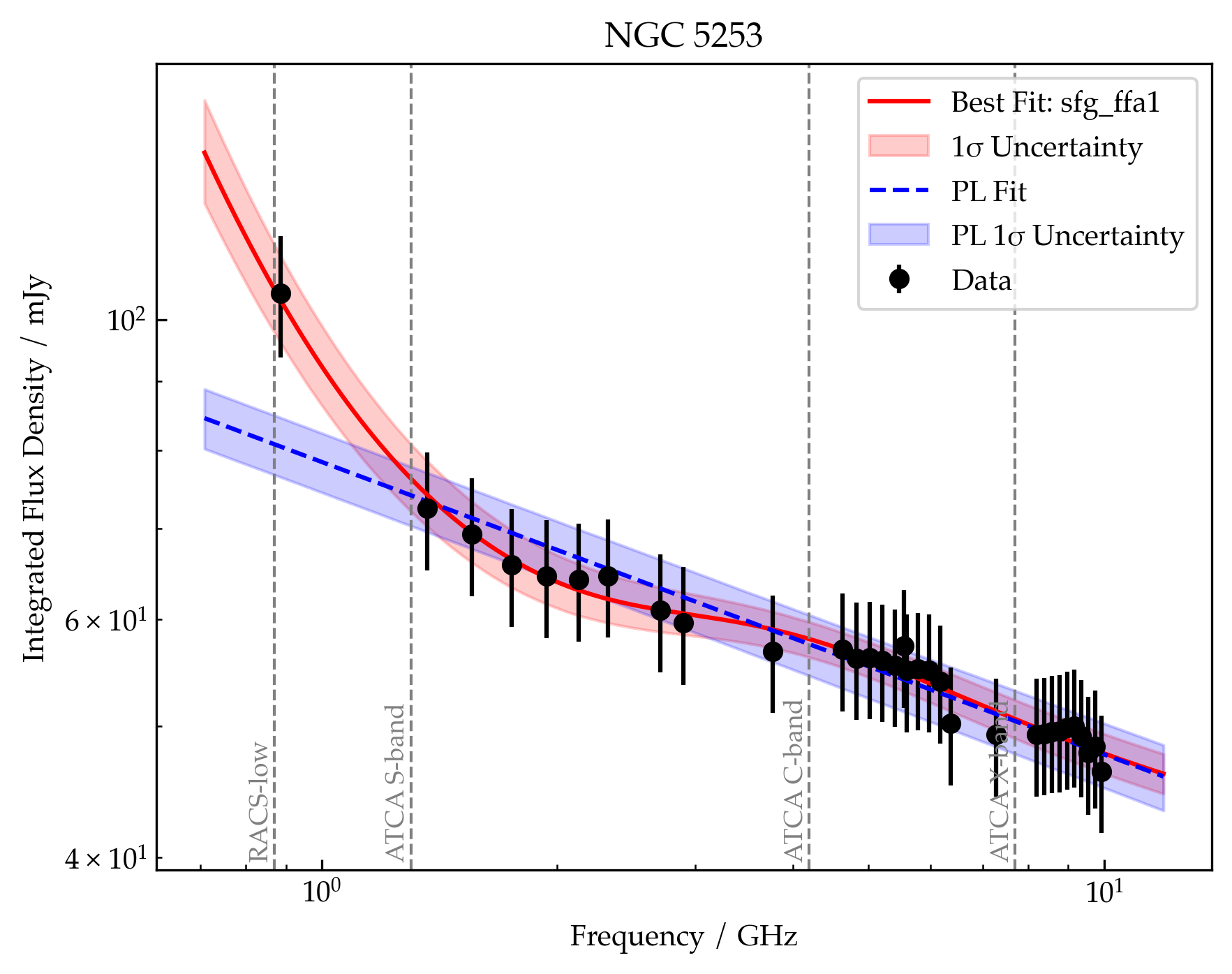}
    \end{minipage}
    \caption{Spectral energy distribution of the RC detected CHILLING sample. The plot displays total intensity data with different observations: GLEAM-X data \citep[][]{gleamx} only for Fairall\,301 and NGC\,625, RACS-Low \citep[][]{racs} and our ATCA S-band, C-band and X-band. The frequency range of the telescope data is indicated by a grey dotted line. Additionally, we show the best-fit in red for each galaxy out of these 19 different models, explained in the Appendix~\ref{model}. The highlighted red regions represent the $1\,\sigma$ uncertainties sampled by \texttt{EMCEE}. For comparison to the best-fit, we show the simple power-law (PL) fit with its uncertainty range in a blue dotted line.}
    \label{sed}
\end{figure*}

\subsection{Spectral models}
In the study of \citet{klein_2018}, four different basic models were fitted to a small sample of dwarf galaxies as well as more massive ones. This has been further developed by \citet[][]{Galvin_2018} and \citet{Grundy_2025}, who focused solely on starbursts and star-forming galaxies, respectively, and developed the spectral fitting models.  Our aim is to improve the spectral fitting for dwarf galaxies and identify which components dominate at different parts of the spectrum. We modified the models presented by \citet[][]{Grundy_2025} by splitting the absorption part into an internal component and an external component. All 19 models are detailed and presented in Appendix~\ref{model}, but these models are essentially combinations of five basic assumptions.

The first assumption involves a standard power-law with only synchrotron emission, given by

\begin{equation}
    S(\nu) = A \left( \frac{\nu}{\nu_0} \right)^{\alpha}
    \label{pl}
\end{equation}
\noindent
where $A$ represents the non-thermal synchrotron emission and $\alpha$ is the non-thermal spectral index at a given frequency $\nu_0$. 
The second model accounts for a combination of synchrotron emission and thermal free–free emission, which can be written as a superposition

\begin{equation}
    S(\nu) = A \left( \frac{\nu}{\nu_0} \right)^{\alpha} + B \left( \frac{\nu}{\nu_0} \right)^{-0.1}
    \label{sfg}
\end{equation}
\noindent
where $A$ represents the synchrotron emission, $\alpha$ is the non-thermal spectral index, and $B$ is the free--free emission at a given frequency $\nu_0$.

Next, we model two individual free–free absorption components at lower frequencies, one for internal absorption and one for external absorption, following the assumptions outlined in \citet{Tingay_2003}. For external free–free absorption, the model is:

\begin{equation}
    S(\nu) = \left( 1 - e^{-\tau_1} \right) \cdot B \cdot \left( \frac{\nu}{\nu_1} \right)^2
\end{equation}
\noindent
while for internal free–free absorption, the model is:

\begin{equation}
    S(\nu) = \left( \frac{1 - e^{-\tau_2}}{\tau_2} \right) \cdot B \cdot  \left( \frac{\nu}{\nu_2} \right)^{2}
\end{equation}
\noindent
where the free--free optical depth parameter is given by  $\tau_i = \left(\nu/\nu_i \right)^{-2.1}$. Finally, the last model component is relevant when inverse Compton losses dominate at higher frequencies, causing CREs to cool rapidly, leading to a break in the spectrum. This model can be written as

\begin{equation}
    S(\nu) = \frac{A \left( \frac{\nu}{\nu_0} \right)^{\alpha}}{1 + \left( \frac{\nu}{\nu_{\rm b}} \right)^{\Delta \alpha}}
\end{equation}
\noindent
where $\nu_{\rm b}$ is the break frequency in the spectrum and $\Delta \alpha$ represents the change in the non-thermal spectral index due to synchrotron and inverse Compton losses, assuming continuous electron injection from massive SF. 
The full explanation of all 19 models is detailed in Appendix~\ref{model}.

\subsection{Model selection and spectral fitting}
To fit all the 19 models to the data used for each dwarf galaxy, we use an affine invariant Markov chain Monte Carlo ensemble sampler \citep[][]{mcmc} implemented as the \texttt{EMCEE PYTHON} package \citep[][]{mcmc_python} using the differential evolution optimisation method.
We choose physically motivated priors to constrain our models, such as the parameters A, B, C and D remain always positive and the spectral index value $\alpha$ should be between $-3$ and $0.5$. We choose for the frequencies $\nu_1$ and $\nu_2$, as well as for the spectral break frequency $\nu_{\rm b}$, a value which should be between $0.01$ and $10$\,GHz. 
For each model, the fit quality is evaluated based on the chi-square and Bayesian statistic. The model with the lowest chi-square and Bayesian information criterion (BIC) is selected as the best representation of the data, and its fit parameters are reported. Models that fail to converge are skipped.

\subsection{Spectral fitting results}
Fig.~\ref{sed} show the spectral energy distribution of the diffuse RC detected CHILLING galaxies. We show here the GLEAM-X data \citep[][]{gleamx}, if available (here only for Fairall\,301, NGC\,625), RACS-Low \citep[][]{racs} and the observed ATCA L/S-,C- and X-band data. The best-fit model have been shown in red, while for comparison the simple power-law is shown in blue.

For ESO\,572-G025 and Fairall\,301 we observe a turnover to lower frequency due to internal or external free--free absorption and probably inverse Compton losses at higher frequencies, according to the fit-model called \texttt{FFA2\_SFG\_SIC} and \texttt{FFA1\_SFG\_SIC}, respectively. 
Inverse Compton losses at higher frequencies are also seen in NGC\,244, fitted by the best fit-model called \texttt{FFA1\_SFG\_SIC} with a slight turnover to lower frequencies. To confirm this turnover associated with free--free absorption, we need even lower frequencies, such as GLEAM-X data. 
For Fairall\,301, we have GLEAM-X data, so we see this turnover happening at approximately 150\,MHz, while for ESO\,572-G025, we see this turnover already happening at 1\,GHz. This turnover to lower frequency is also observed by \citet[][]{Gajovic_2024} in more massive galaxies at frequencies of approximately 200\,MHz. That we observe a turnover at such high frequencies, could be due to the limited surface brightness sensitivity of RACS-Low, as the galaxy shows extended emission at higher frequencies seen in the 2.1\,GHz ATCA images. Deeper ASKAP images taken as part of the Evolutionary Map of the Universe \citep[EMU;][]{emu_pilot, emu_main} survey will allow us to extend this investigation. 
A similar spectral shape, corresponding to the \texttt{SFG\_FFA1} model, is observed in ISZ\,399, IC\,4662, NGC\,625, and NGC\,5253.  
The model describes synchrotron emission from relativistic electrons mixed uniformly with a single volume of thermal free--free plasma. Building on this, and following \citet{Galvin_2018}, we propose that the electron population is mixed inhomogeneously with two distinct star-forming regions, which differ in their optical depths, with only one component becoming optically thick across the observed frequency range and external free--free absorption being observed. Although we only have GLEAM-X data for NGC\,625 regarding this fitting model, we observe a spectral behaviour that resembles findings by \citet{Grundy_2025} in some of their galaxies, e.g. NGC\,491. In both NGC\,625 and NGC\,491, there is a slight turnover around 400\,MHz followed by an increase in flux density at lower frequency. Aside from this similarity in the spectrum, the two galaxies do not share any other notable characteristics.

\section{RC - far infrared correlation}
\label{rcfir}
There is a tight astrophysical relationship between the RC and FIR luminosities in all star formation galaxies, the so-called RC--FIR correlation \citep[][]{Condon_1992}. The empirical luminosity scaling between these two frequency regimes is likely a relation between radio synchrotron emission and the FIR emission of cool dust heated by massive stars. \citet[][]{Yun_2001} derive the following RC--FIR luminosity correlation, 
\begin{equation}
    \log(L_{1.4\rm{GHz}}) = (0.99\pm0.01) \log(L_{60\upmu \rm{m}}/L_\odot)+(12.07\pm0.08)
    \label{rcfir_eq}
\end{equation}
where $\log(L_{1.4\rm{GHz}}[\rm{WHz^{-1}}]) = 20.08 +2\log(D) +\log(S_{1.4\rm{GHz}}[\rm{Jy}])$ and $\log(L_{60\upmu \rm{m}}[L_\odot]) = 6.014 +2\log(D) +\log(S_{60\upmu \rm{m}}[\rm{Jy}])$; $D$ is the luminosity distance in Mpc. The radio flux density  $S_{1.4\rm{GHz}}$ is the extrapolated value of the integrated flux density at 1.4\,GHz, which we have fitted in Sect.~\ref{sed}.

\begin{figure}
    \centering
    \includegraphics[width=\linewidth]{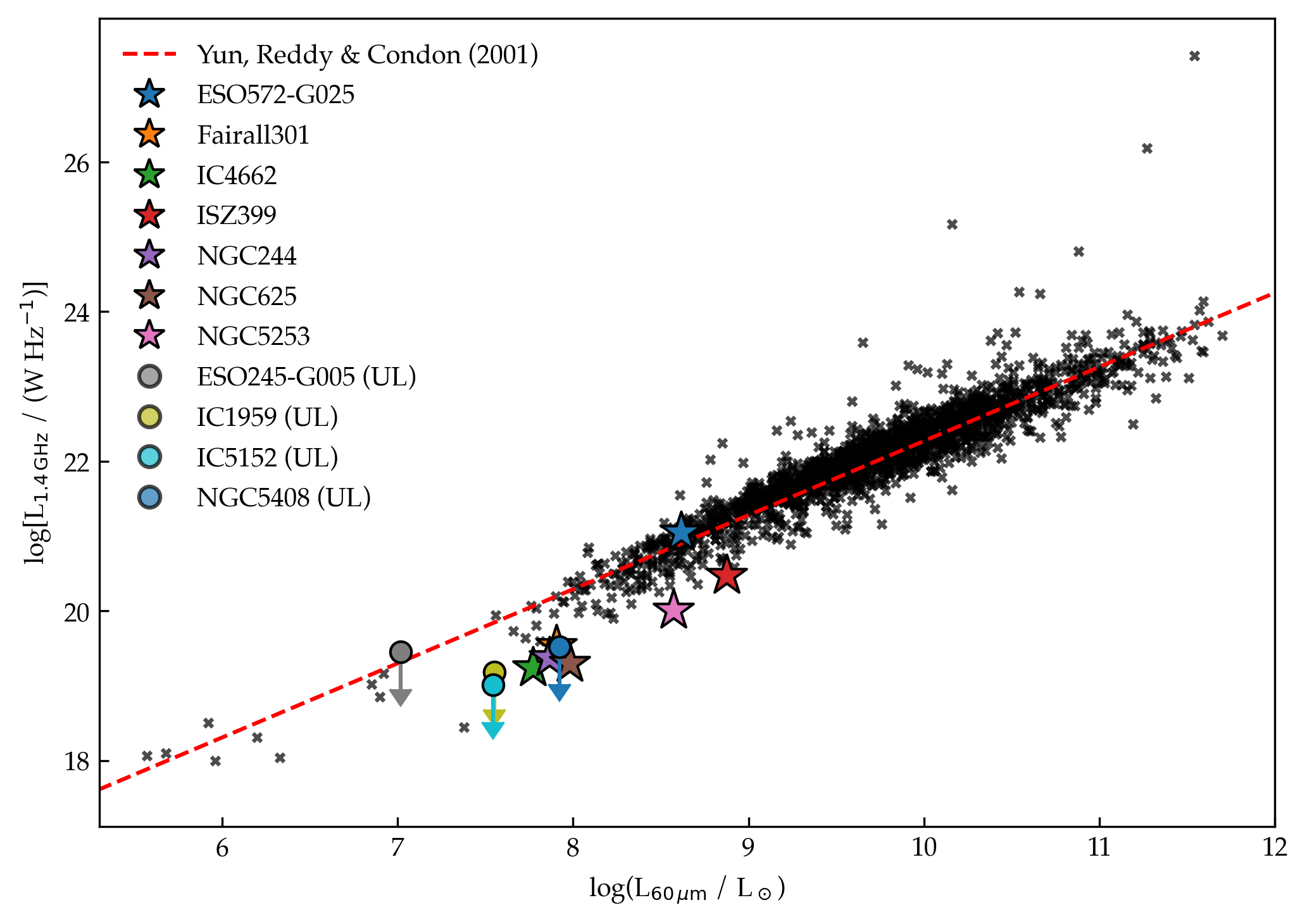}
    \caption{RC--FIR correlation from \citet[][]{Yun_2001}, showing the CHILLING galaxies overlaid. The diffuse RC detected galaxies are shown as stars, while the compact RC detected galaxies are marked as upper limits (UL).}
    \label{rcfir_plot}
\end{figure}

Fig.~\ref{rcfir_plot} presents the FIR–-RC relation for dwarf galaxies in the CHILLING sample that exhibit either diffuse or compact RC emission. For galaxies with diffuse RC emission, we use the SED analysis to extrapolate the 1.4\,GHz flux density and derive the corresponding non-thermal spectral index, indicated by the best-fit (see Table~\ref{sed_test} in Appendix~\ref{stats}). For galaxies with compact RC emission, we calculate the 1.4\,GHz flux density as an upper limit and adopt a fixed non-thermal spectral index of $-0.7$. These compact detections are therefore shown as upper limits in Fig.~\ref{rcfir_plot}.

In Fig.~\ref{rcfir_plot}, we find that nearly all dwarf galaxies show lower RC emission than expected based on their FIR luminosity, when compared to the reference relation from \citet[][]{Yun_2001}. Two main groupings are evident, one around a FIR luminosity of $\sim10^8\,L_\odot$ and another near $\sim10^9\,L_\odot$. 
Dwarf galaxies that show extended diffuse emission, such as IC\,4662, NGC\,625, or NGC\,5253, do not deviate from the general trend of lying below the reference line. On the contrary, these galaxies consistently exhibit a RC deficit.
The dwarf galaxy ESO\,572-G025 lies above the expected relation, likely due to its relatively high SFR and larger physical extent. This suggests that factors such as SFR and galaxy size may help retain CREs within the galaxy, allowing them to emit synchrotron radiation before escaping via galactic winds or outflows.
We also observe a general trend where higher SFRs correspond to higher FIR luminosities. However, this increase does not necessarily bring galaxies closer to the established FIR–-RC relation. Most dwarf galaxies appear under-luminous in the RC, which could either reflect the presence of very extended, diffuse emission that remains undetected by ATCA due to observational limitations, or more likely a true deficit in radio emission caused by CRE losses, as discussed in Sect.~\ref{rcfir_disc}. The latter scenario is favored given that our sample was observed for multiple hours and only a few galaxies show extended emission into the halo. For example, estimating the expected RC flux for NGC\,625 and IC\,4662 to lie on the \citet[][]{Yun_2001} correlation, we find that their missing flux would correspond to factors of 4.7 and 3.3, respectively. Such large discrepancies suggest that the observed deficit cannot be explained solely by sensitivity limits or missing zero spacing flux.

\section{Discussion}
\label{discussion}
\subsection{What triggers RC emission in dwarf galaxies?}
As in their more massive counterparts RC emission in dwarf galaxies is closely linked to star formation activity  \citep[][]{hindson_radio_2018}. This connection also holds for the CHILLING sample, where galaxies with higher SFRs are more likely to exhibit detectable RC emission. 
As shown in Fig.~\ref{detection}, other galaxy properties also influence the presence of RC emission. Absolute magnitude, linked to SFR of the galaxies, and near-infrared brightness ($L_{3.4\,\upmu\text{m}}$), used as proxy for stellar mass, show a strong correlation, reflecting the fact that more massive galaxies generally host more active star formation and more frequent core-collapse supernovae. In contrast, the total neutral atomic hydrogen (H{\sc i}) mass does not show a significant correlation with RC detectability. Although many dwarf galaxies contain substantial H{\sc i} reservoirs, this alone is not sufficient to generate RC emission. Star formation and the associated feedback from young stars is the key factor, as it drives the processes responsible for both thermal and non-thermal RC components.
Dwarf galaxies show in general lower RC as they often have large gas reservoirs but lack significant star formation as they are much smaller in size and also possibly due to low gas densities or internal feedback processes that inhibit the collapse of gas into stars. 
Similarly, rotational velocity, which is commonly used as an indicator of a galaxy’s dynamical mass, does not directly influence RC emission. While rotational velocity traces the overall gravitational potential, it does not necessarily reflect the physical conditions needed for star formation. This is especially true in low-mass systems like dwarf galaxies, where internal feedback (e.g., stellar winds, supernova outflows) can disturb gas dynamics, suppress ordered rotation, and limit star-forming activity. Additionally, non-circular motions in such systems make rotational velocity an unreliable proxy for the energetic processes that drive synchrotron emission.
In summary, neither the total H{\sc i} gas mass nor the rotational velocity alone can account for the presence of RC emission. Instead, RC emission primarily traces recent star formation activity, which depends on localised physical conditions such as gas density, turbulence, and thermal stability. Therefore, while the availability of gas and the galaxy’s dynamics provide important context, they are not sufficient on their own to trigger the RC emission observed in star-forming galaxies.

\subsection{Physical implications of the different RC classification types}
The `diffuse', `compact' and `no/marginal' RC classification reveals systematic differences that suggest distinct physical regimes. Dwarf galaxies with diffuse RC emission tend to have higher integrated SFRs and substantially larger 60 micron luminosities than the `compact' or `no/marginal' systems (mean $\log({\rm SFR_{H\alpha}}) \approx -0.9$ for `diffuse' vs $\approx -1.5$ for `compact' and `no/marginal' in our sample), indicating more spatially extended CR injection. This elevated star-formation activity plausibly provides both the CR electron source term and the turbulent energy required to amplify magnetic fields on galactic scales, producing volume-filling synchrotron emission \citep[e.g.][]{Chyzy_2011,Beck_2015}. Conversely, `compact' RC detections are associated with bright, localised H{\sc ii} complexes but lack disk-wide emission, consistent with CR electrons being largely confined to their birth sites. This may reflect short propagation lengths (low diffusion coefficients \citep[][]{Heesen_2021}, rapid vertical escape via localised outflows \citep[][]{Wang_2022}, or weak/inhomogeneous large-scale magnetic fields that limit CR diffusion \citet[][]{Heesen_2021}). The compact subsample also shows slightly higher mean rotational velocities, suggesting that ordered shear alone is not sufficient for sustaining large-scale synchrotron halos, instead, the interplay of CR injection, turbulent field amplification, and transport efficiency appears to control whether RC emission becomes diffuse. Systems with `no/marginal' RC emission typically have the lowest SFR and FIR luminosity, consistent with insufficient CR production to yield a detectable synchrotron signal given our surface-brightness sensitivity \citep[][]{Yun_2001}. Observational factors such as lack of sensitivity of the ATCA telescopes may further bias against detecting low-surface-brightness emission. 

\subsection{Do we need different models to explain CRE spectra?}

\begin{figure*}
    \centering
    \includegraphics[width=\linewidth]{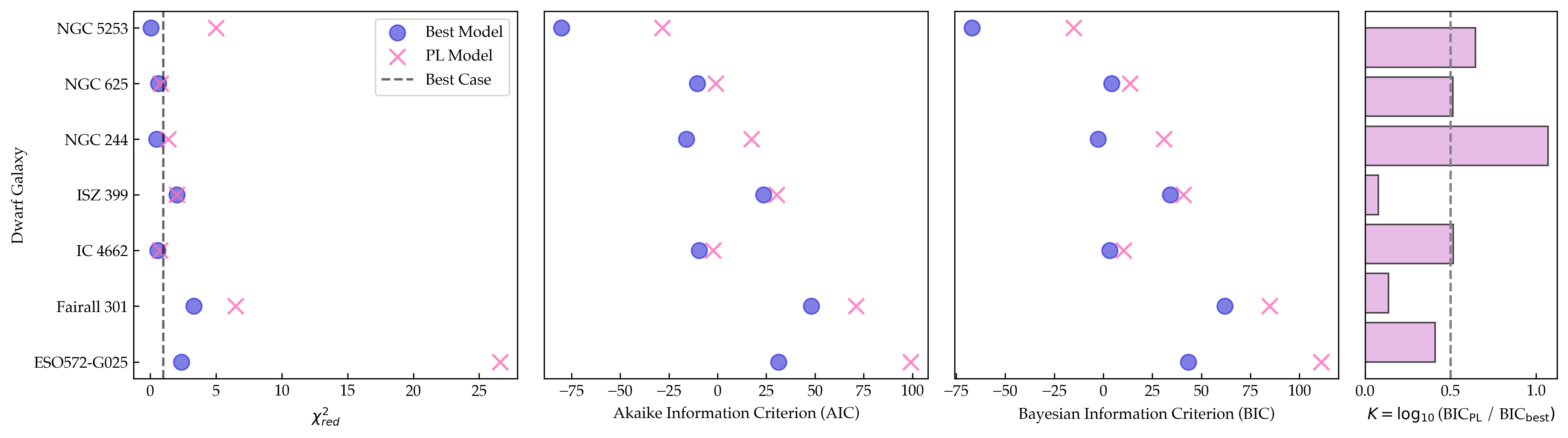}
     \caption{Comparison between the best-fit model and the simple power-law (\texttt{PL}) model using statistical criteria: reduced chi-squared ($\chi^2_{\text{red}}$), Akaike Information Criterion (AIC), Bayesian Information Criterion (BIC), and Jeffreys’ scale \citep[][]{kass_scale}. The black dotted line in the left panel marks the reference value of $\chi^2_{\text{red}} = 1$, while the black dotted line in the right panel indicates the threshold where the the best-fit model are significantly preferred over the simple power-law.}
    \label{statistics}
\end{figure*}

\begin{figure}
    \centering
    \includegraphics[width=\linewidth]{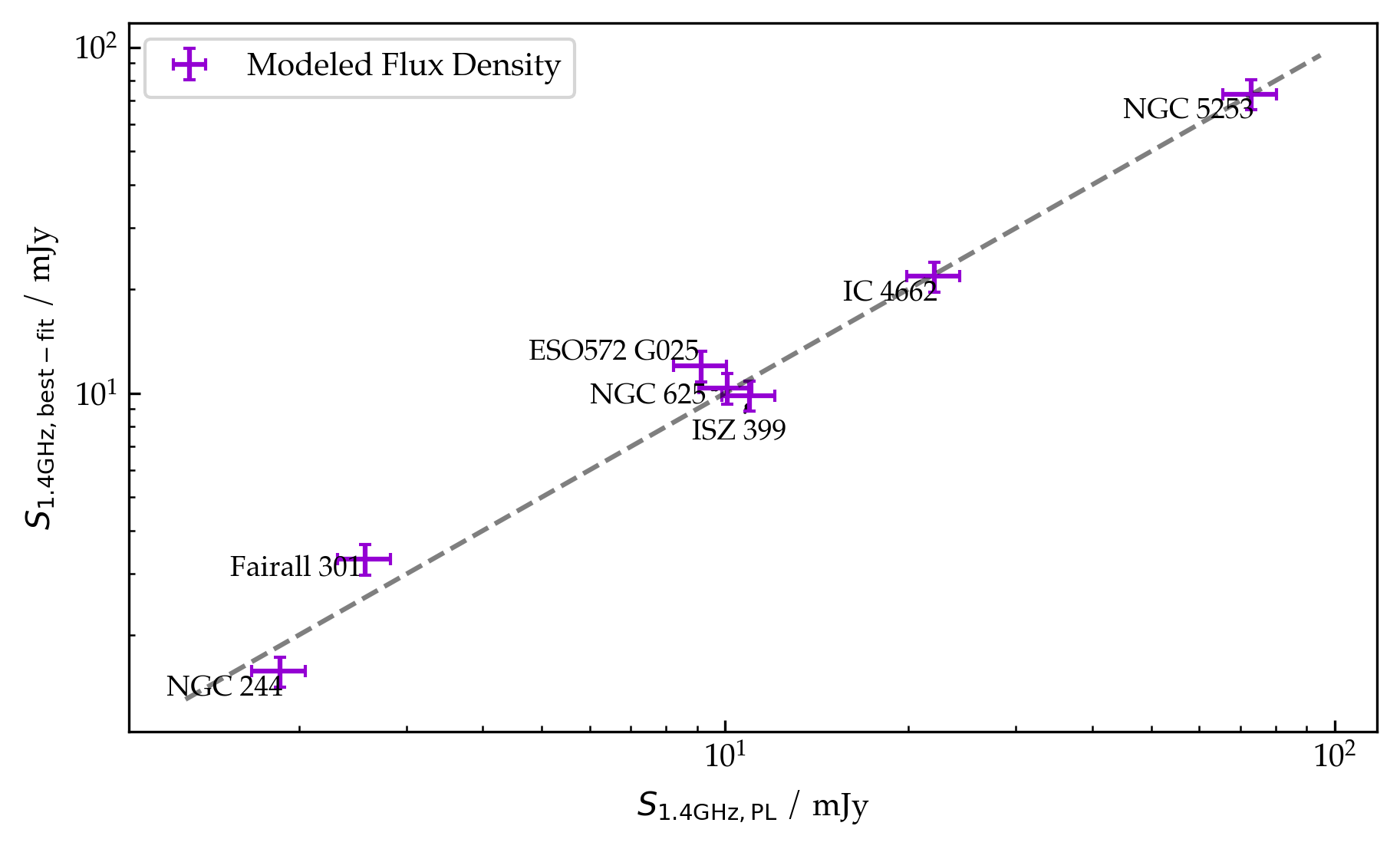}
    \caption{Comparison of the extrapolated flux density $S_{\rm 1.4\,GHz}$ at frequency of 1.4\,GHz, as derived from the best-fit model and the \texttt{PL} model. The uncertainties for both methods are estimated to be 10\,\%.}
    \label{fluxcomp}
\end{figure}

An important question regarding the RC emission is whether complex models are truly necessary or if a simple power-law (\texttt{PL}) is sufficient to explain the spectra. Many dwarf galaxy spectra clearly show curved shapes, with steepening or flattening over several GHz and some show a turnover at low frequencies. 
For a statistical relevant analysis, we compare the reduced chi-square values, the Akaike information criterion \citep[AIC;][]{aic} and the Bayesian information criterion \citep[BIC;][]{bic} of the best-fit models with those of the simple power-law. The AIC and BIC are model selection metrics that evaluate the relative quality of statistical models by balancing goodness of fit with model complexity, where AIC applies a lighter penalty and is more suitable for predictive accuracy, while BIC imposes a stronger penalty, particularly favoring simpler models as sample size increases. Lower values of the AIC or BIC indicate a better balance between model fit and complexity. Therefore, when comparing models, the preferred model is the one with the lowest AIC or BIC value, regardless of whether the values are negative or positive.
In Fig.~\ref{statistics}, we show the comparison of these values of the best-fit models with those of the simple power-law and it reveals that more complex models are indeed required to capture the spectral behaviour and underlying physics. In few cases, such as IC\,4662, NGC\,625, ISZ\,399 and NGC\,244, the reduced chi-squared of the \texttt{PL} is closer to the best value but we see that for each galaxy the AIC and BIC of the best-fit is smaller then the simple power-law, suggesting that we are in need of a more complex model to explain the CRE spectra.
Taking now the Jeffreys' scale \citep[][]{kass_scale}, where the logarithm of the Bayes factor quantifies the strength of evidence in favor of one model over another, we observe in the right panel of Fig.~\ref{statistics} that only for NGC\,244, IC\,4662, NGC\,625 and NGC\,5253 the difference is significant and we need to take a more complex model into account. Looking at the difference of the best-fit model and the simple PL model in Fig.~\ref{sed}, we observe already by eye that we definitely need a complex model for ESO\,572-G025 and Fairall\,301, which is not visible through the statistical test.
The full model parameters for the best-fit of each galaxy are listed in Table~\ref{sed_test} in Appendix~\ref{stats}. Comparing the flux density obtained from the best-fit model and the \texttt{PL}, we observe in Fig.~\ref{fluxcomp}, that the difference is minimal. Only for a few galaxies, such as ESO\,572-G025 or Fairall\,301, the difference is larger than 10\,\%. For these two galaxies, especially for  ESO\,572-G025, the statistical tests differ most in between the simple power-law and the complex model.

\subsection{Does SFR trigger free--free absorption in dwarf galaxies?}
Normally, a high SFR suggest an optically thick regime in H$\alpha$, which leads to free--free absorption, but we do not observe a correlation between the different model components and other characteristics of the galaxies, such as mass, FIR or SFR.
Unfortunately, our sample is too small to draw conclusion if the corresponding model components depend on some of these characteristics.

\subsection{Do dwarf galaxies obey the RC--FIR correlation?}
\label{rcfir_disc}
For galaxies with a FIR luminosity of $\log(L_{60\upmu \rm{m}}/L_\odot)<10^9$, \citet[][]{Yun_2001} already observed an increase of asymmetric scatter, such that sources appear to be under-luminous in radio luminosity $\log(L_{1.4\rm{GHz}})$, compared to the best fit. Such a deviation can occur if the FIR or RC is not directly proportional to the star formation activity. Dust heated by low-mass stars may contribute only to the FIR. Low luminosity, low mass galaxies may lose more radio emission than high luminosity objects to cosmic ray diffusion. \citet[][]{Roychowdhury_2012} studied the RC--FIR correlation in faint irregular galaxies by stacking images of individual galaxies together and finds that both the Spitzer $70\,\upmu$m and 1.4\,GHz fluxes are generally lower than expected from what is expected in spiral galaxies, however the ratio of RC to FIR appears to be consistent with the larger spiral galaxies. \citet[][]{hindson_radio_2018} analysed the RC--FIR correlation with a small dwarf galaxies sample and found that dwarf galaxies contain less dust than spiral galaxies, making them fainter in the FIR for a given level of radio emission, in contrast to our study. 

The low mass sample, analysed by \citet[][]{shao_local_2018}, exhibit lower radio emissions compared to FIR emissions. This is indicated by the analysis of the RC deficiency in dwarf galaxies, which is primarily attributed to factors like a lower stellar mass surface density and a higher H{\sc i}-to-stellar mass ratio. These findings align with our results, that dwarf galaxies exhibit lower radio emission than FIR. The lower radio emission in dwarf galaxies, compared to their FIR emissions, is attributed to several factors. \citet[][]{shao_local_2018} explains it that primarily, dwarf galaxies have lower gas surface densities, which affects the production of cosmic rays that contribute to radio emission. Furthermore, the lower stellar mass surface density in these galaxies results in less efficient conversion of star formation activity into radio emissions. Additionally, in regions of low gas surface density, the escaping CRs may lead to diminished radio output, while the low dust content leads to lower FIR emission, creating a compensatory effect. 
We attribute the observed radio deficit in the RC--FIR correlation to significant energy losses experienced by cosmic ray electrons. Losses, such as synchrotron radiation and inverse Compton scattering which we can observe in the breaks in the SEDs in Fig.~\ref{sed}, affect the total energy of CRs available to generate radio emissions. In environments with lower gas surface densities, such as those typical for dwarf galaxies, these losses can be more pronounced, leading to a relative deficit in the radio output compared to FIR emissions.

\section{Conclusions} 
\label{concl}
We present a RC study of the CHILLING sample, consisting of 11 dwarf irregular galaxies and 4 BCDs spanning a broad range of stellar masses and star formation histories. The combination of multi-band observations with ATCA provides the opportunity to study the full RC spectra and leads us to the following conclusions. 
\begin{enumerate}
    \item We detect RC emission in 11 of the 15 dwarf galaxies, with 7 exhibiting more diffuse emission and 4 displaying more compact emission. The detectability of RC emission in dwarf galaxies is closely linked to recent star formation activity, as all 4 BCDs are detected. Galaxies with higher star formation rates are more likely to exhibit RC emission, which also correlates with absolute magnitude and stellar mass. Most RC-detected galaxies are classified as starbursts. In contrast, parameters such as H{\sc i} mass and rotational velocity show no significant correlation with RC detectability, indicating that RC emission is more strongly influenced by star formation than by galaxy dynamics.

    \item The RC spectra of many dwarf galaxies in the sample display complex, curved features, such as spectral steepening, flattening, and low-frequency turnovers, that deviate markedly from a simple power-law behavior. Statistical evaluations using reduced chi-squared, AIC, and BIC confirm that single power-law models inadequately describe the observed spectral shapes and fail to capture the underlying physical processes. These findings underscore the need for more sophisticated models incorporating mechanisms such as thermal emission, free--free absorption, and spectral breaks arising from cosmic ray electron energy losses.

    \item In particular, the pronounced steepening of RC spectra at higher frequencies is attributed to substantial energy losses experienced by cosmic ray electrons. These losses, primarily due to synchrotron radiation and inverse Compton scattering, become increasingly significant at higher electron energies, leading to the observed decline in flux density.

    \item Nearly all dwarf galaxies in the sample show a deficit in RC emission relative to their FIR luminosity when compared to the correlation seen in more massive galaxies. This deficit is best explained by significant energy losses of cosmic ray electrons, primarily via synchrotron radiation and inverse Compton scattering, which are more effective in the low-density environments typical of dwarf galaxies. 

    \item The possibility that extended low-radio-brightness diffuse emission remains undetected due to the sensitivity limitations of ATCA may partially reflect these instrumental constraints.
    However, the spectral properties observed across the sample more strongly indicate that energy losses of CREs are the primary drivers of both the spectral steepening and the RC deficit.
    
\end{enumerate}

\begin{acknowledgements}
ST, BA, DJB and MS acknowledge the support from the DFG via the Collaborative Research Center SFB1491 \textit{Cosmic Interacting Matters - From Source to Signal}. PK acknowledge the support of the BMBF project 05A23PC1 for D-MeerKAT.
The Australia Telescope Compact Array is part of the Australia Telescope National Facility (https://ror.org/05qajvd42) which is funded by the Australian Government for operation as a National Facility managed by CSIRO. We acknowledge the Gomeroi people as the Traditional Owners of the Observatory site.
\end{acknowledgements}

\bibliographystyle{aa} 
\bibliography{literatur} 

\begin{appendix}
\onecolumn
\section{Radio continuum properties}
\label{RCmaps}
\begin{figure*}[h!]
    \centering
    \includegraphics[width=0.49\linewidth]{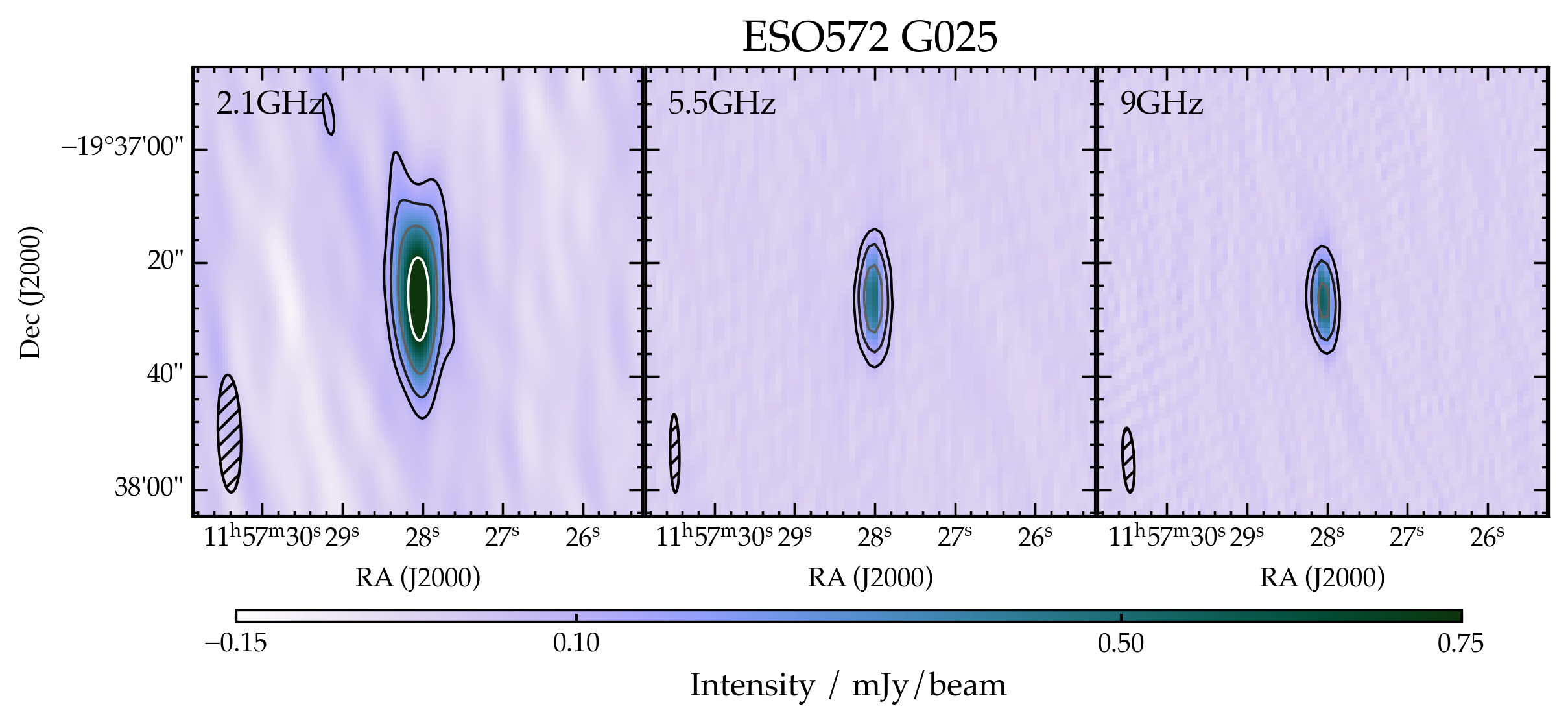}
    \includegraphics[width=0.49\linewidth]{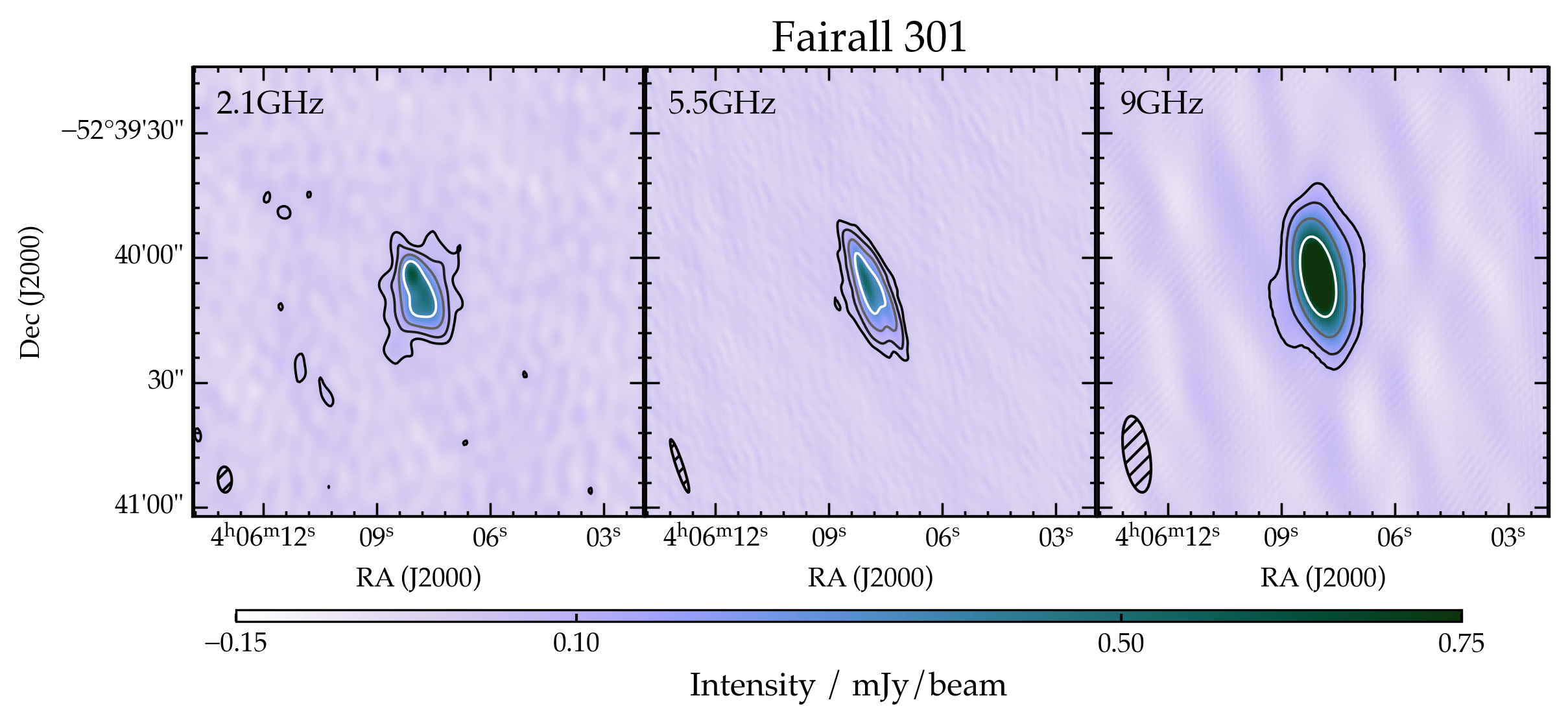}
    \includegraphics[width=0.49\linewidth]{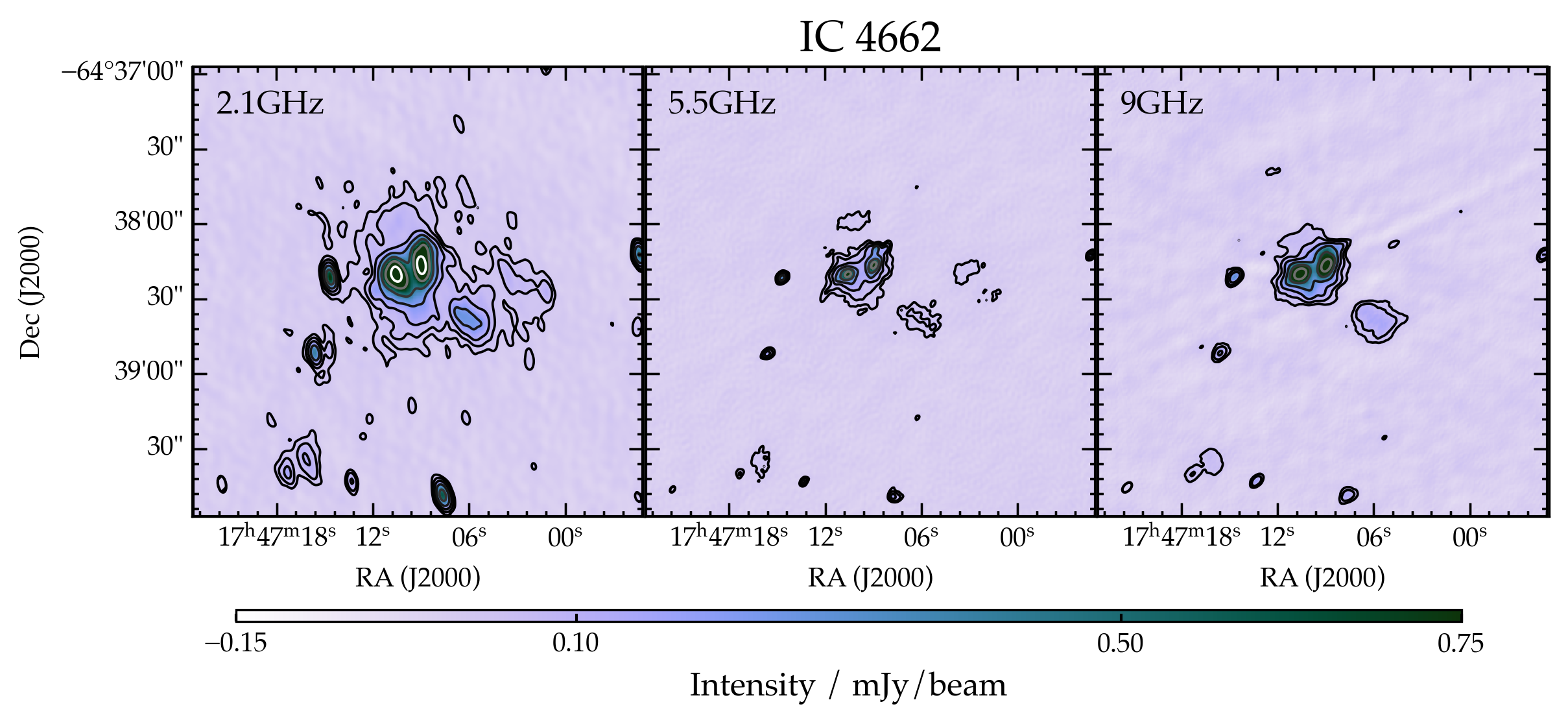}
    \includegraphics[width=0.49\linewidth]{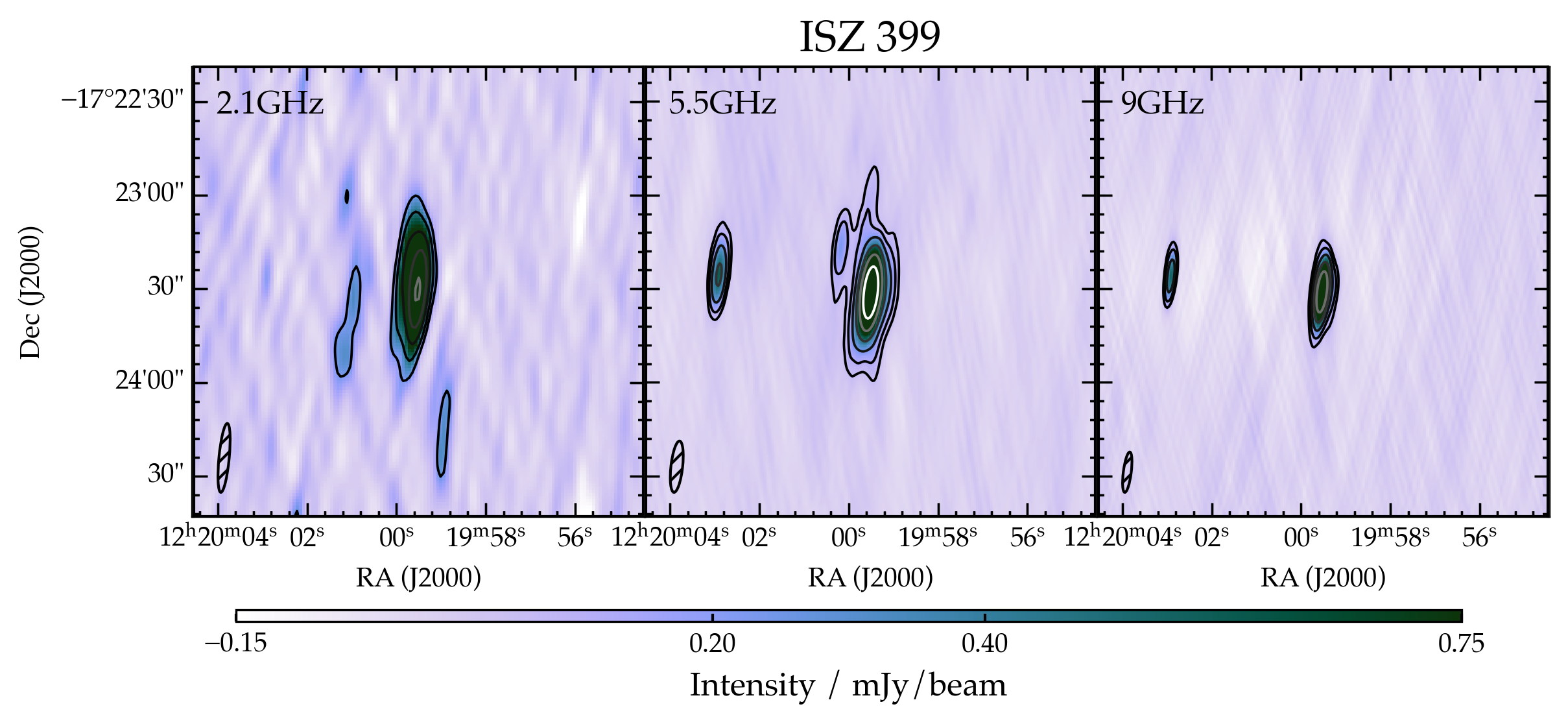}
    \includegraphics[width=0.49\linewidth]{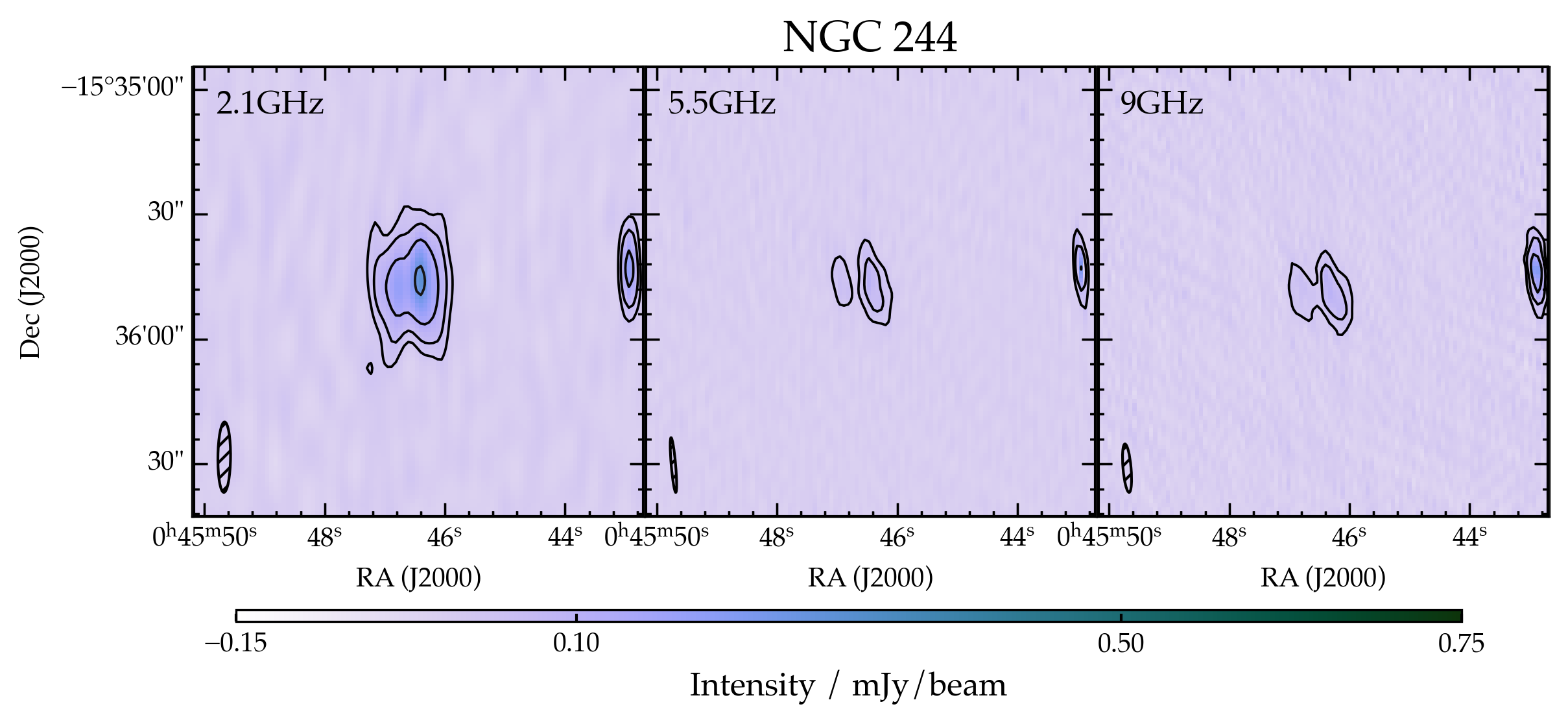}
    \includegraphics[width=0.49\linewidth]{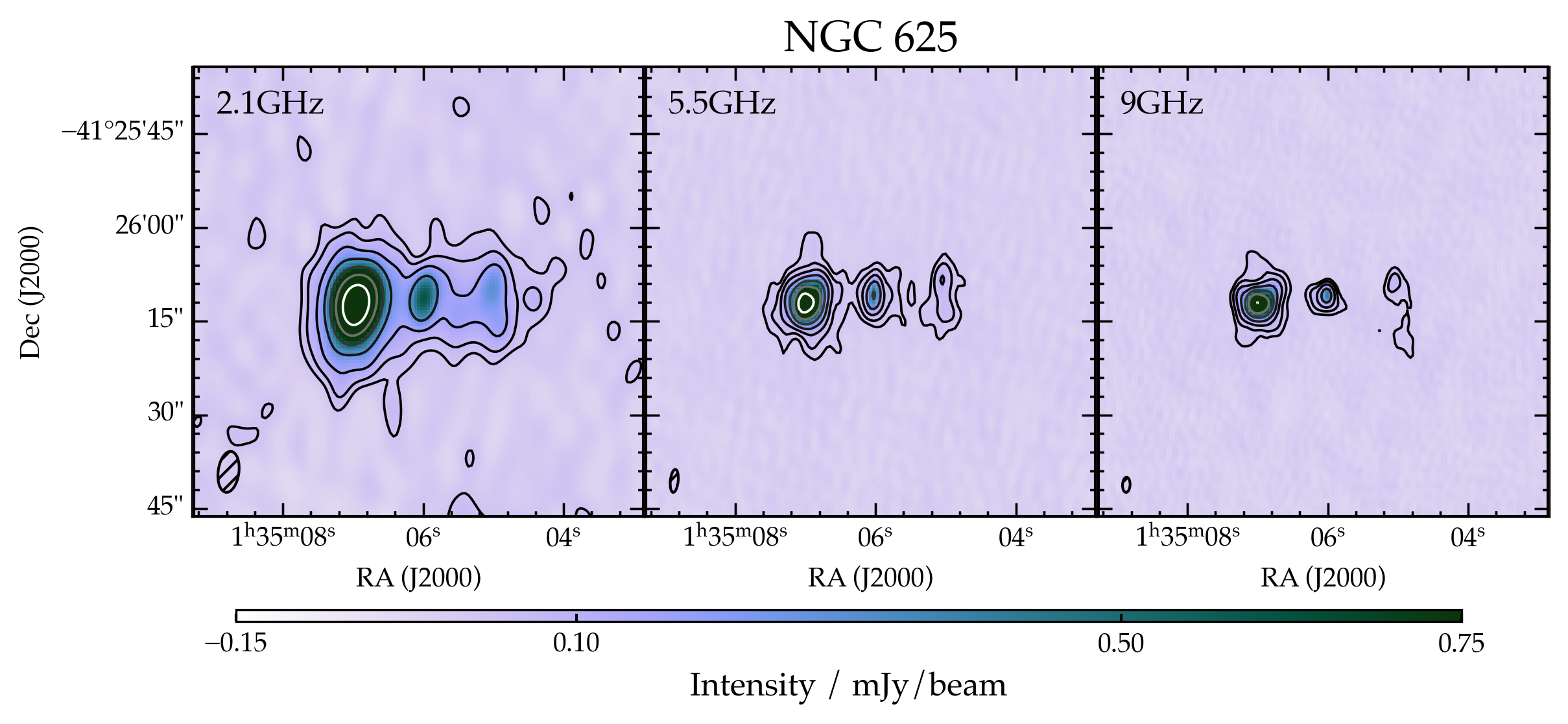}
    \includegraphics[width=0.49\linewidth]{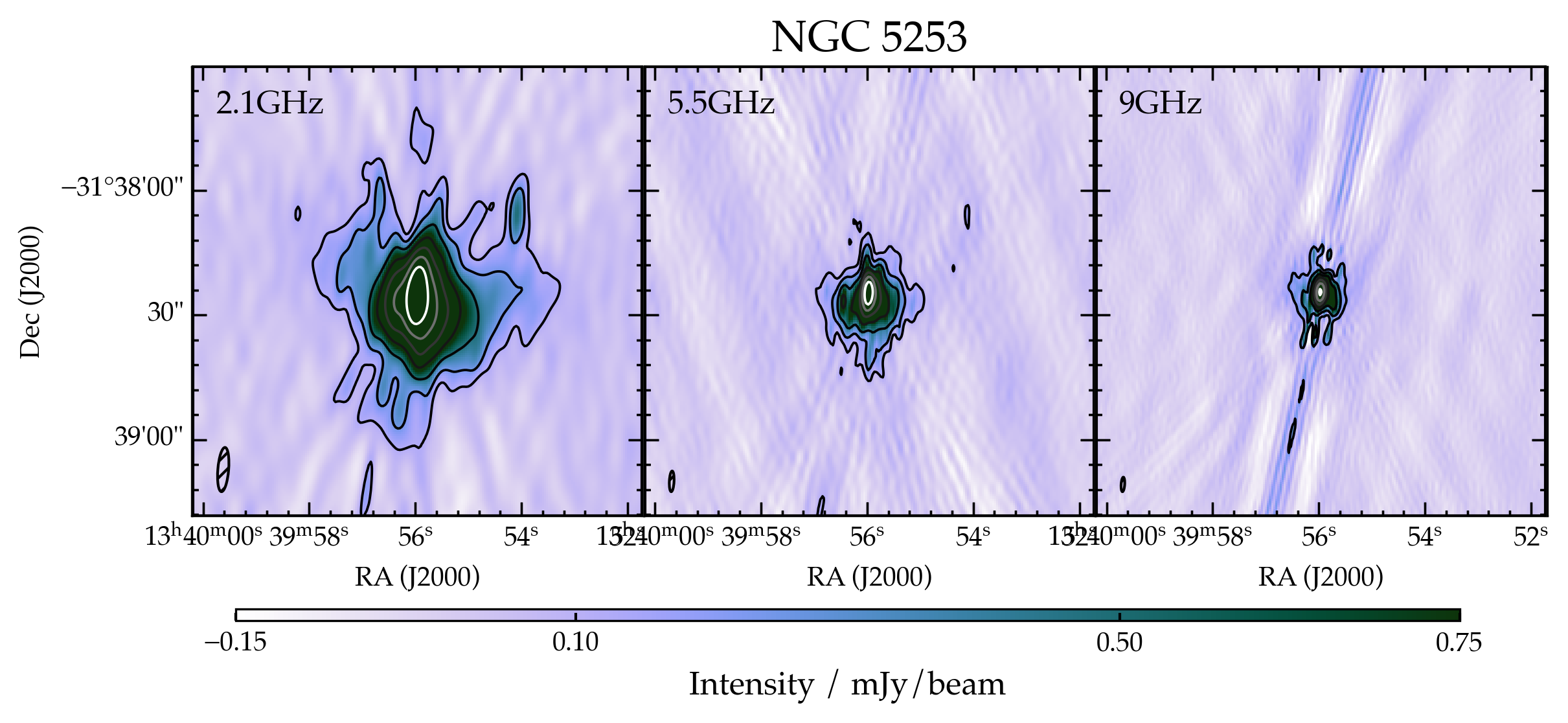}
    \caption{\textit{First row left panel:} Total intensity emission of ESO\,572\,G\,025 at the central frequency of 2.1 ($\sigma=30\,\upmu$Jy/beam), 5.5 ($\sigma=20\,\upmu$Jy/beam) and 9\,GHz ($\sigma=30\,\upmu$Jy/beam) with overlaid contours starting at $3\,\sigma$ and increasing by factor of two. 
    \textit{First row right panel:} Total intensity emission of Fairall\,201 at the central frequency of 2.1 ($\sigma=14\,\upmu$Jy/beam), 5.5 ($\sigma=12\,\upmu$Jy/beam) and 9\,GHz ($\sigma=25\,\upmu$Jy/beam) with overlaid contours starting at $3\,\sigma$ and increasing by factor of two.
    \textit{Second row left panel:} Total intensity emission of IC\,4662 at the central frequency of 2.1 ($\sigma=9\,\upmu$Jy/beam), 5.5 ($\sigma=8\,\upmu$Jy/beam) and 9\,GHz ($\sigma=12\,\upmu$Jy/beam) with overlaid contours starting at $3\,\sigma$ and increasing by factor of two.
    \textit{Second row right panel:} Total intensity emission of ISZ\,399 at the central frequency of 2.1 ($\sigma=73\,\upmu$Jy/beam), 5.5 ($\sigma=20\,\upmu$Jy/beam) and 9\,GHz ($\sigma=25\,\upmu$Jy/beam) with overlaid contours starting at $3\,\sigma$ and increasing by factor of two.
    \textit{Third row left panel:} Total intensity emission of NGC\,244 at the central frequency of 2.1 ($\sigma=10\,\upmu$Jy/beam), 5.5 ($\sigma=8\,\upmu$Jy/beam) and 9\,GHz ($\sigma=8\,\upmu$Jy/beam) with overlaid contours starting at $3\,\sigma$ and increasing by factor of two.
    \textit{Third row right panel:} Total intensity emission of NGC\,625 at the central frequency of 2.1 ($\sigma=12\,\upmu$Jy/beam), 5.5 ($\sigma=8\,\upmu$Jy/beam) and 9\,GHz ($\sigma=8\,\upmu$Jy/beam) with overlaid contours starting at $3\,\sigma$ and increasing by factor of two.
    \textit{Fourth row panel:} Total intensity emission of NGC\,5252 at the central frequency of 2.1 ($\sigma=48\,\upmu$Jy/beam), 5.5 ($\sigma=35\,\upmu$Jy/beam) and 9\,GHz ($\sigma=50\,\upmu$Jy/beam) with overlaid contours starting at $3\,\sigma$ and increasing by factor of two. The beam is shown in the left corner.}
    \label{detections_diffuse}
\end{figure*}

\begin{figure*}[h!]
    \centering
    \includegraphics[width=0.49\linewidth]{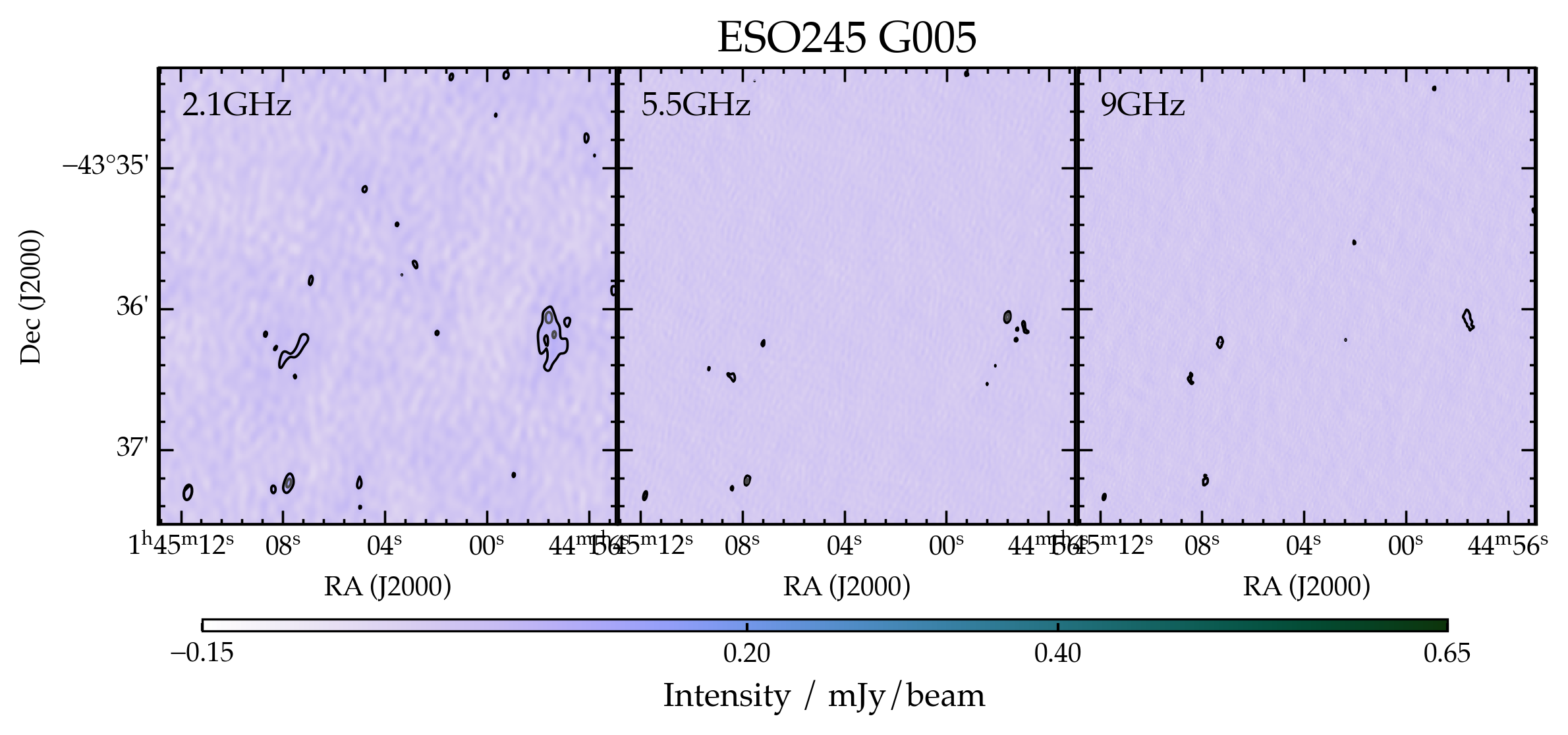}
    \includegraphics[width=0.49\linewidth]{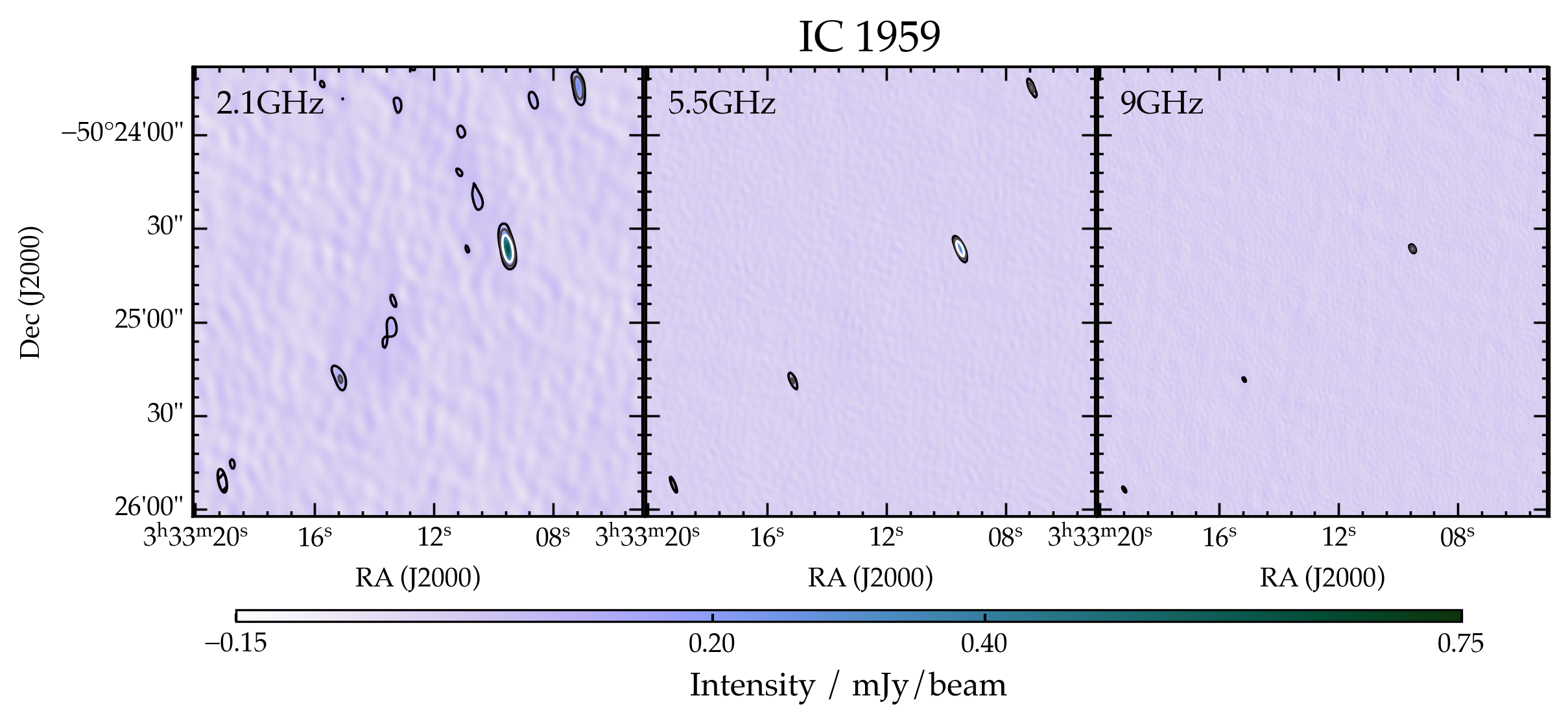}
    \includegraphics[width=0.49\linewidth]{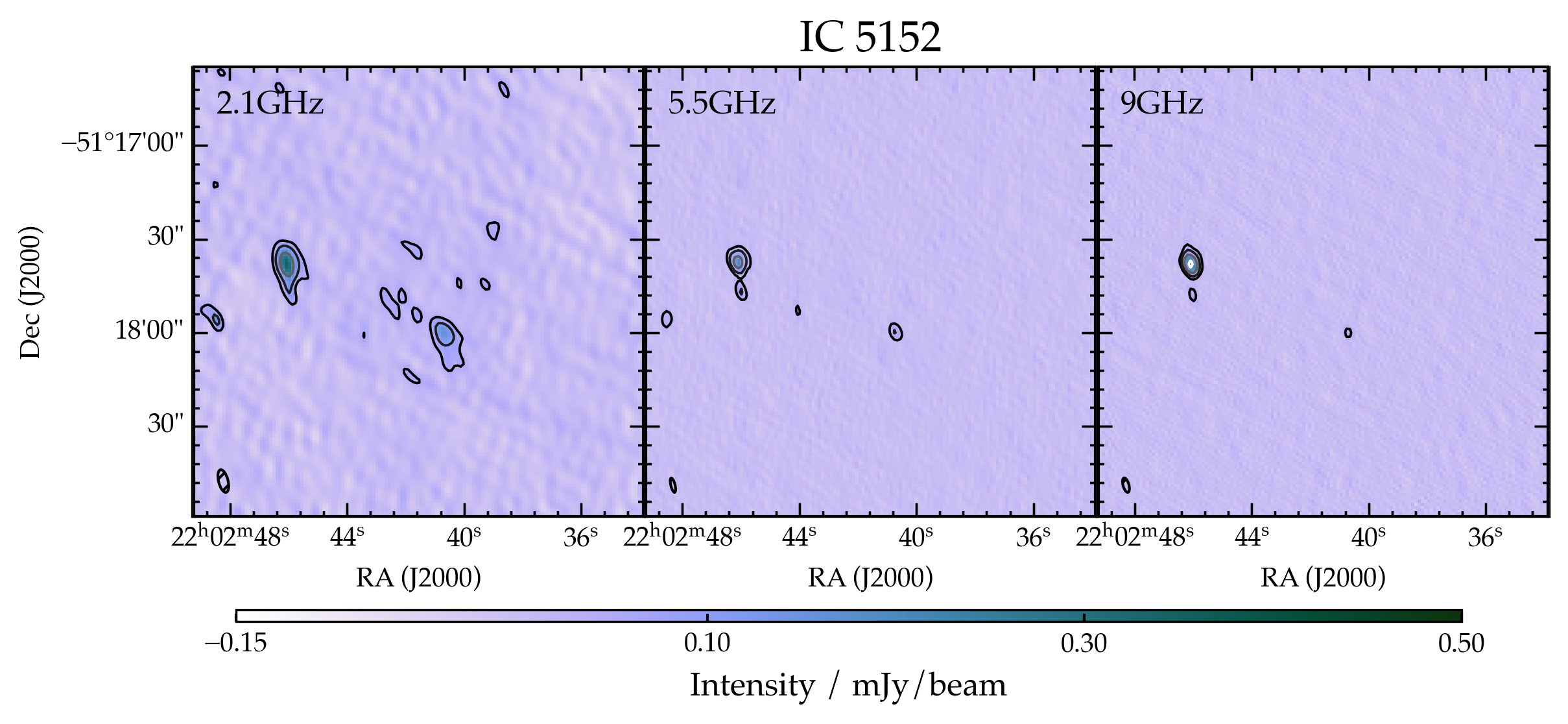}
    \includegraphics[width=0.49\linewidth]{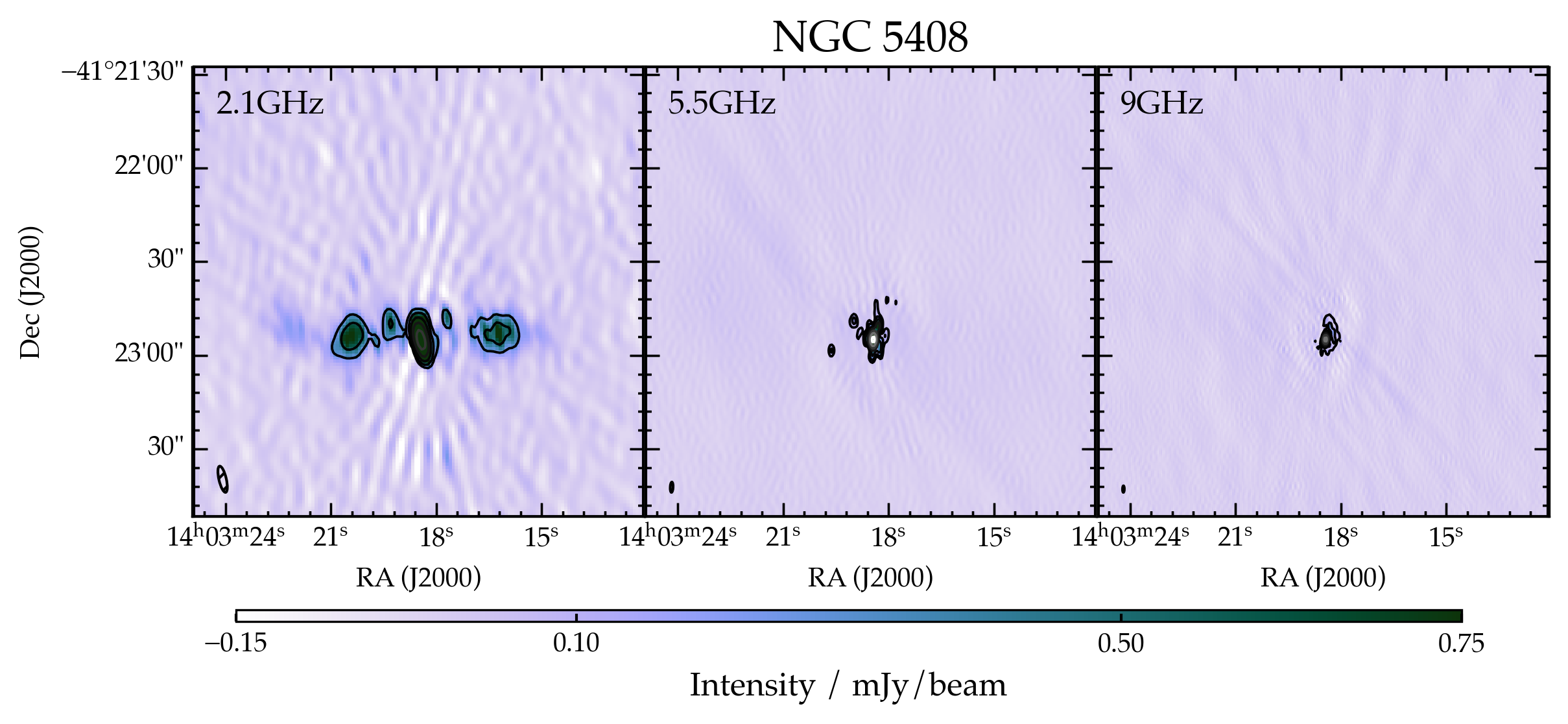}
    \caption{
    \textit{First row left panel:} Total intensity emission of ESO\,245\,G\,005 at the central frequency of 2.1 ($\sigma=15\,\upmu$Jy/beam), 5.5 ($\sigma=6\,\upmu$Jy/beam) and 9\,GHz ($\sigma=6\,\upmu$Jy/beam) with overlaid contours starting at $3\,\sigma$ and increasing by factor of two.
    \textit{First row right panel:} Total intensity emission of IC\,1959 at the central frequency of 2.1 ($\sigma=20\,\upmu$Jy/beam), 5.5 ($\sigma=10\,\upmu$Jy/beam) and 9\,GHz ($\sigma=10\,\upmu$Jy/beam) with overlaid contours starting at $3\,\sigma$ and increasing by factor of two.
    \textit{Second row left panel:} Total intensity emission of IC\,5152 at the central frequency of 2.1 ($\sigma=15\,\upmu$Jy/beam), 5.5 ($\sigma=8\,\upmu$Jy/beam) and 9\,GHz ($\sigma=8\,\upmu$Jy/beam) with overlaid contours starting at $3\,\sigma$ and increasing by factor of two.
    \textit{Second row right panel:} Total intensity emission of NGC\,5408 at the central frequency of 2.1 ($\sigma=85\,\upmu$Jy/beam), 5.5 ($\sigma=17\,\upmu$Jy/beam) and 9\,GHz ($\sigma=18\,\upmu$Jy/beam) with overlaid contours starting at $3\,\sigma$ and increasing by factor of two.
    The beam is shown in the left corner.}
    \label{detections_compact}
\end{figure*}

\begin{figure*}[h!]
    \centering
    \includegraphics[width=0.49\linewidth]{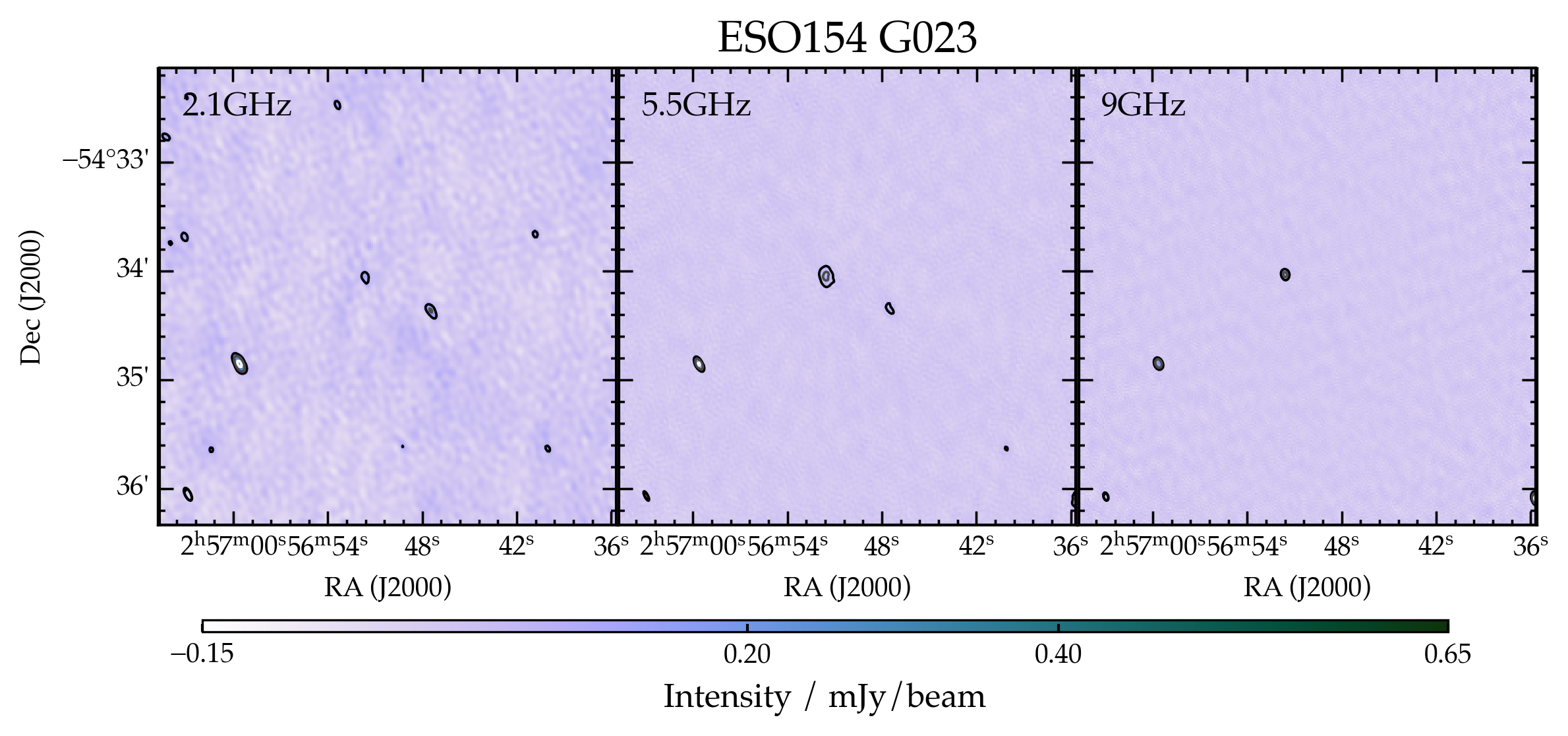}
    \includegraphics[width=0.49\linewidth]{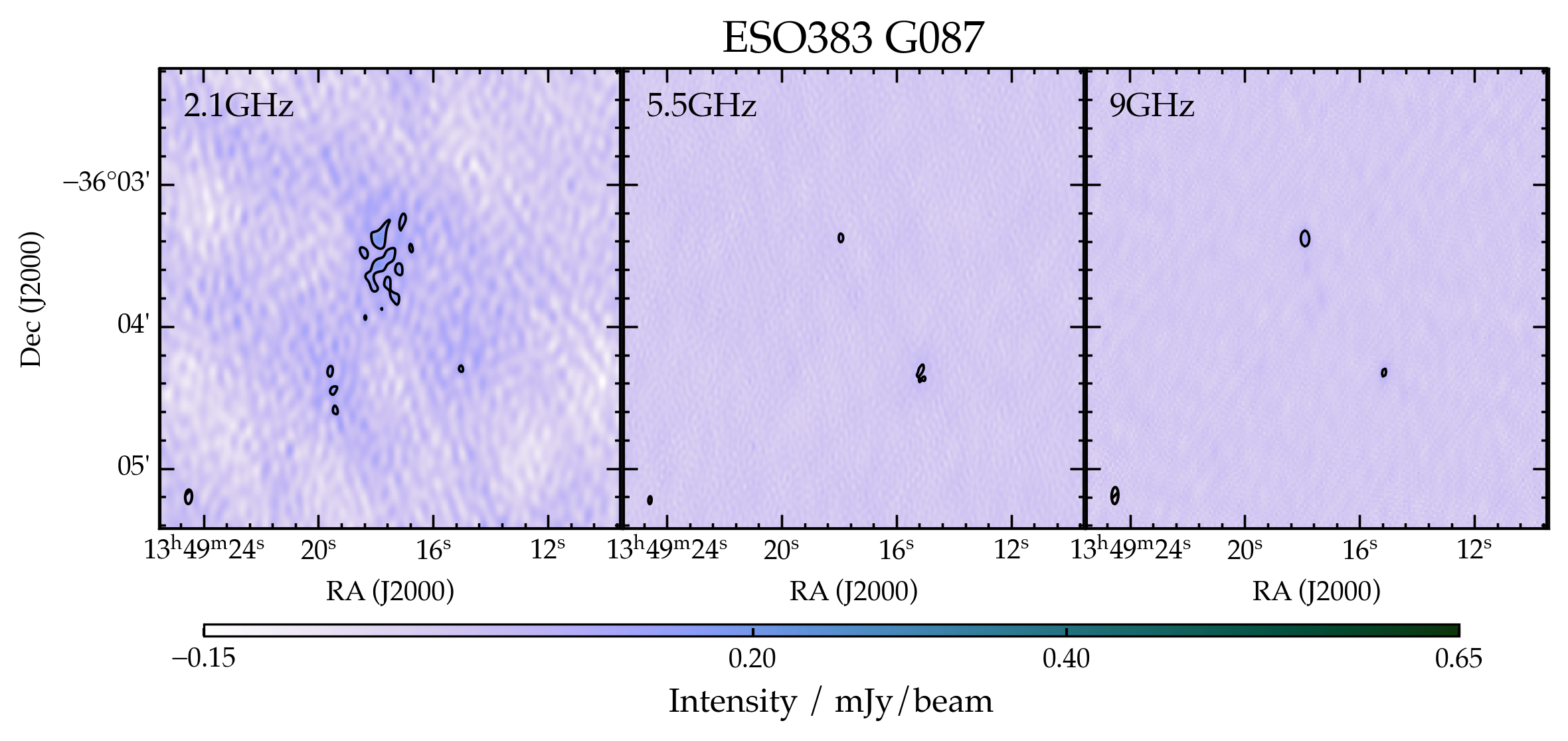}
    \includegraphics[width=0.49\linewidth]{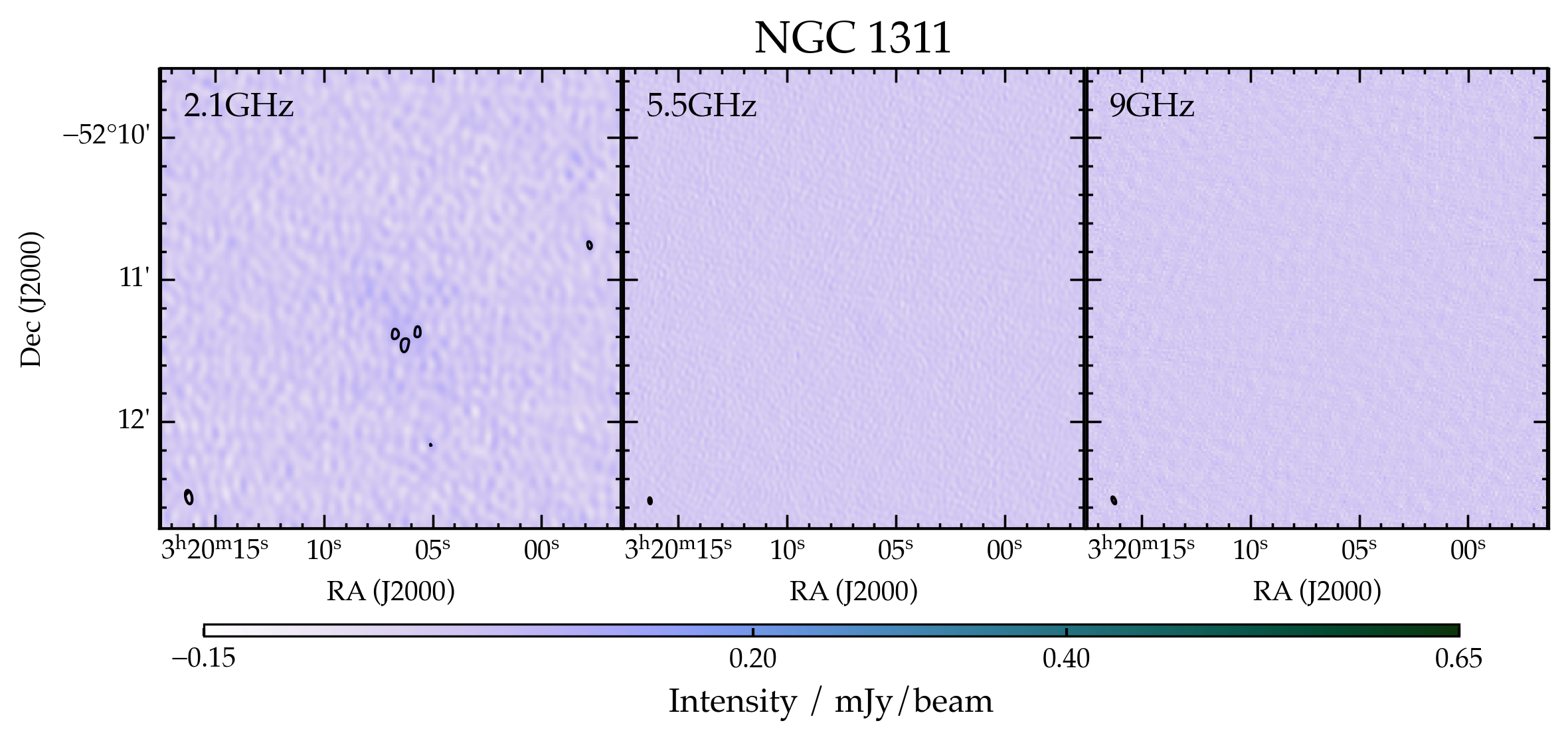}
        \includegraphics[width=0.49\linewidth]{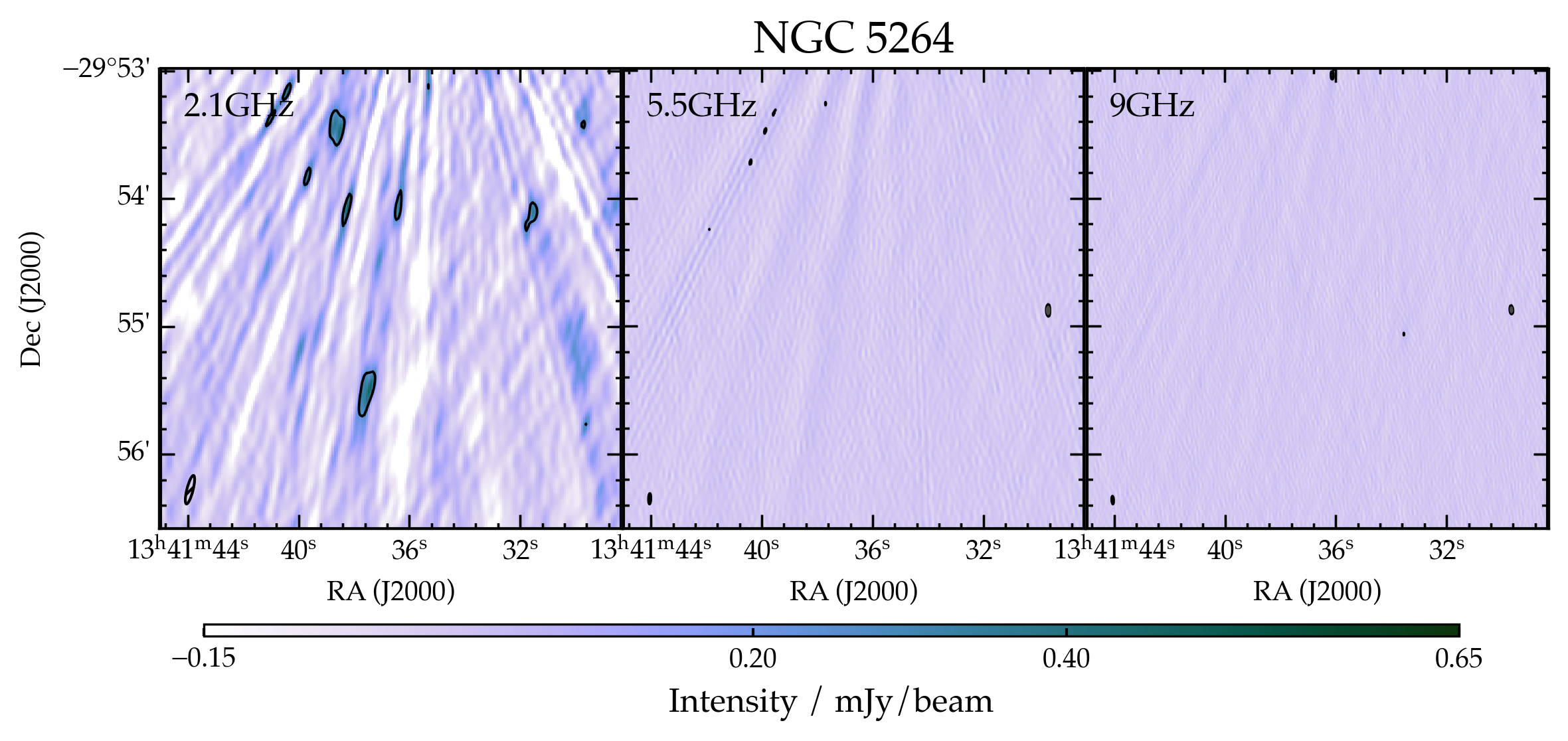}
    \caption{\textit{First row left panel:} Total intensity emission of ESO\,154\,G\,023 at the central frequency of 2.1 ($\sigma=40\,\upmu$Jy/beam), 5.5 ($\sigma=10\,\upmu$Jy/beam) and 9\,GHz ($\sigma=12\,\upmu$Jy/beam) with overlaid contours starting at $3\,\sigma$ and increasing by factor of two. 
    \textit{First row right panel:} Total intensity emission of ESO\,383\,G\,087 at the central frequency of 2.1 ($\sigma=40\,\upmu$Jy/beam), 5.5 ($\sigma=15\,\upmu$Jy/beam) and 9\,GHz ($\sigma=20\,\upmu$Jy/beam) with overlaid contours starting at $3\,\sigma$ and increasing by factor of two.
    \textit{Second row left panel:} Total intensity emission of NGC\,1311 at the central frequency of 2.1 ($\sigma=27\,\upmu$Jy/beam), 5.5 ($\sigma=15\,\upmu$Jy/beam) and 9\,GHz ($\sigma=10\,\upmu$Jy/beam) with overlaid contours starting at $3\,\sigma$ and increasing by factor of two.
    \textit{Second row right panel:} Total intensity emission of NGC\,5264 at the central frequency of 2.1 ($\sigma=80\,\upmu$Jy/beam), 5.5 ($\sigma=20\,\upmu$Jy/beam) and 9\,GHz ($\sigma=13\,\upmu$Jy/beam) with overlaid contours starting at $3\,\sigma$ and increasing by factor of two.
    The beam is shown in the left corner.}
    \label{non-detections}
\end{figure*}

In this section, we present the RC maps of the ATCA data of L/S-band, C-band and X-band. Fig.~\ref{detections_diffuse} presents the RC maps of dwarf galaxies with diffuse radio emission. Fig.~\ref{detections_compact} shows the RC maps of galaxies with compact RC emission. Fig.~\ref{non-detections} displays the RC maps of galaxies with no clear or only weak RC emission, which are classified as no/marginal detections.

\section{Spectral fitting models}
\label{model}
In this section, we present the 19 different models, which have been fitted to the dataset for each dwarf galaxy. These models build upon the work of \citet[][]{Galvin_2018} and \citet[][]{Grundy_2025}, incorporating further refinements to the absorption component by accounting for both internal and external free–free absorption.
\subsection{Base Model}
The first two models are the base for all the other models and have already been described in Sect.\ref{radio}. The first is a basic power-law model for synchrotron emission (Eq.\ref{pl}), while the second is a constant model representing the combined contribution of synchrotron and free–free emission (Eq.~\ref{sfg}).
\paragraph{Power-low (\texttt{PL})}
\begin{equation}
S_{\nu} =  A \left( \frac{\nu}{\nu_0} \right)^{\alpha}
\end{equation}

\paragraph{Superposition of synchrotron and free--free emission (\texttt{SFG})}
\begin{equation}
S_{\nu} =  A \left( \frac{\nu}{\nu_0} \right)^{\alpha} + B \left( \frac{\nu}{\nu_0} \right)^{-0.1}
\end{equation}

\subsection{Prefix: Free-free Absorption (\texttt{FFA\_})}
Synchrotron emission can be attenuated by free-free absorption (FFA) when coexisting with ionised gas that produces free-free emission, leading to spectral curvature at low frequencies. The optical depth for absorption is:
\begin{equation}
    \tau_i = \left( \frac{\nu}{\nu_i} \right)^{-2.1}
\end{equation}
We distinguish between internal and external free-free absorption \citep[][]{Tingay_2003}. For internal free-free absorption, the free-free emission and synchrotron emission are both within the same region, while external free-free absorption occurs when synchrotron emission passes through a foreground layer of ionised gas that emits free-free radiation. In this case, the absorbing material is physically separate from the synchrotron-emitting region. The absorption leads to a low-frequency turnover in the spectrum, with the degree of attenuation depending on the density and path length of the ionized gas.  

\paragraph{Free--Free external Absorption + Synchrotron (\texttt{FFA1\_SFG})}
\begin{equation}
S_{\nu} = (1 - e^{-\tau_1}) \left( B + A \left( \frac{\nu}{\nu_1} \right)^{0.1 + \alpha} \right) \left( \frac{\nu}{\nu_1} \right)^2
\end{equation}
Parameters: $A, B, \alpha, \nu_1$

\paragraph{Free--Free internal Absorption + Synchrotron (\texttt{FFA2\_SFG})}
\begin{equation}
S_{\nu} = \frac{(1 - e^{-\tau_1})}{\tau_2} \left( B + A \left( \frac{\nu}{\nu_1} \right)^{0.1 + \alpha} \right) \left( \frac{\nu}{\nu_2} \right)^2
\end{equation}
Parameters: $A, B, \alpha, \nu_2$\\\\
By assuming that the observed emission arises purely from synchrotron radiation affected by free-free absorption, without any contribution from free--free emission, the parameter $B$ can be set to 0. This yields a simplified version of the full model, while still distinguishing between external and internal free-free absorption scenarios.

\paragraph{Free--Free external Absorption (\texttt{FFA1\_PL})}
\begin{equation}
S_{\nu} = (1 - e^{-\tau_1}) A \left( \frac{\nu}{\nu_1} \right)^{0.1 + \alpha} \left( \frac{\nu}{\nu_1} \right)^2
\end{equation}
Parameters: $A, \alpha, \nu_1$

\paragraph{Free--Free internal Absorption (\texttt{FFA2\_PL})}

\begin{equation}
S_{\nu} = \frac{(1 - e^{-\tau_2})}{\tau_2} A \left( \frac{\nu}{\nu_2} \right)^{0.1 + \alpha} \left( \frac{\nu}{\nu_2} \right)^2
\end{equation}
Parameters: $A, \alpha, \nu_1$

\subsection{Suffix: Free-free Absorption (\texttt{\_FFA})}
These models represent synchrotron emission modified by free--free absorption in ionised gas, but without any contribution from free--free emission itself (i.e., $B = D = 0$). The models assume a single population of relativistic electrons embedded in or obscured by thermal plasma.
This first model (\texttt{PL\_FFA1}) describe synchrotron emission passing through an external ionised region with foregound thermal plasma. In the second model (\texttt{PL\_FFA2}), we assume internal free-free absorption, where the synchrotron-emitting and absorbing regions are co-spatial.

\paragraph{Free--Free external Absorption + only Synchrotron (\texttt{PL\_FFA1})}
\begin{equation}
S_{\nu} = A \left( \frac{\nu}{\nu_0} \right)^{\alpha} + (1 - e^{-\tau_1}) \left( C \left( \frac{\nu}{\nu_1} \right)^{0.1 + \alpha} \right) \left( \frac{\nu}{\nu_1} \right)^2
\end{equation}
Parameters: $A, C, \alpha, \nu_1$

\paragraph{Free--Free internal Absorption + only Synchrotron (\texttt{PL\_FFA2})}
\begin{equation}
S_{\nu} = A \left( \frac{\nu}{\nu_0} \right)^{\alpha} + \frac{(1 - e^{-\tau_2})}{\tau_2} \left( C \left( \frac{\nu}{\nu_2} \right)^{0.1 + \alpha} \right) \left( \frac{\nu}{\nu_2} \right)^2
\end{equation}
Parameters: $A, C, \alpha, \nu_2$

This model extends the simple synchrotron–free--free absorption framework by including a free-free emission component in addition to synchrotron radiation and external free--free absorption. The total emission consists of a synchrotron power law, a free-free component, and an externally (\texttt{SFG\_FFA1}) or internally (\texttt{SFG\_FFA2}) absorbed component consisting of synchrotron and thermal emission
\paragraph{Power-Law + Free--Free external Absorption (\texttt{SFG\_FFA1})}
\begin{equation}
S_{\nu} = A \left( \frac{\nu}{\nu_0} \right)^{\alpha} + B \left( \frac{\nu}{\nu_0} \right)^{-0.1}  + (1 - e^{-\tau_2}) \left( D + C \left( \frac{\nu}{\nu_2} \right)^{0.1 + \alpha} \right) \left( \frac{\nu}{\nu_2} \right)^2
\end{equation}
Parameters: $A, B, C, D, \alpha, \nu_2$

\paragraph{Power-Law + Free--Free internal Absorption (\texttt{SFG\_FFA2})}
\begin{equation}
S_{\nu} = A \left( \frac{\nu}{\nu_0} \right)^{\alpha} + B \left( \frac{\nu}{\nu_0} \right)^{-0.1}  + \frac{(1 - e^{-\tau_2})}{\tau_2} \left( D + C \left( \frac{\nu}{\nu_2} \right)^{0.1 + \alpha} \right) \left( \frac{\nu}{\nu_2} \right)^2
\end{equation}
Parameters: $A, B, C, D, \alpha, \nu_2$

This following models introduce two distinct synchrotron-emitting regions, each experiencing different types of free-free absorption. The first component is attenuated by external free--free absorption (\texttt{FFA1}), while the second experiences internal free--free absorption (\texttt{FFA2}). Both components share the same synchrotron spectral index, representing a single electron population. This setup captures systems with multiple star-forming regions or geometrically distinct zones, where only one region becomes optically thick within the observed frequency range.

This dual-FFA model is inspired by studies like \citet[][]{Galvin_2018}, which show that unresolved blends of emission from multiple regions with different physical conditions can reproduce complex radio SEDs with multiple spectral turnovers or inflection points. It is best used when the data suggest two absorption turnovers at distinct frequencies, possibly corresponding to low- and mid-frequency spectral curvature.
\paragraph{Dual Free--Free Absorption + Power Law (\texttt{FFA1\_PL\_FFA2})}
\begin{equation}
S_{\nu} = (1 - e^{-\tau_1}) A \left( \frac{\nu}{\nu_1} \right)^{0.1 + \alpha} \left( \frac{\nu}{\nu_1} \right)^2 + \frac{(1 - e^{-\tau_2})}{\tau_2} C \left( \frac{\nu}{\nu_2} \right)^{0.1 + \alpha} \left( \frac{\nu}{\nu_2} \right)^2
\end{equation}
Parameters: $A, C, \alpha, \nu_1, \nu_2$

The \texttt{FFA1\_SFG\_FFA1} model is a further extension in which both components, each consisting of synchrotron and thermal emission, are absorbed by two distinct external free-free absorption regions. This model accounts for a scenario in which both components undergo absorption, possibly due to layers of ionised gas in complex starburst geometries.

The more advanced model, \texttt{FFA1\_SFG\_FFA2}, allows each component to have an independent synchrotron spectral index, denoted by $\alpha_1$ and $\alpha_2$. This is physically motivated in systems such as post-mergers or interacting galaxies, where newly formed relativistic electrons may have a flatter spectrum while older populations have steeper slopes due to energy losses. This model provides the highest flexibility and is most appropriate when fitting SEDs that require distinct spectral indices to explain high- and low-frequency behavior.

\paragraph{Dual Free--Free Absorption + Power Law (FFA1-SFG-FFA1)}
\begin{equation}
S_{\nu} = (1 - e^{-\tau_1}) \left( B + A \left( \frac{\nu}{\nu_1} \right)^{0.1 + \alpha} \right)\left( \frac{\nu}{\nu_1} \right)^2 + (1 - e^{-\tau_2}) \left( D +C \left( \frac{\nu}{\nu_2} \right)^{0.1 + \alpha} \right) \left( \frac{\nu}{\nu_2} \right)^2
\end{equation}
Parameters: $A, C, \alpha, \nu_1, \nu_2$

\paragraph{Dual Free--Free Absorption + Power Law (FFA1-SFG-FFA2)}
\begin{equation}
S_{\nu} = (1 - e^{-\tau_1}) \left( B + A \left( \frac{\nu}{\nu_1} \right)^{0.1 + \alpha_1} \right)\left( \frac{\nu}{\nu_1} \right)^2 + \frac{(1 - e^{-\tau_2})}{\tau_2} \left( D +C \left( \frac{\nu}{\nu_2} \right)^{0.1 + \alpha_2} \right) \left( \frac{\nu}{\nu_2} \right)^2
\end{equation}
Parameters: $A, B, C, D, \alpha_1, \alpha_2, \nu_1, \nu_2$

\subsection{Suffix: Inverse-Compton Losses (\texttt{\_SIC})}
At high frequencies, synchrotron spectral steepening can occur not due to free-–free absorption, but as a result of energy losses experienced by cosmic ray electrons. These losses are primarily caused by synchrotron radiation and inverse Compton scattering, which occurs when cosmic ray electrons scatter off ambient photons, including those from the cosmic microwave background or the far-infrared radiation field.
Under a continuous injection model, the effect of these losses is a gradual steepening of the synchrotron spectral index by $\Delta \alpha = -0.5$, beginning around a characteristic break frequency $\nu_b$. This model is useful for describing systems where high-frequency spectral curvature is likely due to energy losses rather than thermal absorption.

\paragraph{Synchrotron with Inverse Compton Losses (\texttt{PL\_SIC})}
\begin{equation}
S_{\nu} = \frac{A \left( \frac{\nu}{\nu_0} \right)^{\alpha}}{1 + \left( \frac{\nu}{\nu_b} \right)^{\Delta \alpha}}
\end{equation}
Parameters: $A, \alpha, \nu_b, \Delta \alpha$

This model extends \texttt{PL\_SIC} by including a thermal free--free emission component. It is intended for galaxies where thermal emission contributes at higher frequencies, in addition to synchrotron radiation affected by inverse Compton or synchrotron losses. 

\paragraph{Synchrotron + Inverse Compton (\texttt{SFG\_SIC})}
\begin{equation}
S_{\nu} = \frac{A \left( \frac{\nu}{\nu_0} \right)^{\alpha}}{1 + \left( \frac{\nu}{\nu_b} \right)^{\Delta \alpha}} + B \left( \frac{\nu}{\nu_0} \right)^{-0.1}
\end{equation}
Parameters: $A, B, \alpha, \nu_b, \Delta \alpha$

The following two models combines external (\texttt{FFA1\_PL\_SIC}) and internal (\texttt{FFA2\_PL\_SIC}) free--free absorption with synchrotron emission undergoing high-frequency steepening due to losses.
\paragraph{Free--Free external Absorption + Synchrotron + Inverse Compton (\texttt{FFA1\_PL\_SIC})}
\begin{equation}
S_{\nu} = (1 - e^{-\tau_1}) \left( A \left( \frac{\nu}{\nu_1} \right)^{0.1 + \alpha} \left( \frac{1}{1 + \left( \frac{\nu}{\nu_b} \right)^{\Delta \alpha}}\right) \right) \left( \frac{\nu}{\nu_1} \right)^2
\end{equation}
Parameters: $A, B, \alpha, \nu_1, \nu_b, \Delta \alpha$

\paragraph{Free--Free internal Absorption + Synchrotron + Inverse Compton (\texttt{FFA2\_PL\_SIC})}
\begin{equation}
S_{\nu} = \frac{(1 - e^{-\tau_2})}{\tau_2} \left( A \left( \frac{\nu}{\nu_2} \right)^{0.1 + \alpha} \left( \frac{1}{1 + \left( \frac{\nu}{\nu_b} \right)^{\Delta \alpha}}\right) \right) \left( \frac{\nu}{\nu_2} \right)^2
\end{equation}
Parameters: $A, B, \alpha, \nu_2, \nu_b, \Delta \alpha$

This model extends the \texttt{SFG\_SIC} model by including external \texttt{FFA1\_} and internal \texttt{FFA2\_} free-–free absorption. 

\paragraph{Free--Free external Absorption + Synchrotron + Inverse Compton (\texttt{FFA1\_SFG\_SIC})}
\begin{equation}
S_{\nu} = (1 - e^{-\tau_1}) \left( B + A \left( \frac{\nu}{\nu_1} \right)^{0.1 + \alpha} \frac{1}{1 + \left( \frac{\nu}{\nu_b} \right)^{\Delta \alpha}} \right) \left( \frac{\nu}{\nu_1} \right)^2
\end{equation}
Parameters: $A, B, \alpha, \nu_1, \nu_b, \Delta \alpha$

\paragraph{Free--Free internal Absorption + Synchrotron + Inverse Compton (\texttt{FFA2\_SFG\_SIC})}
\begin{equation}
S_{\nu} = \frac{(1 - e^{-\tau_2})}{\tau_2} \left( B + A \left( \frac{\nu}{\nu_2} \right)^{0.1 + \alpha} \frac{1}{1 + \left( \frac{\nu}{\nu_b} \right)^{\Delta \alpha}} \right) \left( \frac{\nu}{\nu_2} \right)^2
\end{equation}
Parameters: $A, B, \alpha, \nu_2, \nu_b, \Delta \alpha$

The inclusion of both free–free absorption and synchrotron or inverse Compton losses allows for modeling galaxies whose radio SEDs exhibit both low-frequency turnover and high-frequency steepening-features commonly observed in luminous infrared galaxies and starbursts. However, as simultaneous modeling of multiple synchrotron components, each with their own breaks and absorption features, is currently limited by spectral resolution and coverage. Still, models like \texttt{FFA\_SFG\_SIC} provide significant flexibility for interpreting breaks and curvature in observed radio SEDs.

\section{Spectral comparison}
\label{stats}
In this section, we present the parameter values obtained from the spectral models. We provide a comparison between the parameters of the best-fit model and those of a simple power-law model.

\begin{sidewaystable*}[h!]
\centering
\caption{Model comparison between the best-fit model and a simple power-law (\texttt{PL}) model for the CHILLING dwarf galaxies. Reported are $\chi^2$, reduced $\chi^2$, Akaike information criterion (AIC), Bayesian information criterion (BIC), fit parameters (A–D, $\alpha$, $\nu_1$, $\nu_2$, $\nu_b$, $\Delta\alpha$), and the modeled flux at 1.4\,GHz.}
\label{sed_test}
\begin{tabular}{lccccccccccccccc}
\toprule
Name & Model & $\chi^2$ & $\chi^2_\mathrm{red}$ & AIC & BIC & A & B & C & D & $\alpha$ & $\nu_1$ & $\nu_2$ & $\nu_b$ & $\Delta\alpha$ & Flux / mJy \\
\midrule
ESO\,572-G025 & \texttt{FFA2\_SFG\_SIC} & 45.07 & 2.37 & 31.33 & 43.32 & 40.00 & 0.00 & 10.66 & 25.37 & -1.14 & 4.27 & 1.63 & 1.71 & 2.79 & 12.00 \\
              & \texttt{PL}             & 504.98 & 26.58 & 98.98 & 110.97 & 9.14 & 0.25 & 1.67 & 39.20 & -0.67 & 1.32 & 9.16 & 9.22 & 0.01 & 9.14 \\
Fairall\,301  & \texttt{FFA1\_SFG\_SIC} & 82.50 & 3.30 & 48.14 & 61.88 & 15.97 & 1.58 & 38.16 & 27.28 & -0.85 & 0.22 & 9.90 & 6.62 & 0.30 & 3.31 \\
              & \texttt{PL}             & 162.06 & 6.48 & 71.09 & 84.83 & 2.57 & 32.84 & 36.02 & 25.33 & -0.28 & 3.04 & 0.80 & 0.32 & 1.07 & 2.57 \\
IC\,4662      & \texttt{SFG\_SIC}       & 12.02 & 0.57 & -9.45 & 3.16 & 3.25 & 0.00 & 12.07 & 16.68 & -2.39 & 3.19 & 5.21 & 4.16 & 0.03 & 21.76 \\
              & \texttt{PL}             & 15.28 & 0.73 & -2.24 & 10.37 & 22.04 & 13.84 & 6.43 & 16.02 & -0.21 & 9.59 & 2.17 & 0.45 & 4.46 & 22.04 \\
ISZ\,399      & \texttt{SFG\_FFA1}      & 30.29 & 2.02 & 23.58 & 34.19 & 4.71 & 0.00 & 5.06 & 1.76 & -2.09 & 5.31 & 9.24 & 7.52 & 0.33 & 9.84 \\
              & \texttt{PL}             & 39.97 & 2.66 & 30.24 & 40.84 & 10.98 & 1.00 & 39.36 & 1.53 & -0.69 & 6.91 & 0.85 & 2.87 & 1.46 & 10.98 \\
NGC\,244      & \texttt{FFA1\_SFG\_SIC} & 11.74 & 0.49 & -16.10 & -2.63 & 1.15 & 0.32 & 20.45 & 15.35 & -2.36 & 5.29 & 0.63 & 6.46 & 2.00 & 1.57 \\
              & \texttt{PL}             & 32.41 & 1.35 & 17.41 & 30.88 & 1.86 & 3.54 & 20.55 & 37.42 & -0.86 & 7.47 & 1.29 & 8.92 & 1.05 & 1.86 \\
NGC\,625      & \texttt{SFG\_FFA1}      & 17.20 & 0.61 & -10.34 & 4.15 & 0.49 & 0.00 & 9.01 & 9.33 & -1.33 & 0.45 & 1.33 & 5.93 & 0.47 & 10.34 \\
              & \texttt{PL}             & 22.16 & 0.79 & -0.96 & 13.54 & 10.07 & 12.47 & 2.76 & 29.77 & -0.19 & 8.18 & 6.51 & 1.78 & 0.00 & 10.07 \\
NGC\,5253     & \texttt{SFG\_FFA1}      & 1.49 & 0.065 & -80.06 & -66.87 & 17.69 & 29.48 & 23.33 & 21.75 & -2.14 & 5.33 & 5.86 & 6.10 & 1.97 & 73.15 \\
              & \texttt{PL}             & 7.52 & 0.33 & -28.35 & -15.15 & 72.94 & 29.45 & 25.24 & 6.72 & -0.22 & 8.27 & 1.75 & 9.40 & 0.38 & 72.94 \\
\bottomrule
\end{tabular}
\end{sidewaystable*}

\end{appendix}

\label{lastpage}
\end{document}